\documentclass[a4paper, oneside]{article}

\usepackage[utf8]{inputenc}
\usepackage[T1]{fontenc}
\usepackage[english]{babel}
\usepackage[a4paper,margin=25mm]{geometry}
\usepackage{caption} 
\usepackage[colorlinks=true]{hyperref}
\usepackage{stmaryrd,amsmath,amssymb,float,xcolor,tikz}

\title{Characterisation of the Set of Ground States of\\
Uniformly Chaotic Finite-Range Lattice Models\thanks{\acknowledgements}}

\author{
Gayral~Léo\footnote{IMT, Université Toulouse III - Paul Sabatier, Toulouse, France}~\footnote{
\href{mailto:leo.gayral@math.cnrs.fr}{\texttt{leo.gayral@math.cnrs.fr}}\hfill
\href{https://lgayral.pages.math.cnrs.fr/en/}{\texttt{lgayral.pages.math.cnrs.fr/en}}}
\and
Sablik~Mathieu\footnotemark[2]~\footnote{
\href{mailto:msablik@math.univ-toulouse.fr}{\texttt{msablik@math.univ-toulouse.fr}}\hfill
\href{https://www.math.univ-toulouse.fr/\~msablik/index.html}
{\texttt{math.univ-toulouse.fr/\~{}msablik}}}
\and
Taati~Siamak\footnote{Department of Mathematics and CAMS, American University of Beirut, Lebanon}~\footnote{
\href{mailto:siamak.taati@gmail.com}{\texttt{siamak.taati@gmail.com}} \hfill
\href{https://siamak.isoperimetric.info/}{\texttt{siamak.isoperimetric.info}}}
}

\newcommand{\makeabstract}{\abstract{
Chaotic dependence on temperature refers to the phenomenon of divergence of Gibbs measures
as the temperature approaches a certain value.
Models with chaotic behaviour near zero temperature have multiple ground states,
none of which are stable.
We study the class of uniformly chaotic models, that is, those in which,
as the temperature goes to zero,
every choice of Gibbs measures accumulates on the entire set of ground states.
We characterise the possible sets of ground states
of uniformly chaotic finite-range models up to computable homeomorphisms.

Namely, we show that the set of ground states of every model
with finite-range and rational-valued interactions is topologically closed and connected,
and belongs to the class $\Pi_2$ of the arithmetical hierarchy.
Conversely, every $\Pi_2$-computable, topologically closed and connected set of probability measures
can be encoded (via a computable affine homeomorphism) as the set of ground states
of a uniformly chaotic two-dimensional model with finite-range rational-valued interactions.

\medskip \noindent \textbf{Keywords:} \keywords

\medskip \noindent \href{https://ams.org/mathscinet/msc/msc2020.html}{
\textbf{MSC2020 subject classifications:}} \msc
}}

\newcommand{\keywords}{Gibbs Measures, Chaotic Temperature Dependence, Ground States, Aperiodic Tiles,
Computability, Arithmetical Hierarchy.}

\newcommand{\msc}{Primary 82B20, 37D35, 68Q17; Secondary 68Q04, 68Q87, 37B51, 05B45.}

\newcommand{\acknowledgements}{%
We thank Aernout van~Enter and Jean-René Chazottes for useful comments on an earlier version of this manuscript.
S.~Taati would also like to thank A.~van~Enter for many helpful discussions.
M.~Sablik acknowledges the support of the ANR “Difference” project (ANR-20-CE40-0002) and the CIMI LabEx “Computability of asymptotic properties of dynamical systems” project (ANR-11-LABX-0040).
S.~Taati acknowledges the support of the Université Toulouse III - Paul Sabatier
and the Center for Advanced Mathematical Sciences, AUB.
}

\date{}

\newcommand{\ie}{\emph{i.e.}\ }
\newcommand{\eg}{\emph{e.g.}\ }

\DeclareMathOperator*{\Acc}{Acc}
\newcommand{\acc}[2][]{\Acc_{#1}\left(#2\right)}

\newcommand{\diam}[1]{\mathrm{diam}\left(#1\right)}

\renewcommand{\epsilon}{\varepsilon}
\renewcommand{\phi}{\varphi}

\usepackage{amssymb}
\newcommand{\A}{\mathcal{A}}
\newcommand{\F}{\mathcal{F}}
\newcommand{\M}{\mathcal{M}}
\newcommand{\G}{\mathcal{G}_\sigma}

\usepackage{dsfont}
\newcommand{\N}{\mathds{N}}
\newcommand{\Z}{\mathds{Z}}
\newcommand{\Q}{\mathds{Q}}
\newcommand{\R}{\mathds{R}}

\newcommand{\dense}{\mathfrak{D}}
\newcommand{\per}{\mathfrak{P}}

\newcommand{\btot}{b_{\mathrm{total}}}
\newcommand{\bread}{b_{\mathrm{read}}}

\newcommand{\perim}{\mathrm{per}}
\newcommand{\freq}[2]{\mathrm{freq}_{#1}^{#2}}

\usepackage{mathtools}
\DeclarePairedDelimiter\abs{|}{|}
\DeclarePairedDelimiter\norm{\Vert}{\Vert}

\newcommand{\symb}[1]{\mathtt{#1}}
\newcommand{\pmo}{ {\mathrm{\normalfont\texttt{{±}}}}\symb1 }
\newcommand{\po}{ {\mathrm{\normalfont\texttt{+}}}\symb1 }
\newcommand{\mo}{ {\mathrm{\normalfont\texttt{-}}}\symb1 }

\newcommand{\eqnsp}{\,}

\usepackage{amsthm,thmtools}
\theoremstyle{plain}
\newtheorem*{maintheorem}{Theorem}
\newtheorem{theorem}{Theorem}
\newtheorem{proposition}[theorem]{Proposition}
\newtheorem{corollary}[theorem]{Corollary}
\newtheorem{lemma}[theorem]{Lemma}
\theoremstyle{definition}
\declaretheorem[sibling=theorem,name=Remark,qed=$\blacksquare$]{remark}
\declaretheorem[sibling=theorem,name=Definition,qed=$\blacksquare$]{definition}

\newlength{\indexalign}
\newlength{\pagealign}
\begin{document}

\maketitle
\makeabstract

\section{Introduction}

\subsection{Background}

In equilibrium statistical mechanics, \emph{chaotic dependence on temperature} refers
to the phenomenon of divergence of Gibbs measures
as the temperature approaches a certain value.
In a thermodynamic system,
chaotic behaviour near a temperature $T_0$ would correspond to volatile variations
in the macroscopic state of the system
(\eg non-vanishing changes in the color of a chemical solution,
or in the magnetisation of an alloy) 
when the temperature is varied towards $T_0$.
The possibility of such a behaviour has been mathematically demonstrated
in a number of lattice models.

The first example of a model with chaotic temperature dependence near zero temperature
was found by van~Enter and Ruszel~\cite{EntRus07}.
Their model consisted of \emph{continuous spins} (taking values from the unit circle)
on a $d$-dimensional lattice with $d\geq 1$,
interacting via a \emph{nearest-neighbour} potential with a pair of infinitely nested wells.
They showed that, as the temperature goes to zero,
the Gibbs measures of the model cannot converge.
More specifically, as $T\searrow 0$,
the system undergoes an infinite sequence of phase transitions,
alternating between ferromagnetic and antiferromagnetic states.

The second example, upon which is based the current paper,
is due to Chazottes and Hochman~\cite{ChaHo10}.
They constructed a \emph{one-dimensional} model with discrete spins
(in the binary alphabet $\{\symb0,\symb1\}$) and \emph{infinite-range}
but \emph{exponentially decaying} interactions
(\ie a Lipschitz-continuous potential function)
for which the shift-invariant Gibbs measures do not converge as $T\searrow 0$.
Using a technique from multidimensional symbolic dynamics~\cite{Hochman09},
they were able to ``simulate'' the long-range one-dimensional model
with a \emph{three-dimensional} model,
thus obtaining a model with a \emph{finite} alphabet and \emph{finite-range} interactions
in which the shift-invariant Gibbs measures do not converge.

Other examples of chaotic temperature dependence were given by
Coronel and Rivera-Letelier~\cite{CorRiv15},
and by Bissacot, Garibaldi and Thieullen~\cite{BiGaThi18}.
Coronel and Rivera-Letelier also studied the related phenomenon
of \emph{sensitive dependence on temperature}~\cite{CorRiv15,CoRi19}.

In \emph{one-dimensional} models with \emph{finite} ``spin'' alphabets and
\emph{finite-range} interactions, the Gibbs measures are known to always converge
as $T\searrow 0$~\cite{Bremont03,Leplaideur05,ChaGamUg11}.
This result holds even in the presence of some hard constraints,
namely when the configurations are restricted to a mixing subshift of finite type
rather than a full shift.
Contrasting the one-dimensional case with
the three-dimensional construction of Chazottes and Hochman, 
one may then wonder whether chaotic temperature dependence
can occur in a two-dimensional model
with a finite alphabet and finite-range interactions.
This question was recently settled independently by Chazottes and Shinoda~\cite{ChaShi20}
and Barbieri, Bissacot, Vedove and Thieullen~\cite{BaBiVeThi22},
who managed to adapt the construction of Chazottes and Hochman to two dimensions.

The three-dimensional example of Chazottes and Hochman
relies on a result of Hochman~\cite{Hochman09} 
from symbolic dynamics that states that every one-dimensional effectively closed subshift
can be simulated by a three-dimensional subshift of finite type.
A \emph{subshift} refers to a configuration space that can be specified
by forbidding a collection of finite patterns.
If the defining forbidden patterns can be enumerated by an algorithm,
the subshift is said to be \emph{effectively closed}.
A subshift \emph{of finite type} is a subshift that can be defined
by a finite set of forbidden patterns.
The notion of \emph{simulation} in the latter result simply means that
the configurations of the simulated 1D subshift are precisely the rows of symbols
appearing in the configurations of the simulating 3D subshift.
The power of this result stems from the fact that, in practice,
most mathematically constructed subshifts
(\eg the one used in the 1D construction of Chazottes and Hochman)
can be easily shown to be effectively closed.

Shortly after Hochman's paper appeared, Durand, Romashchenko and Shen~\cite{DuRoShe10}
and Aubrun and Sablik~\cite{AuSa13} independently managed to improve his result
by proving that every effectively closed 1D subshift
can be simulated by a 2D subshift of finite type.
The recent constructions of two-dimensional chaotic models
rely on these improved simulation results, 
with Chazottes and Shinoda using the construction by Durand, Romashchenko and Shen,
and Barbieri, Bissacot, Vedove and Thieullen using the construction by Aubrun and Sablik.

Here, we use the machinery of aperiodic Wang tiles to give a direct construction
of a family of chaotic two-dimensional models
with finite alphabets and finite-range interactions.
While in the earlier constructions, chaos appears in the form
of oscillation between two distinct phases
(\eg between ferromagnetic and antiferromagnetic phases)
as the temperature is decreased to zero,
our method allows for much more complex trajectories.
Indeed, the computational power of the tiles makes it possible to directly control
the distribution of patterns within overlapping temperature intervals,
hence providing a lot of flexibility with regards to the set of ground states.
In fact, our construction is flexible enough that
it allows us to characterise the topological
and computational nature of the set of ground states in \emph{uniformly} chaotic models.

\subsection{Main Result}

The focus of this paper is on two-dimensional models with finite ``spin'' alphabet
and shift-invariant finite-range interactions,
although our construction would work equally well in higher dimensions
and our upper bounds apply to all computable
(finite- or infinite-range) interactions in any dimension.

The equilibrium states of the system at inverse temperature $\beta:=1/T$,
are represented by the shift-invariant probability measures $\mu$ that maximise the pressure
\[
h(\mu) - \beta\mu(\phi) \eqnsp ,
\]
where $h(\mu)$ denotes the entropy per site and $\phi$ denotes the potential
associated to the interactions, indicating the energy contribution of the site at the origin.
For the class of absolutely summable interactions (finite-range interactions in particular),
the equilibrium states are precisely
the Gibbs measures that are shift-invariant~\cite[Theorem 4.2]{Ruelle04}.
The set of all such measures will be denoted by $\G(\beta)$.

The model is said to have \emph{chaotic} temperature dependence near zero temperature
if every family $(\mu_\beta)$ of choices $\mu_\beta\in\G(\beta)$
diverges (in the weak-$*$ topology) as $\beta\to\infty$.
We call a family $(\mu_\beta)$ a (cooling) \emph{trajectory}.
Note that, due to compactness, every trajectory has a convergent subsequence.
The limit of any such subsequence is called a \emph{ground state}.
In other words, the ground states of the model are the accumulation points
of $\G(\beta)$ as $\beta\to\infty$.\footnote{
The term ``ground state'' is used inconsistently in the literature,
and we redirect the reader to a comparison
of various related notions~\cite[Appendix~B.2]{EntFerSok93}.
As an example, in our terminology, the all-plus and all-minus measures are considered
to be ground states in the two-dimensional Ising model
but not in the one-dimensional Ising model.
}
The set of ground states will be denoted by $\G(\infty)$.

Characterising the set of ground states, and their stability at positive temperature,
is one of the fundamental problems in equilibrium statistical mechanics
(see some recent articles~\cite{Geo11,FriVel18} for examples).
A classic example is the Ising model of ferromagnetism, which
has two distinct, mutually singular, stable ground states corresponding
to spontaneous magnetisation in upward and downward directions.
We say that a ground state $\nu$ is (weakly) \emph{stable} if
for every open neighbourhood $\mathcal{E}\ni\nu$
and every sufficiently large inverse temperature $\beta$,
there exists a Gibbs measure $\mu_\beta\in\G(\beta)$ that lies in $\mathcal{E}$.
A model thus has chaotic behaviour near zero temperature
if none of its ground states is stable.

Our main result concerns a stronger form of chaos which we term \emph{uniform chaos}.
We say that a model is \emph{uniformly chaotic} near zero temperature if it has
at least two ground states and every cooling trajectory $(\mu_\beta)$ has
the entire set $\G(\infty)$ of ground states as its accumulation points.
Thus, in a uniformly chaotic system, as $\beta\to\infty$,
every trajectory gets arbitrarily close to every ground state.

There are restrictions on the complexity and topological nature of
the set $\G(\infty)$ of ground states in a uniformly chaotic system.
It turns out that in every uniformly chaotic model,
the set $\G(\infty)$ is closed and connected (with respect to the weak-$*$ topology).
Furthermore, if the interactions are computable,
then $\G(\infty)$ is $\Pi_2$-computable (notions from computable analysis will be
introduced in Section~\ref{sec:constraints:computable-framework}).
We show that these obstructions in fact characterise
the possible sets of ground states of uniformly chaotic models.
Namely, every closed, connected and $\Pi_2$-computable set of measures can be realised
(through a computable and affine homeomorphism)
as the set of ground states of a uniformly chaotic model
with computable finite-range interactions.

\begin{maintheorem}
In every uniformly chaotic model with computable finite-range interactions,
the set of ground states $\G(\infty)$ is closed, connected and $\Pi_2$-computable.

Conversely, given a closed, connected and $\Pi_2$-computable set $K$
of probability measures on $\{\pmo\}^\N$,
there exists a two-dimensional uniformly chaotic model with
computable (in fact, zero-one valued) finite-range interactions
for which $\G(\infty)$ is computably and affinely homeomorphic to $K$.  
\end{maintheorem}

The first part of the theorem corresponds to the results of Section~\ref{sec:constraints},
in particular Proposition~\ref{prop:ground-states:connected}
and Theorem~\ref{thm:pi2-upper-bound}.
The computable affine homeomorphism mentioned in the second part
can be chosen independently of $K$. 
More specifically, we construct a subshift of finite type $X_{\F_0}$
whose ergodic measures encode
(in a computable and one-to-one fashion)
the elements of $\{\pmo\}^\N$ (see Proposition~\ref{prop:bicomputable-map}).
Given a closed, connected, and $\Pi_2$-computable set $K'$
of shift-invariant measures on $X_{\F_0}$,
we extend the alphabet of $X_{\F_0}$ and construct an interaction on this extended alphabet
that has $K'$ as its set of ground states (Theorem~\ref{thm:pi2-lower-bound}).
For models that may be chaotic but are not uniform,
we will prove a looser $\Pi_3$ upper bound
on the computational complexity of $\G(\infty)$,
but we are unable to provide a general construction achieving that bound.

The reader may find the requirement of the computability of the interactions to be
artificial and physically unjustified.
We offer two responses to such a criticism:
The short and somewhat evasive response is that the computability of the interactions
can be replaced with the assumption that the interactions have rational values.
While the latter assumption may still be physically unjustified,
it is \emph{a priori} void of any connection with computability.
The better but more technical response is that the computability of the interactions
is only a technical assumption to make its values accessible to algorithms.
An alternative possibility for the upper bound is to make
the values of the interactions accessible via an oracle.
The $\Pi_2$-computability of $\G(\infty)$ would then be relative to such an oracle.
For the lower bound, we do not specifically need computable potentials,
and any potential corresponding to the same finite-range interactions
would give rise to the same ``spontaneous computations'' within Gibbs measures.

While computability theory is already exploited in the three-dimensional construction
of Chazottes and Hochman (and its sequels),
it appears there only as an elegant tool to construct complex examples.
Our result shows that the notion of computability is essential in understanding
the phenomenon of chaotic temperature dependence.

The importance of computability considerations in characterising the possible values
of asymptotic invariants in discrete dynamical systems
has now been established for more than a decade.
A famous result states that the set of entropies of multidimensional subshifts of finite type
is exactly the set of real numbers that are right-recursively enumerable~\cite{HoMe10}
(also called $\Pi_1$-computable, in reference to the arithmetical hierarchy).
Characterisations of similar nature have been obtained for various objects,
including some growth-type invariants
of multidimensional subshifts of finite type~\cite{Meyer11},
the possible sets of periods of multidimensional subshifts of finite type~\cite{JeanVa14},
the sets of limit measures of cellular automata iterated on an initial measure~\cite{HeSa18},
and the $\mu$-limit sets~\cite{BoDePoSaTh15}
and the generic limit sets~\cite{EsNuTor23} of cellular automata.

Let us conclude this section by mentioning two of our side results
that could be of independent interest.
First, we show that, as a function of the inverse temperature,
the topological pressure is computable (Proposition~\ref{prop:top-pressure:computable}).
This is true in any dimension, for models in which the configuration space is a full shift.
In the one-dimensional setting, the computability of the topological pressure
on subshifts of finite type and other classes of subshifts
has already been established~\cite{Spandl08,BuDaWoYa22}.
As a second byproduct of our main result,
we prove that the question of whether
a given finite-range interaction in two dimension is chaotic
is undecidable (Theorem~\ref{thm:chaos:complexity}).
We expect that, in two or more dimensions,
most questions regarding the thermodynamic behaviour of finite-range models
at or near zero temperature are undecidable.
See the recent work of Burr and Wolf~\cite{BurWol20} for results in this direction
but in the setting of one-dimensional subshifts of finite type.

\subsection{Overview of the Paper}

Given the intricate nature of the main result,
each of the following sections will focus on a key idea.
We have made an effort to make each section as self-contained as possible
by using the results of the other sections as black boxes and restating them when possible.

\begin{itemize}
    \item In Section~\ref{sec:framework}, we introduce our notations and terminology.
    In particular, we present the setting of the lattice models we will work in,
    and formulate the notions of ground states, chaos and stability in more details.
    We also recall the notion of Wang tiles, which will be the main tool in our construction.
    
    \item In Section~\ref{sec:constraints},
    we establish some obstructions on the set of ground states $\G(\infty)$.
    First, we prove that for every continuous potential,
    the set of ground states $\G(\infty)$
    is compact and connected in the weak-$*$ topology.
    Then, using basic tools from computable analysis,
    we prove that for every computable potential,
    the pressure function is computable,
    the set $\G(\beta)$ of equilibrium states
    at any given temperature belongs to the class $\Pi_1$,
    and the set of ground states $\G(\infty)$ belongs
    to the class $\Pi_3$ of the arithmetic hierarchy.
    In the case of a uniformly chaotic model,
    we use another approach to establish a $\Pi_2$ upper bound
    on the complexity of $\G(\infty)$.
    
    \item In Section~\ref{sec:uniform-markers}, we translate the key ideas
    of Chazottes and Hochman~\cite{ChaHo10} into a broader framework.
    The key black-box result of this section gives a local description of the Gibbs measures
    of a certain class of finite-range models at moderately low temperatures.
    Namely, it states that if the ground patterns of a model
    satisfy a certain entropy condition,
    then within a certain temperature range,
    every shift-invariant Gibbs measure induces,
    with high probability, a grid of \emph{markers}
    chosen from a set~$Q$,
    the relative frequencies of which are almost uniform over $Q$.

    \item In Section~\ref{sec:robinson}, we construct a class of tilesets
    (with the associated potentials) for which we have a tight control
    on the entropy of a family of marker sets $\left(Q_k\right)$.
    Built upon the famous Robinson tileset~\cite{Rob71},
    our tilesets use embedded unary computations
    (as described in Appendix~\ref{appendix:unary})
    and odometers to create a well-controlled structure,
    and then use a certain class of fast-enough Turing machines
    to force distributions of words through that structure.
    
    \item In Section~\ref{sec:chaos}, we merge the results from the previous two sections
    to conclude that, for this class of tilesets,
    the Gibbs measures accumulate to $\G(\infty)$ uniformly.
    We also identify the Gibbs measures of these model
    with measures on infinite binary strings,
    forced by the simulated Turing machines.
    
    \item In Section~\ref{sec:measures}, we show that every connected $\Pi_2$ set of measures
    on infinite strings can be realised as the set of accumulation points
    of a sequence computed by a fast-enough Turing machine.
    In particular, whenever $\G(\infty)$ is not a singleton,
    this provides an example of uniformly chaotic system.
    Lastly, we use this construction to prove that the question of chaoticity is undecidable.
    
    \item To facilitate the reading of this article, we include a table of notations
    in Appendix~\ref{appendix:notations}.
\end{itemize}

\newpage
\tableofcontents
\newpage

\section{Setting} \label{sec:framework}

Our main interest lies in two-dimensional lattice models
with finite ``spin'' alphabets and finite-range interactions.
Nevertheless, the results of Section~\ref{sec:constraints} work
in any dimension and for a very general class of interactions,
and the construction presented in Sections~\ref{sec:robinson}--\ref{sec:measures}
can be easily adapted to higher dimensions.
Likewise, the results of Section~\ref{sec:uniform-markers} apply
to finite-range models in any dimension.

\subsection{Configurations and Measures}

A microscopic \emph{configuration} of the model
is an assignment $x:\Z^d\to\A$ of a symbol from a finite alphabet $\A$
to each site of the $d$-dimensional lattice $\Z^d$.
The restriction of a configuration $x$ to a set $I\subseteq\Z^d$ is denoted by $x_I$.
A partial configuration $w\in\A^I$
with finite support $I\Subset\Z^d$ is called a \emph{pattern}.
A finite set $I\Subset\Z^d$ is called a \emph{window}.
We will use the notation $I_n=\llbracket 0,n-1\rrbracket^d$ for hypercubic windows.
The \emph{diameter} of a window $I$ is the smallest $r$
such that $I\subseteq k+I_{r+1}$ for some $k\in\Z^d$.

The set $\Omega_\A:=\A^{\Z^d}$ of configurations is equipped with the product topology,
which is compact and metrizable.
A basis for the product topology is given by the \emph{cylinder} sets, that is,
the sets of the form $[w]=\left\{y\in\Omega_\A, y_I=w\right\}$
where $w\in\A^I$ is a pattern on a window $I$.
When $Q$ is a set of patterns,
we use the notation $[Q]$ for the union of the cylinders $[w]$ with $w\in Q$.

A Borel probability measure on $\Omega_\A$ is uniquely determined
by the probabilities it associates to the cylinders.
The set of all Borel probability measures on $\Omega_\A$
will be denoted by $\M\left(\Omega_\A\right)$.
More generally, given a topological space $X$,
we write $\M(X)$ to denote the set of Borel probability measures on $X$.

We endow $\M\left(\Omega_\A\right)$ with the \emph{weak-$*$} topology.
This is a compact and metrizable topology.
In the weak-$*$ topology, a sequence of measures $\mu_n\in\M\left(\Omega_\A\right)$
converges to a limit measure $\mu\in\M\left(\Omega_\A\right)$
if $\mu_n([w])\to\mu([w])$ for every cylinder $[w]$.
A (non-canonical) metric defining the weak-$*$ topology
on $\M\left(\Omega_\A\right)$ is given by:
\[
    d^*(\mu,\nu):=\sum_{n=1}^\infty \frac{1}{2^{n+1}}d_n(\mu,\nu) \eqnsp ,
\]
where $d_n(\mu,\nu):=\sum_{w\in\A^{I_n}}\abs[\big]{\mu([w])-\nu([w])}$.
We will use the notation $B(\mu,r)$ for the open ball of radius $r$ around $\mu$,
and likewise $\overline{B(\mu,r)}$ for the closed ball.

\subsection{Subshifts and Wang Tiles}

We write $\sigma^k(x)$ for the \emph{translation} (or \emph{shift})
of a configuration $x$ by $k\in\Z^d$.
More specifically, $y=\sigma^k(x)$ is
a configuration satisfying $y_i=x_{k+i}$ for every $i\in\Z^d$.
The shift maps are continuous and define an action of $\Z^d$ on $\A^\Z$.

The set $\Omega_\A$
(or rather, the dynamical system defined by the shift action on $\Omega_\A$)
is called a \emph{full shift}.
We may write $\Omega$ instead of $\Omega_\A$
when there is no ambiguity regarding the alphabet $\A$.
A closed and shift-invariant subset of a full shift is called a \emph{subshift}.
Given a set of patterns $\F$, one can define a subshift $X_\F$ that consists of
all configurations that avoid (all translations of) the elements of $\F$, that is:
\[
    X_\F=\left\{x\in\A^\Z,\forall k\in\Z^d,\forall I\Subset\Z^d,
    \sigma^k(x)_I\notin\F\right\} \eqnsp .
\]
The patterns in $\F$ are called the \emph{forbidden} patterns defining $X_\F$.
A subshift \emph{of finite type} (\emph{SFT} for short)
is a subshift that can be defined by a finite set of forbidden patterns.
A pattern $w$ is said to be \emph{locally admissible}
with respect to a set of forbidden patterns $\F$ if
none of its subpatterns is (a translation of) an element of $\F$.
A \emph{globally admissible} pattern with respect to $\F$ is a pattern
that can be extended to a configuration of the subshift $X_\F$.

A convenient way to describe forbidden patterns is using the language of \emph{tilings}.
A (Wang) \emph{tile} is a unit square with possible bumps, dents and decorations.
A finite set of tiles is simply referred to as a \emph{tileset}.
An example of a tileset which will play a key role in our construction
is depicted in Figure~\ref{fig:RobEnhanced}.
Viewed as an alphabet, a tileset comes with a finite set of forbidden patterns,
namely pairs of (horizontally, vertically, or diagonally) adjacent tiles
whose bumps, dents or decorations do not match.
Every tileset thus naturally defines an SFT.
The configurations of this SFT are also referred to as \emph{valid tilings}.

We recall two features of Wang tiles that play crucial roles
in multi-dimensional symbolic dynamics.
First, there exist tilesets (such as the one in Figure~\ref{fig:RobEnhanced})
that are \emph{aperiodic}, meaning that they admit valid tilings but no periodic ones.
Second, one can design the bumps, dents and decorations in such a way that valid tilings
(or locally admissible patterns) simulate Turing machines.
We will further discuss these features in Section~\ref{sec:robinson}
and exploit them in our construction.
We refer the reader to dedicated books~\cite{GrSh87,JeanVa20}
for more background on the theory of Wang tiles.

Given a subshift $X\subseteq\Omega_\A$,
we identify $\M(X)$ with the subset of $\M\left(\Omega_\A\right)$
consisting of the measures supported by $X$.
The restriction of $\M(X)$ to shift-invariant measures will be denoted by $\M_\sigma(X)$.

\subsection{Interactions and Potentials}

An \emph{interaction} is a family $\Phi=(\Phi_I)_{I\Subset\Z^d}$
of maps $\Phi_I:\Omega_\A\to\R$ such that $\Phi_I(x)$ depends only on the restriction $x_I$.
The value $\Phi_I(x)$ is interpreted as the energy contribution of the pattern $x_I$.
The total \emph{energy content} of a window $J\Subset\Z^d$
is the sum $E_J(x)=\sum_{I\subseteq J}\Phi_I(x)$.
We restrict ourselves to interactions that are \emph{shift-invariant},
meaning $\Phi_{k+I}(x)=\Phi_I(\sigma^k (x))$ for every $I\Subset\Z^d$ and $k\in\Z^d$.
For brevity, we drop the adjective \emph{shift-invariant} when talking about interactions.

An interaction $\Phi$ is said to have \emph{finite range} if there exists an $r\in\N$
such that $\Phi_I= 0$ for every $I$ with a diameter larger than $r$.
Such an $r$ is referred to as the \emph{range} of the interaction.
A broader family of interactions are those that are \emph{absolutely summable}, that is:
\[
    \sum_{I\ni 0} \norm{\Phi_I} < \infty \eqnsp ,
\]
where $\norm{\cdot}$ denotes the uniform norm.
The latter condition covers (more or less)
all ``physically relevant'' interactions~\cite{EntFerSok93}.

By a \emph{potential} we shall mean a map $\phi:\Omega_\A\to\R$.
Every absolutely summable interaction $\Phi$
(more generally, every interaction satisfying a weaker summability condition)
defines a continuous potential $\phi$ where
\[
    \phi(x):=\sum_{I\ni 0} \frac{1}{\abs{I}}\Phi_I(x) \eqnsp .
\]
The value of $\phi$ is interpreted as the energy contribution of the site at the origin.
The potential associated to a finite-range interaction is a \emph{local} function,
that is, there exists a finite window $D\Subset\Z^d$
such that the restriction $x_D$ uniquely determines $\phi(x)$.

Given any finite set of forbidden patterns $\F$,
one can define a non-negative finite-range interaction $\Phi=(\Phi_I)_{I\Subset\Z^d}$
that assigns energy $1$ to each \emph{fault}, that is,
\[
    \Phi_I(x)=
    \begin{cases}
        1   & \text{if $x_I$ is a translation of an element of $\F$,} \\
        0   & \text{otherwise.}
    \end{cases}
\]
We call this the \emph{fault interaction} corresponding to $\F$.
The potential associated to $\Phi$ will be denoted by $\phi_\F$
and is called the \emph{fault potential} associated to $\F$.
Fault interactions have earlier been exploited
in models of quasicrystals~\cite{Radin86,Miekisz90,Miekisz97,LeuPar00}.
The interaction we will use in our construction is of this type.

\subsection{Pressure and Equilibrium States}

Let $H(q)=-\sum_{s\in S} q(s)\log_2(q(s))$ be the \emph{entropy}
of a probability distribution $q$ on a finite set $S$.
We will measure entropies in bits, hence the base-$2$ logarithm in the definition of $H(q)$.
The \emph{entropy per site} (\ie the Kolmogorov--Sinai entropy)
of a shift-invariant measure $\mu\in\M_\sigma\left(\Omega_\A\right)$ is:
\[
    h(\mu)=\lim_{n\to\infty}\frac{1}{\abs{I_n}}H\left(\mu_{I_n}\right)
    = \inf_{n\in\N} \frac{1}{\abs{I_n}}H\left(\mu_{I_n}\right) \eqnsp ,
\]
where $\mu_{I_n}$ denotes the projection of $\mu$ on the patterns on $I_n$,
such that $\mu_{I_n}(w)=\mu([w])$ for $w\in\A^{I_n}$.

Consider a model with a continuous potential $\phi$.
According to the variational principle of equilibrium statistical mechanics,
the \emph{equilibrium states} of the model at \emph{inverse temperature} $\beta$
are described by the shift-invariant measures $\mu\in\A^{Z^d}$
that maximise the \emph{pressure}
\[
    p_\mu(\beta) := h(\mu) - \beta\mu(\phi) \eqnsp .
\]
The maximal value, denoted $p(\beta)$, is called the \emph{topological pressure}.
The Dobrushin--Lanford--Ruelle theorem~\cite[Theorem 4.2]{Ruelle04} tells us that,
if $\phi$ is the potential associated to an absolutely summable interaction $\Phi$,
then the equilibrium states of $\phi$
coincide with the shift-invariant Gibbs measures for~$\Phi$.
We will denote the set of all equilibrium states
at inverse temperature $\beta$ by $\G(\beta)$.
We will use this notation even when $\varphi$ is
not associated to an absolutely summable potential.

We refer the reader to classical books~\cite{Walt82,DeGriSi76,Ruelle04}
for a broader background on ergodic theory and the thermodynamic formalism.

\subsection{Ground States, Stability and Chaos} \label{sec:framework:chaos}

An accumulation point of equilibrium states
as $\beta\to\infty$ is called a \emph{ground state}.
In other words, a measure $\nu\in\M_\sigma(\Omega_\A)$ is a ground state
if there exists a sequence of $\beta_n\to\infty$
and equilibrium states $\mu_n\in\G(\beta_n)$ such that $\mu_n\to\nu$.
For instance, the two Dirac measures concentrated
at the all-plus and all-minus configurations are ground states
in the two-dimensional Ising model but not in the one-dimensional model.
We denote the set of all ground states by $\G(\infty)$.
By compactness of $\M\left(\Omega_\A\right)$, the set $\G(\infty)$ is non-empty.

It is easy to see that the ground states attain
the minimum expected energy per site $\nu\mapsto\nu(\varphi)$.
Furthermore, among all the minimum energy measures,
the ground states have the highest entropy per site.
However, as the one-dimensional Ising model illustrates,
not every maximum entropy, minimum energy measure is a ground state.

We say that a ground state $\nu$ is (weakly) \emph{stable} if
one can choose an equilibrium state $\mu_\beta$, for each $\beta>0$,
in such a way that $\mu_\beta\to\nu$ as $\beta\to\infty$.
The all-plus and all-minus measures in the two-dimensional Ising model
are examples of stable ground states.
A family $(\mu_\beta)$ identifies a \emph{trajectory} of the system
as the temperature goes to zero.

A model is said to have \emph{chaotic} temperature dependence near zero temperature
if none of its ground states are stable.
A \emph{uniform} model is one in which every trajectory $(\mu_\beta)$
has the entire set $\G(\infty)$ of ground states as its accumulation points.
A uniform model with at least two ground states is clearly chaotic.
We call such a model a \emph{uniformly chaotic} model.

Not every chaotic model is uniformly chaotic.
As a simple example, consider the direct product of a two-dimensional
(uniformly or non-uniformly) chaotic model with the Ising model.
The chaotic component ensures the new model is chaotic,
while the Ising component ensures that the new model is not uniform.
Namely, the trajectory of the Gibbs measures with the plus boundary condition
converges to the all-plus measure thus does not have the all-minus measure
as an accumulation point, and conversely.\footnote{
We thank A.~van~Enter for pointing out this example.
}

Given the common occurrence of accumulation points in this article,
we will use the following notation.
The set of accumulation points of a sequence $(x_n)$ of points in a metric space
will be denoted by $\acc[n\to\infty]{x_n}$.
Likewise, we write $\acc[n\to\infty]{M_n}$ for the set of the accumulation points
of a sequence $(M_n)$ of subsets of a metric space.
We use a similar notation for families indexed by real numbers rather than integers:
\[
\begin{split}
    \acc[\beta\to\infty]{x_\beta} &=
    \left\{y,\forall\epsilon>0,\forall\beta_0>0,\exists\beta\geq\beta_0,
    d\left(x_\beta,y\right)<\epsilon\right\} \eqnsp ,\\
    \acc[\beta\to\infty]{M_\beta} &=
    \left\{y,\forall\epsilon>0,\forall\beta_0>0,\exists\beta\geq\beta_0,
    \exists x\in M_\beta,d(x,y)<\epsilon\right\} \eqnsp .
\end{split}
\]

With this notation, $\G(\infty)=\acc[\beta\to\infty]{\G(\beta)}$, and the model is uniform
if and only if, for every cooling trajectory $(\mu_\beta)$,
we have $\G(\infty)=\acc[\beta\to\infty]{\mu_\beta}$.

\section{Topology and Computational Complexity of \texorpdfstring{$\G(\infty)$}{G\_σ(∞)}}
\label{sec:constraints}

The goal of this section is two-fold.
First, we describe the topology of $\G(\infty)$.
We prove that, in every model, $\G(\infty)$ is topologically connected,
and in every uniform model, one can find a sequence of inverse temperatures 
$\beta_n\to\infty$ such that $\G(\infty)=\acc{\G\left(\beta_n\right)}$.
Second, we obtain a computational upper bound on $\G(\infty)$.
To this end, we first introduce an appropriate framework
for computational analysis in the current setting.
Assuming the potential $\phi:\Omega_\A\to\R$ is a computable function,
we the prove that $\G(\infty)$ is a $\Pi_3$-computable set.
Moreover, in case of a uniform model,
we sharpen the latter upper bound to $\Pi_2$-computability.

\subsection{Topological Properties of \texorpdfstring{$\G(\infty)$}{G\_σ(∞)}}

\begin{lemma}[Weak Continuity of Gibbs Measures]
\label{lem:weak-continuity}
Let $(\beta_n)$ be a sequence of inverse temperatures and $(\mu_n)$ be
a sequence of Gibbs measures with $\mu_n\in\G\left(\beta_n\right)$.
If $\beta_n\to\beta\in\R^+$ and $\mu_n\to\mu$, then $\mu\in\G(\beta)$.
In particular, for any $\beta>0$:
\[
\bigcap\limits_{\epsilon>0}\overline{\bigcup\limits_{\substack{\abs{\beta'-\beta}
\leq\epsilon\\ \beta'\neq\beta}}\G\left(\beta'\right)}\subseteq \G(\beta) \eqnsp .
\]
In other words, $\beta\mapsto \G(\beta)$ is 
upper semi-continuous with respect to the Hausdorff distance
on closed subsets of $\M_\sigma\left(\Omega_\A\right)$
and the partial order given by inclusion.

\begin{proof}
The first statement is standard, but we include an argument for completeness.
Consider $\mu'\in\G(\beta)$.  Note that, for every $n$,
\[
h(\mu_n) - \beta_n\mu_n(\phi) \geq
h(\mu') - \beta_n\mu'(\phi) \eqnsp .
\]
Taking the limit superior of both sides as $n\to\infty$
and using the upper semi-continuity of $\mu\mapsto h(\mu)$
and the continuity of $\mu\mapsto\mu(\phi)$, we obtain that
\[
h(\mu) - \beta\mu(\phi) \geq
h(\mu') - \beta\mu'(\phi) \eqnsp ,
\]
which means $\mu$ achieves the topological pressure at~$\beta$,
hence $\mu\in\G(\beta)$.

The second statement is a direct consequence of the first.
\end{proof}
\end{lemma}

\begin{proposition}
\label{prop:ground-states:connected}
The set $\G(\infty)$ is compact and connected. 

\begin{proof}
The set $\G(\infty)$ is a closed subset
of the compact set $\M_\sigma\left(\Omega_\A\right)$, so it is itself compact.

Assume now that the set of ground states $\G(\infty)$ is not connected.
Then, there exists two disjoint open sets
$A,B\subseteq\M_\sigma\left(\Omega_\A\right)$ such that
both intersect $\G(\infty)$ and $\G(\infty)\subseteq A\sqcup B$.
By compactness, there is a threshold $\beta_0$ such that for any $\beta\geq\beta_0$,
we have $\G(\beta)\subseteq A\sqcup B$ --
or else we would have a sequence of Gibbs measures
$\left(\mu_{\beta_n}\right)_{n\in\N}$ with $\beta_n\to\infty$
in the compact space $(A\sqcup B)^c$,
thus some accumulation point outside of $\G(\infty)$.

For any such $\beta\geq\beta_0$,
since $\G(\beta)$ itself is convex (thus connected),
one has either $\G(\beta)\subseteq A$ or $\G(\beta)\subseteq B$.
If all the sets were included in $A$ (resp. $B$) after a rank
then we would have $B\cap \G(\infty)=\emptyset$, a contradiction.
Thus, we have two inverse temperatures $\beta_A,\beta_B\geq\beta_0$
such that $\G\left(\beta_A\right)\subseteq A$ and $\G\left(\beta_B\right)\subseteq B$.
Without loss of generality, we assume $\beta_A<\beta_B$.

Consider now the partition
$\left[\beta_A,\beta_B\right]=I_A\sqcup I_B$
such that $G(\beta)\subseteq A$ if and only if $\beta \in I_A$.
As both sets are non-empty, one of them must have
an accumulation point in the other.
Without loss of generality, assume we have a sequence
$\left(\beta_n\right)$ in $I_A$ that converges to a point $\beta\in I_B$.
By passing to a subsequence if needed, we also have a sequence of Gibbs measures
$\left(\mu_{\beta_n}\right)$ in $A$ that converges to a measure $\mu$.
Since $A\subseteq B^c$, and $B^c$ is compact, we must have $\mu\notin B$.
However, using the previous lemma, $\mu\in\G(\beta)\subseteq B$, which is a contradiction.
We conclude that $\G(\infty)$ is indeed connected.
\end{proof}
\end{proposition}

Let us emphasize that the latter topological constraint holds
for every model with a continuous potential.
We now turn our attention to the case of a uniform model
and obtain a preliminary result that will be used
in the following section to derive computational upper bounds on $\G(\beta)$.

We will denote the diameter of a compact set 
$K\subseteq\M_\sigma\left(\Omega_\A\right)$
by $\diam{K}=\max_{\mu,\mu'\in K} d^*\left(\mu,\mu'\right)$.
Given $\epsilon>0$, let $A_\epsilon:=\{\beta>0,\diam{\G(\beta)}<\epsilon\}$.
The weak continuity lemma implies that
$\beta\mapsto\diam{\G(\beta)}$ is upper semi-continuous,
so the sets $A_\epsilon$ are all open.

\begin{proposition}\label{lem:uniform-diameter}
Consider a uniform model.
For every $\epsilon>0$, we have $\sup A_\epsilon = \infty$,
\ie the set $A_\epsilon$ contains arbitrarily big values of $\beta$.
Moreover, $\G(\infty)=\acc[\beta\to\infty]{(\G(\beta))_{\beta\in A_\epsilon}}$.

\begin{proof}
First, assume that $A_\epsilon$ is bounded by $\beta_0$,
\ie $\diam{\G(\beta)}\geq\epsilon$ for any $\beta > \beta_0$.
Consider a measure $\mu\in\M_\sigma\left(\Omega_\A\right)$.
For every $\beta>\beta_0$, we can pick some $\mu_\beta\in\G(\beta)$ such that 
$d^*\left(\mu,\mu_\beta\right)\geq\frac{\epsilon}{2}$.
Thus, $\mu\notin\acc{\mu_\beta}$.
This holds in particular for each $\mu\in\G(\infty)$,
contradicting the uniformity of the model.  Therefore, $A_\epsilon$ is not bounded.

It follows that the set of accumulation points
$\acc[\beta\to\infty]{(\G(\beta))_{\beta\in A_\epsilon}}$ is well-defined.
To prove the second claim, note that the inclusion $\supseteq$ trivially holds.
Suppose that the other inclusion does not hold,
and consider a ground state $\mu\notin\acc{(\G(\beta))_{\beta\in A_\epsilon}}$.
Let us choose a trajectory $(\mu_\beta)$ as follows.
For each $\beta$,
if $\beta\in A_\epsilon$, we pick an arbitrary $\mu_\beta\in\G(\beta)$,
and if $\beta\notin A_\epsilon$,
we can pick $\mu_\beta$ such that
$d^*\left(\mu,\mu_\beta\right)\geq\frac{\epsilon}{2}$ as before.
Then $\mu\notin\acc{\mu_\beta}$, which once again contradicts the uniformity of the model.
\end{proof}
\end{proposition}

\subsection{Notions of Computable Analysis}
\label{sec:constraints:computable-framework}

Before examining the computability constraints on $\G(\infty)$,
we review what computability in the current setting means.
For convenience, we will restrict ourselves to the case of computable potentials
(see Remark~\ref{rmq:computable-potential}), although,
as pointed out in the introduction,
this restriction is made only to make the values of the potential accessible to algorithms. 
An alternative formulation is to make the values available via an oracle.
On the other hand, we do not limit ourselves to local potentials.
Computability automatically ensures that the potential is continuous.

Since there are only countably many computable potentials (at most one per algorithm),
it follows that there are only countably many corresponding sets
of ground states $\G(\infty)$.
The objective is describe the algorithmic complexity of such sets.
However, not only measures are real-valued functions,
for which the usual notion of computability for functions $\N\to\N$ is inadequate,
but also we are concerned with describing $\G(\infty)$,
which is (often) an uncountable set of measures.
In this section,
we introduce a minimal framework for \emph{computable analysis} sufficient for our purposes 
and collect some preliminary results to be used in the following sections.
An interested reader may consult broader introductions
to the general framework~\cite{Wei00,BraHerWei08}
or the implementations of the framework in other settings such as
the study of Julia sets~\cite{BraYa09},
shift-invariant measures~\cite{GaHoRo10} and Cantor sets~\cite{GaHeRoSa20}.

As is standard in computer science,
the informal idea of algorithm will be formalised
as a \emph{Turing machine}~\cite{Turing36}.
We will call a function $f:\N\to\N$ \emph{computable}
if there exists an algorithm which, upon input $n$, outputs $f(n)$.
This notion naturally extends to countable sets such as $\Q$
that can be explicitly encoded as integers. 
A real number $x$ is called \emph{computable} if there is a
computable function $f:\N\to\Q$ such that $\abs{f(n)-x}<2^{-n}$ for every $n$.

\begin{definition}[Computable Metric Space]
A \emph{computable metric space} is a metric space $(X,d)$
together with a countable dense family $\dense=\left(z_n\right)_{n\in\N}$
and a computable $f:\N^3\to\Q$ such that,
for every $i,j,n\in\N$, $\abs{d\left(z_i,z_j\right)-f(i,j,n)}\leq 2^{-n}$.
An element $x\in X$ is said to be computable if there is
a computable $f:\N\to\N$ such that $d\left(x,z_{f(n)}\right)\leq 2^{-n}$ for every $n$.
In this case, we say that the algorithm $f$ \emph{computes} $x$.
More generally, $x$ is \emph{limit-computable} if there is an algorithm $f:\N\to\N$
such that $d\left(x,z_{f(n)}\right)\to 0$ but without any control on the convergence speed.
\end{definition}

In principle, the notion of computability thus obtained depends
not only on the set~$\dense$ but also on the order chosen
to enumerate the elements of~$\dense$ (\ie the \emph{encoding} of $\dense$ in $\N$).
Nonetheless, different \emph{natural} choices for $\dense$ and its enumeration
often lead to equivalent notions of computability,
in the sense that one can translate one encoding into another using a computable function.
We will assume that such a natural encoding is available,
and to simplify the notations, consider maps directly defined
on the set $\dense$ instead of its encoding into~$\N$.

It follows from the definition that, given $x,y\in\dense$ and $c\in\Q$,
the question of whether $d(x,y)<c$ (resp., $d(x,y)>c$) is \emph{semi-decidable}:
there is an algorithm $f(x,y,c)$ that answers the question in finite time
\emph{whenever the answer is positive},
but may loop without halting \emph{when the answer is negative}.
(Here, the infinite loop occurs when $d(x,y)=c$.)
In particular, given $x,y\in\dense$ and $r,s\in\Q$,
checking if $B(x,r)\cap B(y,s)\neq\emptyset$
(equivalently, $B(x,r)\cap \overline{B(y,s)}\neq\emptyset$) is semi-decidable,
whereas checking $\overline{B(x,r)}\cap \overline{B(y,s)}\neq\emptyset$ is 
\emph{co-semi-decidable}
(\ie there is an algorithm that answers in finite time whenever the answer is negative).

By extension, if $(X,d,\dense)$ is a computable space,
then so is the space of \emph{compact subsets} of $X$ equipped
with the Hausdorff distance (\ie
$d_H\left(K,K'\right):=\max\left(\max_{x\in K} d\left(x,K'\right),
\max_{y\in K'}d(K,y)\right)$ for compact $K,K'\subseteq X$)
and the countable dense basis of \emph{finite subsets} of $\dense$.
Indeed, for every compact set $K\subseteq X$ and every $\epsilon>0$,
there exists a finite (thus compact) subset $C\Subset\dense$
such that $d_H(K,C)\leq\epsilon$.

An equivalent (and somewhat more intuitive) way of defining computability
on compact subsets is to use intersecting closed rational balls.\footnote{
We refer the reader to an article by Cenzer and Remmel~\cite{CeRe02}
for a discussion on the different ways to define the computability of compact sets.
}
To do so, we identify a compact set $K\subseteq X$ with the set of
intersecting neighbourhoods $\mathcal{N}(K)\subseteq \dense\times\Q$, defined as:
\[
\mathcal{N}(K):=\left\{(x,r)\in\dense\times\Q:d(x,K)\leq r\right\} \eqnsp .
\]
In particular, as the space $\dense\times\Q$ is countable,
we can study $\mathcal{N}(K)$ using the usual framework for computable subsets of $\N$,
and define a compact set $K$ to be \emph{computable} if the associated indicator function 
$\mathds{1}_{\mathcal{N}(K)}:\dense\times\Q\to\{0,1\}$ is.
By analogy with subsets of $\R^2$,
this means that $K$ can be ``displayed'' on a computer screen at any zoom level,
using an algorithm to decide whether each pixel
(as a ``ball'' of a given radius) must be lighted
(which happens if and only if it intersects $K$).

Using this formalism, we can then naturally extend the study
of computable compact sets $K$
to the study degrees of (un)computability and the arithmetical hierarchy.
More precisely, a compact $K\subseteq X$ is said to be \emph{$\Pi_k$-computable}
if there exists
a computable Boolean function $f:\mathfrak{D}\times\Q\times\N^k\to \{0,1\}$ such that:
\[
(x,r)\in\mathcal{N}(K) \Leftrightarrow \underset{\text{$k$ alternating quantifiers}}
{\underbrace{\forall y_1,\exists y_2,\forall y_3,\dots}}
f\left(x,r,y_1,\dots,y_k\right)=1 \eqnsp .
\]
The notion of \emph{$\Sigma_k$-computability} is defined analogously
except that the first quantifier is required to be existential rather than universal.

For the first levels in the arithmetical hierarchy,
one can often obtain equivalent and seemingly more natural characterisations.
For instance, $K$ is $\Pi_1$-computable if and only if there is
a computable sequence $(x,r):\N\to\dense\times\Q$ such that
$K=\left(\bigcup\limits_{n\in\N} B\left(x_n,r_n\right)\right)^c$.

A result of interest in the current paper is the fact that
$K$ is $\Pi_2$-computable if and only if
$K=\acc{x_n}$ for some computable sequence $x:\N\to\dense$.
In the case of \emph{connected} $\Pi_2$ sets,
we can refine the result to also require $d\left(x_n,x_{n+1}\right)\to 0$.
This was already proven in the specific context of 1D invariant measures
on a subshift~\cite[Proposition 6]{HeSa18}.
In the proposition below, we provide a generalised statement with a more concise proof.

We say that a subset $K\subseteq X$ of a computable metric space $(X,d,\dense)$
is \emph{recursively compact} if $K$ is compact
and it is semi-decidable whether a given finite set of open balls
(encoded as elements of $\dense\times\Q$) covers $K$.
We say that a computable metric space $(X,d,\dense)$ is \emph{locally recursively compact}
if all closed balls $\overline{B(x,r)}$ with $x\in\dense$
and $r\in\Q$ are recursively compact, uniformly in $x$ and $r$
(\ie we can use the same algorithm with $x$ and $r$ as inputs).\footnote{
This notion does not perfectly correlate with the notion of local compactness,
but will prove to be a sufficient simplification for our purpose.
}
Note that if a space $X$ is recursively compact,
then it is also locally recursively compact.

\begin{proposition}[Connected $\Pi_2$ Compacts as Accumulations Sets] 
\label{prop:pi2-connected-accumulation}
Let the computable metric space $(X,d,\dense)$ be locally recursively compact.
Then, the \emph{connected} $\Pi_2$-computable compact subsets $K\subseteq X$
are exactly those that can be obtained as accumulation sets $K=\acc{x_n}$
with $n\in\N\mapsto x_n\in\dense$ a computable map and $d\left(x_n,x_{n+1}\right)\to 0$.

\begin{proof}
Without loss of generality, assume the space $X$ itself is compact,
or else we can replace $X$ by $B:=\overline{B_d\left(x_0,r\right)}$
such that $K\subseteq B$.
Since $X$ is then \emph{recursively} compact,
it is possible to \emph{compute} a non-decreasing sequence of
finite sets $\dense_{k}\subseteq\dense_{k+1}\subseteq\dense$ such that 
$X=\bigcup_{x\in\dense_k}B_d\left(x,\frac{1}{k}\right)$ for every $k\in\N$.

Let $K$ be a connected $\Pi_2$-computable compact set,
and $f:\dense\times\Q^+\times\N^2\rightarrow\{0,1\}$
the associated computable function, such that $d(x,K)\leq \frac{1}{k}$ if and only if
$\forall n\in\N,\exists t\in\N, f(x,k,n,t)=1$.

For $T,k\in\N$, define:
\[
V_k^T:=\left\{\left(x_i\in\dense_i\right)_{i\in\llbracket 1,k\rrbracket} ,
\forall i,j\in\llbracket 1,k\rrbracket,d\left(x_i,x_j\right)<\frac{2}{\min(i,j)},
\forall n\leq k,\exists t\leq T,f\left(x_i,i,n,t\right)=1 \right\} \eqnsp .
\]
Thus defined, $V_k^T$ uses the computable distance function $d:\dense^2\to\R$.
Without loss of generality, we can replace $d\left(x,y\right)$
by a computable rational upper bound $\overline{d}(x,y,k+T)$
at distance at most $\frac{1}{2^{k+T}}$
(with $\overline{d}:\dense^2\times\N\to\Q$ computable, non-decreasing in its third variable),
so that $(k,T)\mapsto V_k^T$ is a computable map too.

In particular, $V_k^T\subseteq V_k^{T+1}\subseteq\prod_{i=1}^k\dense_i$
is a non-decreasing sequence,
and $\prod_{i=1}^k\dense_i$ is finite, so the sequence is stationary after a rank $t_k$.
Without loss of generality, by inductively replacing $t_k$ by $\max\left(t_k,t_{k-1}\right)$,
we can have $k\mapsto t_k$ non-decreasing.
Likewise, if $\left(x_i\right)_{i\leq k+1}\in V_{k+1}^T$,
then its prefix satisfies $\left(x_i\right)_{i\leq k}\in V_k^{T+1}$
(we must increase $T$ by one to compensate the loss of precision for $\overline{d}$,
to compute the same $\overline{d}\left(x_i,x_j,k+T+1\right)$ in both cases),
so that in particular
$\pi_{\llbracket 1,k\rrbracket}\left(V_{k+1}^{t_{k+1}}\right)\subseteq V_k^{t_k}$
(with $\pi_{\llbracket 1, k\rrbracket}$ the projection on the first $k$ coordinates).

Let $x\in\dense_k$ such that $B_d\left(x,\frac{1}{k}\right)\cap K\neq\emptyset$.
Then there is $\left(x_i\right)\in V_k^{t_k}$ such that $x_k=x$.
Indeed, let $y\in B_d\left(x,\frac{1}{k}\right)\cap K$, and fix $i<k$.
Because the $\frac{1}{i}$-neighbourhood of $\dense_i$ covers $K$,
it follows that there is some $x_i\in\dense_i$
such that $y\in B_d\left(x_i,\frac{1}{i}\right)$.
We have $B_d\left(x_i,\frac{1}{i}\right)\cap K\neq\emptyset$,
so the formula
\[
\forall n\leq k, \exists t\in\N,f\left(x_i,i,n,t\right)=1
\]
holds in particular.
By the triangle inequality, we have
$d\left(x_i,x_j\right)< \frac{1}{i}+\frac{1}{j}\leq \frac{2}{\min(i,j)}$
(and this likewise holds for $\overline{d}\left(x_i,x_j,k+T\right)$
for $T$ big-enough, thus for $t_k$).
Thence $\left(x_i\right)_{i\leq k}\in V_k^{t_k}$.
As the $\frac{1}{k}$-neighbourhood around $\dense_k$ covers $K$
and $B_d\left(x_k,\frac{1}{k}\right)\cap K=\emptyset$
when $x\in \dense_k\backslash \pi_k\left(V_k^{t_k}\right)$,
we conclude that:
\[
K \subseteq \bigcup\limits_{x\in \pi_k\left(V_k^{t_k}\right)}
B_d\left(x,\frac{1}{k}\right) \eqnsp .
\]
However, we only still have a rough covering of $K$,
and may have outliers $x\in \pi_k\left(V_k^{t_k}\right)$
such that $B_d\left(x,\frac{1}{k}\right)\cap K=\emptyset$.
Nevertheless, for a fixed scale $\ell\in\N$, the sequence
$\left(\pi_{\llbracket 1,\ell\rrbracket}\left(V_k^{t_k}\right)\right)_{k\geq \ell}$
is non-increasing, thus stationary.
Denote $\phi(\ell)$ the finite rank at which the stationary limit is reached,
and $W_\ell:=\pi_\ell\left(V_{\phi(\ell)}^{t_{\phi(\ell)}}\right)$
the corresponding projection.

Now, for $x\in\dense_\ell$,
we have $B_d\left(x,\frac{1}{\ell}\right)\cap K\neq\emptyset$ if and only if $x\in W_\ell$.
The direct implication works as before,
using a sequence $\left(x_k\in\dense_k\right)_{k\in\N}$
such that $x=x_\ell$ and $\bigcap_{k\in\N}B_d\left(x_k,\frac{1}{k}\right)=\{y\}\subseteq K$.
Conversely, if $B_d\left(x,\frac{1}{\ell}\right)\cap K=\emptyset$,
then there is a rank $n\in\N$ such that $\forall t\in\N,f(x,\ell,n,t)=0$,
and thus $x\notin\pi_\ell\left(V_k^{t_k}\right)$ for $k\geq n$.

To obtain the desired computable sequence $\left(x_n\right)_{n\in\N}$,
we first define the computable sequence of finite sets
$U_{T+1}=\bigsqcup_{k\leq T+1}\pi_k\left(V_k^{T+1}\backslash V_k^T\right)$
(we use a disjoint union to underline that $U_T$ is a \emph{multiset},
and a single element $x\in\dense$ may appear once for each value of $k\leq T$).

Let us fix a scale $\ell\in\N$.
Whenever $T>t_\ell$ and $k\leq \ell$, then $V_k^{T}\backslash V_k^{T-1}=\emptyset$,
so $U_T$ only uses elements for scales $\ell< k \leq T$,
while the first sets of the disjoint union are empty.
Thus, cumulatively,
$\bigsqcup_{i\leq\ell}\pi_i\left(V_i^{t_i}\right)\subseteq\bigsqcup_{t\leq T} U_t$.
As the sets $V_i^{t_i}$ are all non-empty
(their $\frac{1}{i}$-neighbourhood covers $K\neq\emptyset$),
it follows that $\sum_{T\in\N} \abs{U_T} = \infty$,
so any sequence $\left(x_n\right)$ induced by ``enumerating''
the sets $\left(U_T\right)$ will be infinite indeed.
Furthermore, it will contain (at least) an occurrence
of each element of $\left(V_k^{t_k}\right)_{k\in\N}$,
which can approximate any element of $K$ with precision $\frac{1}{k}$,
so that $K\subseteq \acc{x_n}$.

Now, let us compute a path traversing $U_T$.
To do so, we compute the biggest scale $i\leq T$ that satisfies the following property:
there exists an intermediate scale $j\in\llbracket i,T\rrbracket$ such that,
using only elements of $\pi_i\left(V_j^T\right)$ in-between two elements of $U_T$,
we can traverse $U_T$ with distances of less than $\frac{2}{i}$ between consecutive elements
(and if no such scale exists we simply enumerate $U_T$ instead).

When $T>t_{\phi(\ell)}$,
the set $W_\ell=\pi_\ell\left(V_{\phi(\ell)}^T\right)$ gives us such a scale $i$.
In one hand, $W_\ell$ forms an optimal covering of the connected set $K$,
with intersecting balls of radius $\frac{1}{\ell}$,
so we can traverse $W_\ell$ itself with jumps at most $\frac{2}{\ell}$.
On the other hand, any element of $x\in U_T$ is actually in a set $\pi_i\left(V_i^T\right)$
with $i\geq \phi(\ell)\geq \ell$, thus at distance at most $\frac{2}{\ell}$
of $x_\ell\in W_\ell$ for the corresponding sequence $\left(x_j\right)_{j\leq i}\in V_i^T$,
so we can visit each element of $U_T$ while traversing $W_\ell$.
To summarise, when $T>t_{\phi(\ell)}$, we can compute a path traversing $U_T$,
using intermediate elements in $\pi_i\left(V_j^T\right)$ (with $\ell\leq i \leq j\leq T$),
with distance at most $\frac{2}{\ell}$ between consecutive elements.
For any such element $x$ (either in $U_T$ or an intermediate element),
we have a corresponding finite sequence $\left(x_i\right)\in V_j^T$
(for some $\ell\leq j \leq T$),
and thus in particular $x_\ell\in W_\ell$,
so that $d(x,K)\leq \frac{3}{\ell}$ by triangle inequality.

Likewise, to link the last visited element $y\in U_T$ to the first element $z\in U_{T+1}$
(or more generally to the next non-empty set $U_{T'}$ instead),
we may use the same algorithm, and conclude that we can link $y$ to $z$
with distances at most $\frac{2}{\ell}$ between elements
and while staying at distance at most $\frac{3}{\ell}$ of $K$.
Overall, by sticking the traversals of each $U_T$ with these links,
we finally obtain a well-defined computable sequence $\left(x_n\right)\in\dense^\N$,
such that $\acc{x_n}\subseteq K$ and $d\left(x_n,x_{n+1}\right)\to 0$,
which concludes this implication.

For the other direction, any set $K:=\acc{x_n}$ is obviously $\Pi_2$,
using the fact that $B_d\left(x,\frac{1}{k}\right)\cap K\neq \emptyset$ if and only if
$\forall n\in\N,\exists t\geq n, d\left(x,x_t\right)<\frac{1}{k}+\frac{1}{n}$,
with the elements $x_t$ corresponding to a subsequence
converging to an element in $B_d\left(x,\frac{1}{k}\right)\cap K$.
Furthermore, $K$ must be connected using an argument similar
to that of Proposition~\ref{prop:ground-states:connected}.
\end{proof}
\end{proposition}

Without the connectedness assumption, we can
follow the same argument to obtain a similar characterisation of the $\Pi_2$-computable sets:

\begin{corollary}
Let $(X,d,\dense)$ be a locally recursively compact computable metric space.
Then, the $\Pi_2$-computable compact subsets $K\subseteq X$
are exactly those that can be obtained as accumulation sets $K=\acc{x_n}$
with $n\in\N\mapsto x_n\in\dense$ a computable map.
\end{corollary}

The above proposition offers an intuitive approach
to prove the $\Pi_2$-computability of the set of ground states $\G(\infty)$
as a compact subset of $\M_\sigma\left(\Omega_\A\right)$
by writing it in terms of accumulation points.
However, before doing that, we need a computable structure
on $\M_\sigma\left(\Omega_\A\right)$ and its compact subsets.

\begin{remark}[Computations on Measures] \label{rmq:computable-measures}
In order to turn $\left(\M_\sigma\left(\Omega_\A\right),d^*\right)$
into a computable metric space,
we use the countable dense set $\per$ consisting of the orbit averages
of the periodic configurations in~$\Omega_\A$.
More specifically, $\per$ consists of measures
of the form $\widehat{\delta_w}:=\frac{1}{\abs{I_n}}\sum_{k\in I_n} 
\delta_{\sigma^k\left(w^{\Z^d}\right)}$
for some $n\in\N$ and $w\in \A^{I_n}$,
where $w^{\Z^d}$ refers to the configuration obtained
by periodically repeating $w$ in all directions.
It is known that $\per$ is dense in $\M_\sigma\left(\Omega_\A\right)$.
Notice that $(x,y)\in\per^2\mapsto d^*(x,y)$ is computable,
\ie there is a computable $f:\per^2\times\N\to\Q$
such that $\abs{d^*(x,y)-f(x,y,n)} \leq 2^{-n}$.
Hence, $\left(\M_\sigma\left(\Omega_\A\right),d^*,\per\right)$ is a computable metric space.

A measure $\mu$ is computable, as an \emph{element} of the computable space,
if there is an algorithm $f:\N\to\per$ such that $d^*(\mu,f(n))\leq 2^{-n}$ for every $n$.  
Note that this implies that $\mu:w\mapsto\mu([w])$ is computable
as a real-valued function on cylinder sets.

As a computable metric space, $\M_\sigma\left(\Omega_\A\right)$ is recursively compact
(thus locally recursively compact).
We refer the reader to the existing literature~\cite{GaHoRo10}
for an overview of these computability properties of the set of probability measures.
\end{remark}

\begin{definition}[Computable Maps on Computable Spaces] \label{def:computable-map}
Let $\left(X,d,\dense\right)$ and $\left(X',d',\dense'\right)$
be two computable metric spaces.
The map $\psi:X\to X'$ is said to be \emph{computable} if there exists a computable function
$f:\N\times\dense'\times\Q\to\dense\times\Q$ such that, for every $y\in\dense'$ and $r\in\Q$:
\[
\psi^{-1}\left(B_{d'}(y,r)\right) = \bigcup_{n\in\N} B_{d}(f(n,y,r)) \eqnsp .
\]
In particular, $\psi$ is continuous.
Likewise, a sequence $\left(\psi_k\right)_{k\in\N}$ of maps is (uniformly) \emph{computable}
if there is a computable function $f:\N^2\times\dense'\times\Q\to\dense\times\Q$ such that
$\psi_k^{-1}\left(B_{d'}(y,r)\right) = \bigcup_{n\in\N} B_{d}(f(k,n,y,r))$
for every $k\in\N$, $y\in\dense'$ and $r\in\Q$.

When $X$ and $X'$ are locally recursively compact,
the computability of a map $\psi:X\to X'$ becomes equivalent
to the existence of computable map $(f,s):\dense\times\Q\to\dense'\times\Q$ such that
\begin{enumerate}
    \item $\psi(B_d(x,r))\subseteq B_{d'}(f(x,r),s(x,r))$
    for every $x\in\dense$ and $r\in\Q$,
    \item $\sup_{x\in K\cap\dense} s(x,r) \underset{r\to 0}{\longrightarrow} 0$
    for every compact $K\subseteq X$.
\end{enumerate}
The first condition tells us that we can map
an approximation of $x\in X$ to an approximation of $\psi(x)$,
and the second condition tells us that this process converges
(in a uniformly continuous way on compact sets),
so that whenever $x\in X$ is computable, so is $\psi(x)\in X'$.
\end{definition}

\begin{figure}
    \centering
    \includegraphics[width=.5\textwidth]{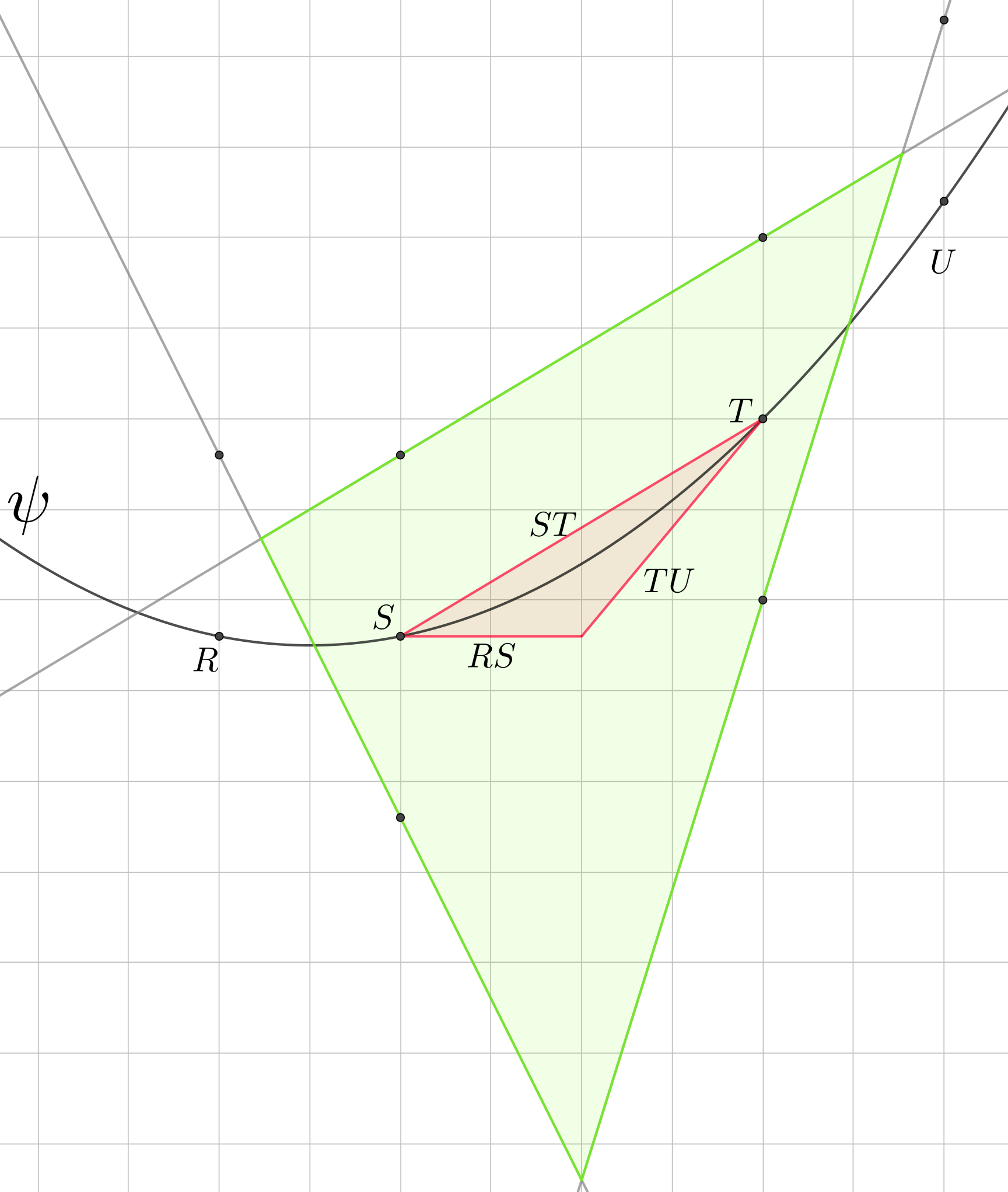}
    \caption{Inclusion of the graph of a convex map $\psi$ into a computable triangle.}
    \label{fig:convex-triangle}
\end{figure}

The next lemma shows that, in order for a convex function on $\R$ to be computable, it is enough that it is uniformly computable over rationals.
To turn $\R$ into a computable metric space, we use the metric $d(x,y)=\abs{y-x}$ and the countable set~$\Q$.

\begin{lemma} \label{lem:convex-computable}
Let $\psi:\R\to\R$ be a \emph{convex} map. Then $\psi$ is computable if and only if
there is a computable $g:\Q\times\N\to\Q$
such that $\abs{g(x,n)-\psi(x)}\leq 2^{-n}$ for every $x\in\Q$ and $n\in\N$
(\ie $\psi|_\Q$ is uniformly computable).

\begin{proof}
The forward implication holds for every computable map.
Namely, in order to find $g(x,n)$, we can estimate $\psi(]x-\epsilon,x+\epsilon[)$
for smaller and smaller values of $\epsilon$ until we reach precision $2^{-n}$ for~$\psi(x)$.

For the opposite direction, given $s,t\in\Q$, we estimate $\psi(]s,t[)$
with the help of the function $g$ and exploiting the convexity of~$\psi$.
The idea is illustrated in Figure~\ref{fig:convex-triangle}.
For convenience, we replace $g$ with another function $\tilde{g}:\Q\times\Q\to\Q$
such that $\abs{\tilde{g}(x,\epsilon)-\psi(x)}<\epsilon$ for every $x,\epsilon\in\Q$.
Computing such a~$\tilde{g}$ using~$g$ is straightforward.

Denote by $\rho=\frac{\abs{t-s}}{2}\in\Q$ the radius of the ball $]s,t[$,
and consider $r=s-\rho$ and $u=t+\rho$.
Consider the point $S=(s,\psi(s))$ and similarly $R$, $T$ and $U$.
On the interval $[s,t]$, we know that the graph of $\psi$ is below the line $ST$,
but above the lines $RS$ and $TU$, thus included in the (small red) triangle.
Note that, although $s$, $t$, $r$ and $u$ are rationals,
their images under $\psi$ may not be rationals.
To circumvent this, as shown in Figure~\ref{fig:convex-triangle},
we can replace $ST$ by the 
``higher'' line crossing the points $(s,\tilde{g}(s,\rho)+\rho)$
and $(t,\tilde{g}(t,\rho)+\rho)$,
$RS$ by the ``lower'' line crossing $(r,\tilde{g}(r,\rho)+\rho)$
and $(s,\tilde{g}(s,\rho)-\rho)$,
and likewise $TU$ by the line crossing $(t,\tilde{g}(t,\rho)-\rho)$
and $(u,\tilde{g}(u,\rho)+\rho)$.
This new (big green) triangle, defined using only computable rational coordinates,
contains the previous one, hence
its vertical range $]m(s,t),M(s,t)[$ covers $\psi(]s,t[)$.

Now, note that as long as $x\in\Q$ stay well bounded in a compact set $K$,
by convexity the slopes of the previous lines behave well, so that
$\sup_{x\in\Q\cap K} M(x-r,x+r)-m(x-r,x+r) \underset{r\to 0}{\longrightarrow} 0$.
Therefore, $]m(x-r,x+r),M(x-r,x+r)[$ provides a convergent estimate
for $\psi(x)$, which means $\psi$ is computable in the sense of the above definition
for locally recursively compact spaces.
\end{proof}
\end{lemma}

The next lemma states that applying a computable map to a set cannot increase its complexity.

\begin{lemma}
Let $\psi:X\to X'$ be a computable map between locally recursively compact metric spaces 
$(X,d,\dense)$ and $(X',d',\dense')$.
If $K\subseteq X$ is $\Pi_2$-computable, then so is $\psi(K)$.

\begin{proof}
By Proposition~\ref{prop:pi2-connected-accumulation},
$K=\acc{x_n}$ for a computable sequence $x\in\dense^\N$.
As $\psi$ is continuous, we have $\psi(K)=\acc{\psi\left(x_n\right)}\subseteq X'$.

Using the computable map $(f,s):\dense\times\Q\to\dense'\times\Q$ that computes~$\psi$,
we then define the sequence $y\in\dense'^\N$ by $y_n:=f\left(x_n,2^{-n}\right)$.
By compactness of (the $1$-neighbourhood of) $K$, we know in particular that
$d'\left(y_n,\psi\left(x_n\right)\right)\to 0$.
It follows that $\psi(K)=\acc{y_n}$ is the accumulation set of a computable sequence~$y$,
hence it is $\Pi_2$-computable.
\end{proof}
\end{lemma}

We cannot hope for any kind of converse result in the general case,
as a constant map $\psi$ that maps an entire compact set $K$
into a single computable point $x$ erases all the complexity of~$K$.
However, under stronger assumptions on $\psi$ (injectivity in particular),
we obtain the following equivalence:

\begin{corollary} \label{cor:injection-pi2-equivalence}
Let $\psi:X\to X'$ be an injective computable map
between locally recursively compact metric spaces $(X,d,\dense)$ and $(X',d',\dense')$,
and let $\psi':X'\to X$ be a (surjective) computable map
such that $\psi'\circ\psi=\mathrm{Id}_X$.
Then the images of (connected) $\Pi_2$-computable compact subsets of $X$ are exactly
the (connected) $\Pi_2$-computable compact subsets of $X'$ included $\psi(X)$.

\begin{proof}
As $\psi$ is a continuous injection, the set $K$ is connected if and only if $\psi(K)$ is.
Using the previous lemma, we already established that if $K\subseteq X$ is $\Pi_2$,
then so is $K'=\psi(K)$.
Conversely, any compact $K'\subseteq \psi(X)$ is the image of a compact $K$,
and if $K'$ is $\Pi_2$, then using the previous lemma for $\psi'$,
it follows that $K=\psi'\left(K'\right)$ is $\Pi_2$ too,
so $K'$ is indeed the image of a $\Pi_2$ compact set.
\end{proof}
\end{corollary}

\subsection{Computability Obstructions on \texorpdfstring{$\G(\infty)$}{G\_σ(∞)}}
\label{sec:constraints:computable-analysis}

In order to obtain a general upper bound on the complexity of $\G(\infty)$,
a first step will be to obtain a computable description of $\G(\beta)$
(\eg for computable values of $\beta>0$).
More generally, we will obtain a description of the sets
$\G\left(\left[\beta^-,\beta^+\right]\right):=
\bigcup_{\beta\in\left[\beta^-,\beta^+\right]}\G(\beta)$.

\begin{lemma}
For any $0\leq\beta^-\leq\beta^+$,
the set $\G\left(\left[\beta^-,\beta^+\right]\right)$ is compact.

\begin{proof}
Let $\left(\mu_n\right)\in\G\left(\left[\beta^-,\beta^+\right]\right)^\N$ be
a sequence of measures converging to a measure $\mu$, and
let $\left(\beta_n\right)\in\left[\beta^-,\beta^+\right]^\N$ be
the corresponding sequence of inverse temperatures.
By passing to a subsequence if necessary,
we can assume that $\beta_n$ also converge to an inverse temperature $\beta$.
Since $\beta$ will inevitably be in $[\beta^-,\beta^+]$,
it follows from Lemma~\ref{lem:weak-continuity} that 
$\mu\in\G(\beta)\subseteq\G\left(\left[\beta^-,\beta^+\right]\right)$.
Therefore, $\G\left(\left[\beta^-,\beta^+\right]\right)$ is a closed subset
of the compact space $\M_\sigma\left(\Omega_\A\right)$ and is thus compact itself.
\end{proof}
\end{lemma}

Since $\G(\beta)$ is by definition the set of measures that realise
the topological pressure $p(\beta)$,
a first step in studying $\G(\beta)$ is to study the pressure itself.
In the one-dimensional case, when the phase space is an SFT $X_\F$
(\ie we maximise the pressure $\mu\mapsto p_\mu(\beta)$ within $\M_\sigma\left(X_\F\right)$,
instead of among all the measures $\mu\in\M_\sigma\left(\Omega_\A\right)$),
it is known that the pressure $\beta\mapsto p(\beta)$ is computable~\cite{BuDaWoYa22}.
In particular, $p(0)$ is the \emph{topological entropy} of the subshift $X_\F$.
On the other hand, in two and higher dimensions, even the topological entropy
of a single SFT can be uncomputable as a real number~\cite{HoMe10}.
Hence, the scheme of proof used to show the computability of $\beta\mapsto p(\beta)$
in one dimension does not directly generalize to higher dimensions.
Here, we obtain a higher-dimensional result
when the phase space is the full-shift $\Omega_\A$.
Note that, in this case, the latter obstacle is not present
as the topological entropy $p(0)$ is simply equal to $\log_2(\abs{\A})$.

\begin{remark}[Computable Potentials] \label{rmq:computable-potential}
Let us take some time to properly convey
what it means for a potential $\phi:\Omega_\A\to\R$ to be \emph{computable}.
Let us translate the general framework of Definition~\ref{def:computable-map}
into something more natural.

As before, to turn $\R$ into a computable space, we use $\Q$ as the countable dense set.
Let $\mathfrak{W}$ denote
the set of patterns $w\in\A^{\llbracket -r,r\rrbracket^d}$ with $r\in\N$.
Recall that the cylinder sets $[w]$ with $w\in\mathfrak{W}$ form
a countable topological basis for $\Omega_\A$.
In this setting, a potential $\phi$ is computable if
there exists a pair of computable functions $f:\mathfrak{W}\to\Q$ and $r:\N\to\N$
such that $\abs[\big]{f\left(x_{\llbracket -r(m),r(m)\rrbracket^d}\right)-\phi(x)}<2^{-m}$ 
for every $x\in\Omega_\A$ (which can be seen as an oracle) and $m\in\N$.
In particular, a computable potential is necessarily continuous.

If $\Phi$ is an interaction taking only finitely many computable real values,
then the associated potential $\phi_\Phi$ is computable:
as soon as $r$ goes beyond the range of $\Phi$, the set $\Phi_\Phi([w])$ becomes
a singleton containing a computable real number.
If $\Phi$ takes only rational values,
then we can compute the exact rational values for the potential
$\phi_\Phi([w])$ for such big-enough words.
All of this holds in particular for a fault potential $\phi_\F$
associated to a set~$\F$ of forbidden patterns.
\end{remark}

\begin{proposition}
\label{prop:top-pressure:computable}
Let $\phi:\Omega_\A\to\R$ be a computable potential.
Then the pressure $\beta\mapsto p(\beta)$ is computable. 

\begin{proof}
Let $h_n:\mu\in\M_\sigma\left(\Omega_\A\right)\mapsto 
-\frac{1}{\abs{I_n}}\sum_{w\in\A^{I_n}}\mu([w])\log_2(\mu([w]))$
be the finite-window entropy per site on $I_n$, so that
$h(\mu)=\lim_{n\to\infty}h_n(\mu)=\inf_n h_n(\mu)$.
Unlike $h$ which is only upper semicontinuous,
$h_n$ is continuous on $\M_\sigma\left(\Omega_\A\right)$.
In fact, the family $\left(h_n\right)$ is uniformly computable.

Let $\M_n$ denote the set of probability measures on the finite window $\A^{I_n}$.
This set can be viewed as a finite-dimensional simplex.
Given $\nu\in\M_n$, we can construct a measure 
$\widehat{\nu}\in\M_\sigma\left(\Omega_\A\right)$
by partitioning the lattice $\Z^d$ into copies of the window $I_n$,
picking a pattern for each window at random according to~$\nu$
independently of the others, and then averaging over the finite shift orbit
to obtain a shift-invariant measure $\widehat{\nu}$.
More specifically,
\[
\widehat{\nu} = \frac{1}{\abs{I_n}}\sum_{i\in I_n}\sigma^i\left(\nu^{\Z^d}\right)
\]
where $\nu^{\Z^d}$ is the product of infinitely many copies of $\nu$,
one for each copy of $I_n$.
Note that the set $\widehat{\M_n}:=\left\{\widehat{\nu},\nu\in\M_n\right\}$
contains the periodic points $\widehat{\delta_w}$ with $w\in\A^{I_n}$,
hence $\per\subseteq\bigcup_{n\in\N}\widehat{\M_n}$.

Since $h_n(\mu)$ depends only on the marginal of $\mu$ on the window $I_n$,
it makes sense to apply $h_n$ to measures $\nu\in\M_n$.
For each scale $n$, we can then define the finite-window pressure as a function 
$p_n:(\nu,\beta)\in\M_n\times\R\mapsto h_n(\nu)-\beta \widehat{\nu}(\phi)$.
Observe that, for every $\nu\in\M_n$, we have $h_n(\nu)=h(\widehat{\nu})$,
thus we have the bound $p_n(\nu,\beta)=p_{\widehat{\nu}}(\beta)\leq p(\beta)$.

Next, we introduce the finite-window maximal pressure
$p_n^*:\beta\mapsto\max_{\nu\in\M_n}p_n(\nu,\beta)$.
Since $p_n$ is computable on $\M_n\times\R$,
and $\M_n\times\R$ is locally recursively compact space, 
the function $p_n^*$ is also computable.
In fact, since computing $p_n^*$ involves roughly the same algorithm regardless of
the dimension of the simplex (which itself is a computable function of $n$),
the functions $\left(p_n^*\right)$ are uniformly computable.
Since $p_n(\nu,\beta)\leq p(\beta)$ for every $n$, we also have $p_n^*\leq p$.

Consider next a Gibbs measure $\mu\in\G(\beta)$,
so that $p_\mu(\beta)=p(\beta)$, and let $\mu_n=\mu|_{I_n}\in\M_n$.
Since $h(\mu)\leq h_n(\mu_n)$, we have
\[
\begin{split}
p(\beta) &\leq h_n\left(\mu_n\right)-\beta\mu(\phi) \\
&= p_n\left(\mu_n,\beta\right)+\beta\left(\widehat{\mu_n}(\phi)-\mu(\phi)\right) \\
&\leq p_n^*(\beta)+ \beta\abs[\big]{\widehat{\mu_n}(\phi)-\mu(\phi)} \eqnsp .
\end{split}
\]

Let $f:\mathfrak{W}\to\Q$ and $r:\N\to\N$ be computable functions witnessing
the computability of~$\phi$ as in Remark~\ref{rmq:computable-potential}.
Observe that the only part where $\mu$ and $\mu_n^{\Z^d}$ may significantly differ
is close to the boundary of the window~$I_n$.
In particular, for every $m\in\N$ and $n\geq r(m)$, we have
\begin{alignat*}{3}
&\left|\widehat{\mu_n}(\phi)-\mu(\phi)\right| &&\leq
\frac{\big(n-2r(m)\big)^d}{n^d}\times 2^{-m}
&&+\frac{n^d - \big(n-2r(m)\big)^d}{n^d}\times\norm{\phi}_\infty \\
& &&\leq 2^{-m} &&+ \frac{2 d r(m)}{n}\times\norm{\phi}_\infty \eqnsp .
\end{alignat*}
Since $\phi$ is computable,
we can trivially compute (an upper bound on) 
$\norm{\phi}_\infty$.
Clearly, by choosing $m$ and then $n$ large enough in a computable manner,
the above upper bound can be made arbitrarily small.
Therefore, given an inverse temperature $\beta\in\Q$ and a positive $\epsilon\in\Q$,
one can compute an $n\in\N$ such that $\abs{p-p^*_n}<\epsilon$.
In other words, $p$ is uniformly computable on~$\Q$.

Now, recall that $p=\sup p_\mu$ is convex because
it is the supremum of a family of affine (thus convex) functions.
Therefore, using Lemma~\ref{lem:convex-computable},
we can conclude that $p$ is computable on~$\R$.
\end{proof}
\end{proposition}

Remark that we can use the computable potential $\phi$ as a countable parameter too.
Doing so tells us that $\beta\mapsto p(\beta)$ is uniformly computable in $\phi$,
or equivalently that $(\beta,\phi)\mapsto \sup_{\mu} h(\mu)-\beta\mu(\phi)$
is also a computable map.

\begin{proposition} \label{prop:semi-decidable}
Let $\phi$ be a computable potential.
It is semi-decidable to know whether the quadruplet
$\left(x,r,\beta^-,\beta^+\right)\in\per\times\Q^3$ is such that 
$\G\left(\left[\beta^-,\beta^+\right]\right)\cap
\overline{ B\left(x,r\right)}=\emptyset$.

\begin{proof}
Let $h_n$ denote the finite-window entropy per site as introduced
in the proof of Proposition~\ref{prop:top-pressure:computable},
and define $\psi_n(\mu,\beta):=p(\beta)-\left(h_n(\mu)-\beta\mu(\phi)\right)$.
The family $\left(\psi_n\right)$ is uniformly computable.
It is also uniformly bounded, meaning that there is a large enough $A\in\Q$ such that,
for every $n$, the values of $\psi_n$ are in the interval~$[-A,A]$.

Notice that $\sup_n \psi_n(\mu,\beta) = p(\beta)-p_\mu(\beta)\geq 0$,
with equality to $0$ if and only if $\mu\in\G(\beta)$.
Thus:
\[
\bigcup_{\beta\geq 0}\G(\beta)\times\{\beta\} = \bigcap_{n\in\N}\psi_n^{-1}([-A,0]) \eqnsp .
\]
Since $(\psi_n)_n$ is uniformly computable,
the sets $\psi_n^{-1}([-A,0])$ are uniformly $\Pi_1$-computable,
\ie there is a computable $f:\N^2\times\Q^2\times\per\times\Q$ such that:
\[
\left(\overline{B(x,r)}\times\left[\beta^-,\beta^+\right]\right)\cap \psi_n^{-1}([-A,0])
\neq \emptyset\Leftrightarrow \forall k\in\N, f(n,k,\beta^-,\beta^+,x,r)=1 \eqnsp .
\]
It follows that:
\[
\overline{B(x,r)}\cap \G(\left[\beta^-,\beta^+\right])\neq\emptyset
\Leftrightarrow \forall k,n\in\N, f(n,k,\beta^-,\beta^+,x,r)=1 \eqnsp .
\]
Hence, the complementary question is semi-decidable.
\end{proof}
\end{proposition}

In other words,
the compact sets $\G\left(\left[\beta^-,\beta^+\right]\right)$ are
uniformly $\Pi_1$-computable.
This holds in particular for the sets $\G(\beta)$ with $\beta\in\Q^+$.
Consequently, according to Remark~\ref{rmq:computable-measures},
it is semi-decidable to know if $\G(\beta)\subseteq B(x,r)$
for an input $(x,r,\beta)\in\per\times\Q^2$.

\begin{proposition}
Let $\phi$ be a computable potential.
Then the compact set $\G(\infty)$ is $\Pi_3$-computable.

\begin{proof}
Let $(x,r)\in \per\times\Q$. We have:
\[
\begin{array}{rl}
&\G(\infty)\cap\overline{B(x,r)}\neq\emptyset\\
\Leftrightarrow&\forall\beta_0\in\N,\forall\delta\in\Q^{+*},
\exists\beta\in\R_{\geq \beta_0}^+,
\G(\beta)\cap\overline{B\left(x,r+\delta\right)}\neq\emptyset\\
\Leftrightarrow & \forall\beta_0\in\N,\forall\delta\in\Q^{+*},
\exists b\in\N_{\geq\beta_0},
\G([b,b+1])\cap\overline{B\left(x,r+\delta\right)}\neq\emptyset \eqnsp .
\end{array}
\]
The first equivalence comes from the fact that every ground state
in $\G(\infty)\cap\overline{B(x,r)}$ will be approached by Gibbs measures in $\G(\beta)$
at distance $\delta$ (after any rank $\beta_0$ and with any precision $\delta>0$,
which we can both assume to be rational by monotonicity of the properties considered).
For the second one, simply note that such a real number $\beta\geq\beta_0$
\emph{must} be included in an integer interval $[b,b+1]$.

Using Proposition~\ref{prop:semi-decidable}, we deduce that checking if
$\G([b,b+1])\cap\overline{B\left(x,r+\delta\right)}$
is (non-)empty is a (co-)semi-decidable problem.
We can thus reduce the rightmost part to
a formula $\forall k, f(k,b,x,r+\delta)$ with a computable $f$.
Thence, $\G(\infty)$ is a $\Pi_3$-computable compact set.
\end{proof}
\end{proposition}

\begin{theorem} \label{thm:pi2-upper-bound}
Let $\phi$ a computable potential, and assume it induces a uniform model.
Then the set of ground states $\G(\infty)$ is $\Pi_2$-computable.

\begin{proof}
Let $(x,r)\in \per\times\Q$. We have:
\[
\begin{array}{rl}
&\G(\infty)\cap\overline{B(x,r)}=\emptyset \\
\Leftrightarrow&\forall\beta_0\in\N,\forall\delta\in\Q^{+*},
\exists\beta\in\R_{\geq \beta_0}^+,
\G(\beta)\cap\overline{B\left(x,r+\delta\right)}\neq\emptyset\\
\Leftrightarrow&\forall\epsilon\in\Q^{+*},\forall\beta_0\in\N,\forall\delta\in\Q^{+*},
\exists\beta\in\R_{\geq\beta_0},\exists y\in \per,\\
&\G(\beta)\subseteq B\left(y,\epsilon \right)\text{ and }
B\left(y,\epsilon \right)\cap\overline{ B\left(x,r+\delta\right)}\neq\emptyset\\
\Leftrightarrow&\forall\epsilon\in\Q^{+*},\forall\beta_0\in\N,\forall\delta\in\Q^{+*},
\exists\beta\in\Q_{\geq\beta_0},\exists y\in \per,\\
&\G(\beta)\subseteq B\left(y,\epsilon \right)\text{ and }
B\left(y,\epsilon \right)\cap\overline{ B\left(x,r+\delta\right)}\neq\emptyset \eqnsp .
\end{array}
\]
As in the previous proposition, the first equivalence simply comes from
the definition of $\G(\infty)$ as an accumulation set.
The second equivalence comes from the fact that the model is uniform,
thus using Proposition~\ref{lem:uniform-diameter},
we have $\G(\infty)=\acc{\left(\G(\beta)\right)_{\beta\in A_\epsilon}}$,
\ie the accumulation points can be reached
using only the inverse temperatures $\beta\in A_\epsilon$
for which the diameter of $\G(\beta)$ is less than $\epsilon$.
In particular, as $A_\epsilon$ is an open set,
by weak continuity of Gibbs measures (Lemma~\ref{lem:weak-continuity})
we can instead consider $A_\epsilon\cap \Q$ and still reach all the accumulation points.

Since the sets $\G(\beta)$ are uniformly $\Pi_1$-computable for $\beta\in\Q$,
and $\M_\sigma\left(\Omega_\A\right)$ is recursively compact,
we deduce that it is semi decidable to say if $\G(\beta)\subseteq B\left(y,\epsilon \right)$.
Likewise, in any computable space,
checking whether $B\left(y,\epsilon \right)\cap\overline{ B\left(x,r\right)}\neq\emptyset$
is semi-decidable.
In both cases, this only adds more existential quantifiers on the right of the formula,
hence we can conclude that $\G(\infty)$ is $\Pi_2$-computable.
\end{proof}
\end{theorem}

These two upper bounds suggest a profound link between
the computational complexity of the set $\G(\infty)$
and the dynamic properties of the model (\ie whether it is uniform or not).
Ultimately, proving the ``optimality'' of both bounds would allow us to
offer a more affirmative conclusion on this matter.
The rest of this article will be dedicated to a first step in this direction:
the optimality of the $\Pi_2$ bound for uniform models.

\section{Uniform Marker Distribution in Gibbs States} \label{sec:uniform-markers}

In this section, we present a general framework for understanding
the Gibbs measures of a certain class of finite-range models at moderately low temperatures.
The models in question have the property that their ground patterns
consist mostly of distinguishable, non-overlapping blocks we call markers.
The main result of this section (Theorem~\ref{thm:equidistribution}) states that,
under some entropy assumption on markers and ground patterns,
every shift-invariant Gibbs measure within a certain temperature interval induces,
with high probability, a large grid of markers, each chosen uniformly at random.
This result will later be applied, in the context of our construction,
to a hierarchy of marker sets at larger and larger scales.

The key ideas in this section are borrowed from the infinite-range,
one-dimensional construction of Chazottes and Hochman~\cite{ChaHo10},
but here we adapt them to the framework of finite-range models in higher dimensions,
and present them in a more streamlined fashion.

\subsection{Ground Configurations and Ground Patterns}

In this section, we let $\Phi$ be a finite-range interaction with range $r$
that takes only non-negative values (\ie $\Phi_I\geq 0$ for every $I\Subset\Z^d$),
and let $\varphi$ be the associated potential.
For such an interaction, the \emph{total energy}
\[
E(x) = \sum_{I\Subset\Z^d}\Phi_I(x) = \sum_{k\in\Z^d}\phi\left(\sigma^k (x)\right)
\]
of a configuration $x$ is well-defined, though it can possibly take the value $+\infty$.
We let $Z_\Phi$ denote the set of all configurations $x$ with $E(x)=0$.
We shall assume that $Z_\Phi\neq\emptyset$.
An interaction of this type is called a \emph{non-frustrated} interaction~\cite{Miekisz98}. 
The elements of $Z_\Phi$ are called non-frustrated ground configurations,
or (more concisely) \emph{null-energy} configurations.

A \emph{ground pattern} is a pattern with null energy content,
\ie a pattern $w\in\A^I$ with $E_I(w)=0$.
The set of all ground patterns on a window $I$ is denoted by $G_I$.
The elements of $Z_\Phi$ are precisely the configurations
whose restrictions to any finite window is a ground pattern.

The interaction appearing in our construction will be
a fault interaction associated to a tileset.
Observe that for such an interaction, the non-frustrated ground configurations
are precisely the valid tilings, and the ground patterns
are precisely the locally admissible tilings on finite windows.

\subsection{Ubiquity of Ground Markers at Low Temperatures}
\label{sec:uniform-markers:ubiquity}

For the rest of this section, we let
\[
\alpha:=\min\left\{\Phi_I(x), x\in \Omega_\A,I\Subset\Z^d,\Phi_I(x)>0\right\} \eqnsp .
\]
We shall assume that $\Phi$ is not constantly null, thus the latter set is non-empty and we have $\alpha>0$.
In the case of a fault interaction, $\alpha=1$.
Note that, whenever $w\in\A^I\setminus G_I$, we have $E_I(w)\geq\alpha$.

\begin{lemma}[High Probability of Ground Patterns]
\label{lem:ground-patterns}
Let $I\Subset\Z^d$ and $\epsilon>0$, and define
the threshold $\beta_0:=\frac{\log_2(\abs{\A})}{\alpha \epsilon}\abs{I}$.
Then, for every $\beta\geq\beta_0$ and $\mu\in\G(\beta)$,
we have $\mu\left(\left[G_I\right]\right)\geq1-\epsilon$.

\begin{proof}
Let $\beta\geq\beta_0$ and $\mu\in\G(\beta)$.
Since $Z_\Phi$ is a non-empty subshift, it supports at least one shift-invariant measure.
Let $\nu\in\M_\sigma(Z_\Phi)$ be such a measure.
Then, $\nu(\phi)=0$, hence:
\[
0\leq h(\nu) = h(\nu)-\beta \nu(\phi) = p_\nu(\beta)\leq p_\mu(\beta)
=h(\mu)-\beta\mu(\phi)\leq \log_2(\abs{\A})-\beta\mu(\phi) \eqnsp .
\]
It follows $\mu(\phi)\leq \frac{\log_2(\abs{\A})}{\beta}$.
Using the Markov inequality, we then have:
\[
1-\mu\left(\left[G_I\right]\right)=\mu\left(\left\{x\in \Omega_\A,E_I(x)\geq\alpha\right\}\right)
\leq\frac{1}{\alpha}\mu\left(E_I\right)\leq \frac{1}{\alpha}\abs{I}\mu(\phi) \eqnsp ,
\]
which concludes the proof.
\end{proof}
\end{lemma}

We now introduce the notion of markers.
Let us recall the notation $I_n=\llbracket 0,n-1\rrbracket^d$ for a hypercube of size $n$. 

\begin{definition}[Ground Markers] \label{def:marker}
Let $\ell,m\in\N$.
An \emph{$(\ell,m)$-marker set} is a set $Q\subseteq G_{I_\ell}$
of ground patterns on $I_\ell$ that satisfies the following conditions:
\begin{itemize}
    \item \emph{Non-overlapping}: For every $u,v\in Q$ and $k\in\Z^d$,
    if $u_{I_\ell\cap(-k+I_\ell)}=(\sigma^k (v))_{I_\ell\cap(-k+I_\ell)}$,
    then either $I_\ell\cap(-k+I_\ell)=\emptyset$ or $k=0$.
    \item \emph{Covering}: For every $w\in G_{I_m}$,
    there is a translation $k$ such that $(\sigma^k (w))_{I_\ell}\in Q$.
\end{itemize}
If $m=(2+\tau)\ell-1$ for some $\tau\geq 0$, we say that $Q$ has \emph{margin factor} $\tau$.
\end{definition}

\begin{figure}
    \centering
    \includegraphics[scale=.6]{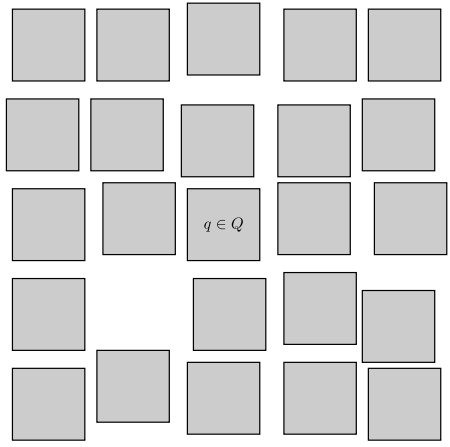}
    \caption{A family of markers tiling a high-density portion of a window.}
    \label{fig:dense-markers}
\end{figure}

To motivate the latter definition, observe that
if $Q\subseteq G_{I_\ell}$ is a marker set with margin factor $\tau=0$,
then every large ground pattern is essentially covered
(except for a margin of width $\ell-1$ along its boundary)
by non-overlapping translations of markers from $Q$.
A non-zero margin factor allows for a non-zero space
and some wiggle room between the markers (see Figure~\ref{fig:dense-markers}).

\begin{lemma}[Covering Lemma] \label{lem:covering}
Let $Q$ be a marker set with margin factor $\tau$ on $I_\ell$.
For $x\in \Omega_\A$, let $f_n(x)$ denote the proportion of $x_{I_n}$
covered by (disjoint) markers from $Q$.
For every $\epsilon,\delta>0$, there exists an inverse temperature $\beta_0$
(given by the previous lemma, corresponding to $I_m$ and $\frac{\epsilon}{2}$)
such that, for every $\beta\geq\beta_0$ and every ergodic $\mu\in\G(\beta)$,
there exists an $n_0\in\N$ such that, for every $n\geq n_0$, we have:
\[
\mu\left(\left\{x\in \Omega_\A,f_n(x)\geq 1-\epsilon'\right\}\right)\geq 1-\delta \eqnsp ,
\]
with $\epsilon'=1-\frac{1-\epsilon}{(1+\tau)^d}$.

\begin{proof}
Let $\beta_0=\frac{2\log_2(\abs{\A})m^d}{\alpha\epsilon}$.
Let $I=I_\ell$ and $J=I_m$.
The previous lemma gives $\mu\left(\left[G_J\right]\right)\geq 1-\frac{\epsilon}{2}$
for any $\beta\geq\beta_0$ and $\mu\in\G(\beta)$.

By the convergence in probability version of the ergodic theorem:
\[
\mu\left(\left\{x\in \Omega_\A,\abs[\bigg]{\mu\left(\left[G_J\right]\right)-\frac{1}{\abs{I_n}}
\sum\limits_{k\in I_{n-m+1}}\mathds{1}_{\left[G_J\right]}\circ\sigma^k(x)}
>\frac{\epsilon}{2}\right\}\right)\underset{n\to\infty}{\longrightarrow} 0 \eqnsp .
\]
It follows that, after some rank $n_0$, with $\mu$-probability of at least $1-\delta$:
\[
\frac{1}{\abs{I_n}}\sum_{j\in I_{n-m+1}}
\mathds{1}_{\left[G_J\right]}\circ\sigma^j(x)\geq 1-\epsilon \eqnsp .
\]
Thus, letting $\mathcal{J}_n(x):=\left\{j\in I_{n-m+1},\sigma^j(x)\in\left[G_J\right]\right\}$,
we have $\frac{\abs{\mathcal{J}_n(x)}}{\abs{I_n}}\geq 1-\epsilon$
with $\mu$-probability at least $1-\delta$.

Consider a configuration $x$ satisfying $\frac{\abs{\mathcal{J}_n(x)}}{\abs{I_n}}\geq 1-\epsilon$.
For every $j\in \mathcal{J}_n(x)$, $x$ contains a translation of a pattern from $Q$ within $J+j$.
That is, there is an $i\in\Z^d$ such that $I+i\subseteq J+j$ and $\sigma^i(x)_I\in Q$.
Such a window $I+i$ can fit in at most
\[
[J:I]:=\abs[\big]{\{j,I\subseteq J+j\}}=\abs{I_{m-\ell+1}} \leq (1+\tau)^d \ell^d
\]
positions inside $J$.

Let $\mathcal{P}_n(x)$ denote the set of pairs $(j,k)$
with $k\in I+i\subseteq J+j\subseteq I_n$ for some $i$,
such that $x_{I+i}$ is a marker from $Q$ and $x_{J+j}$ is a ground pattern.
Let $\mathcal{K}_n(x)$ denote the set of positions $k\in I_n$
that are covered in $x$ by some marker from $Q$,
so that $f_n(x)=\frac{\abs{\mathcal{K}_n(x)}}{\abs{I_n}}$.
By double counting the elements of $\mathcal{P}_n(x)$, we get:
\[
\abs{I}\times\abs{\mathcal{J}_n(x)} \leq \abs{\mathcal{P}_n(x)}
\leq \abs{\mathcal{K}_n(x)} \times [J:I] \eqnsp .
\]
The left inequality comes from the fact that for each position $j\in \mathcal{J}_n(x)$,
$x$ contains a marker from $Q$ on $J+j$,
hence there are at least $\abs{I}$ possible values $k$ for which $(j,k)\in\mathcal{P}_n(x)$.
To see why the inequality on the right holds,
observe that every $k\in\mathcal{K}_n(x)$ is covered by a unique marker in $x$
(recall that markers do not overlap!),
hence a unique $i$ such that $x_{I+i}$ is a marker and $k\in I+i$.
In turn, for every such $i$, there exists at most $[J:I]$ values for $j$
such that $I+i\subseteq J+j\subseteq I_n$ and $x_{J+j}$ is a ground pattern.

We conclude that:
\[
f_n(x)=\frac{\abs{\mathcal{K}_n(x)}}{\abs{I_n}}\geq 
\frac{\abs{I}}{[J:I]}\frac{\abs{\mathcal{J}_n(x)}}{\abs{I_n}}
\geq \frac{1}{(1+\tau)^d}\times (1-\epsilon) =1-\epsilon' \eqnsp ,
\]
for $x$ in a set with $\mu$-probability at least $1-\delta$.
\end{proof}
\end{lemma}

Intuitively, for a Gibbs measure $\mu$ at low temperature,
most of any $\mu$-typical configuration is covered by markers.
The lower the temperature, the higher the density regions covered by markers.
The next lemma formulates the fact that, when $\mu$ is ergodic,
among the occurrences of markers in a $\mu$-typical configuration,
the relative frequency of each marker follows the law of $\mu$.

\begin{lemma}[Relative Frequencies] \label{lem:relative-frequency}
Let $Q$ be a set of patterns on a window $I\Subset\Z^d$,
and $\mu\in\M_\sigma\left(\Omega_\A\right)$ any ergodic measure such that $\mu([Q])>0$.

Then, for every $\delta,\kappa>0$, there exists an $n_0$ such that,
for every $n\geq n_0$, with $\mu$-probability at least $1-\delta$,
the frequency of a pattern $q\in Q$ among all occurrences of $Q$ (completely) inside $I_n$
is in the interval $\left[\mu\left([q]\middle|[Q]\right)-\kappa,\mu([q]|[Q])+\kappa\right]$.

\begin{proof}
By the ergodic theorem,
$\frac{1}{\abs{I_n}}\sum\limits_{i,I+i\subseteq I_n}\mathds{1}_{[Q]}\circ\sigma^i 
\overset{\mu\text{-a.s.}}{\longrightarrow} \mu([Q])>0$.
This convergence also holds for the cylinder $[q]$, so by taking their ratio we conclude that
the frequency of $q\in Q$ satisfies:
\[
\frac{\sum_{I+i\subseteq I_n}\mathds{1}_{[q]}\circ\sigma^i}
{\sum_{I+i\subseteq I_n}\mathds{1}_{[Q]}\circ\sigma^i}
\overset{\mu\text{-a.s.}}{\longrightarrow} \frac{\mu([q])}{\mu([Q])}=\mu([q]|[Q]) \eqnsp .
\]
Almost sure convergence implies convergence in probability,
which happens uniformly in the finite set $Q$, hence the claimed result.
\end{proof}
\end{lemma}

Lemma~\ref{lem:relative-frequency} in particular applies
under the hypotheses of Lemma~\ref{lem:covering}.
The goal of the next two subsections is to provide technical tools
to obtain explicit information on the conditional distribution $\mu_Q:q\mapsto\mu([q]|[Q])$.

\subsection{Tighter Bounds on the Pressure} \label{sec:uniform-markers:pressure-bounds}

The variational characterisation of the Gibbs measures
offers the following approach to deducing information about them:
In order to show that every element of $\G(\beta)$ satisfies a property $\star$,
one can by contraposition show that if $\mu$ does not satisfy $\star$,
then the pressure of $\mu$ is not maximal, that is, $p_\mu(\beta)<p(\beta)$.
Following this general idea, we will later study the property that
the empirical distribution of markers is almost uniform.
Since the exact value of the pressure is hardly ever explicitly known,
using bounds is a reasonable way to proceed.

In Lemma~\ref{lem:ground-patterns}, we used the trivial pressure bounds $p(\beta)\geq 0$
(which is independent of $\beta$) and $p_\mu(\beta)\leq\log_2(\abs{\A})-\beta\mu(\varphi)$.
The following lemmas provide tighter bounds on $p(\beta)$ and $p_\mu(\beta)$.

\begin{lemma}[Lower Bound on the Topological Pressure] \label{lem:pressure-lower-bound}
Let $n\geq 2r$, and let $S\subseteq\A^{I_n}$ be a non-empty set of patterns.
Let $\hat{E}\in\R$ be such that $\max_{q\in S} E_{I_n}(q)\leq\hat{E}$.
Then, for every $\beta>0$:
\[
p(\beta) \geq \frac{\log_2(\abs{S}) - \beta\hat{E}}{\abs{I_n}}
-\beta\frac{\abs{I_n}-\abs{I_{n-2r}}}{\abs{I_n}}\norm{\phi} \eqnsp ,
\]
and the term on the right is of order $O\left(\frac{\beta}{n}\right)$.

\begin{proof}
We use the fact that $p(\beta)\geq p_\mu(\beta)$ for every measure $\mu\in\M_\sigma(\Omega_\A)$.
Consider the measure $\mu$ defined as follows.
First, partition the lattice $\Z^d$ into blocks $\left(I_n+nk\right)_{k\in\Z^d}$.
Then, for each block,
pick a pattern from $S$ uniformly at random and independently of the other blocks.
This defines a periodic measure $\nu\in\M\left(\Omega_\A\right)$.
Finally, average this measure $\nu$ over its finite translation orbit
to obtain $\mu\in\M_\sigma(\Omega_\A)$.

The entropy per site of $\mu$ is $h(\mu)=\frac{1}{\abs{I_n}}\log_2(\abs{S})$.
The expected energy per site under $\mu$ can be bounded as follows:
\[
\begin{split}
\mu\left(\phi\right) &=
\frac{1}{\abs{I_n}}\sum_{i\in I_n}\nu\left(\phi\circ\sigma^i\right)\\
&=\frac{1}{\abs{I_n}}\sum_{i\in I_n}\sum_{C\ni i}\frac{1}{\abs{C}}\nu\left(\Phi_C\right)\\
&=\frac{1}{\abs{I_n}}\left(\sum_{C\subseteq I_n}\nu\left(\Phi_C\right)
+\sum_{C\not\subseteq I_n}\frac{\abs{C\cap I_n}}{\abs{C}} \nu\left(\Phi_C\right)\right)\\
&\leq\frac{1}{\abs{I_n}}\left(\nu\left(E_{I_n}\right)+\sum_{i\in I_n \backslash
\llbracket r,n-r-1\rrbracket^d}\nu\left(\phi\circ\sigma^i\right)\right)\\
&\leq\frac{1}{\abs{I_n}}\left(\hat{E}
+\left(\abs{I_n}-\abs{I_{n-2r}}\right)\norm{ \phi}\right) \eqnsp .
\end{split}
\]
The claimed lower bound for $p(\beta)$ follows.
\end{proof}
\end{lemma}

\begin{lemma}[Upper Bound on the Pressure] \label{lem:pressure-upper-bound}
Let $\mu\in\M_\sigma(\Omega_\A)$ be an ergodic measure.
Let $n_1,n_2,\ldots$ be a sequence with $n_i\nearrow\infty$, and for each $i$,
consider a set of patterns $R_i\subseteq\A^{I_{n_i}}$.
Let $\gamma>0$ and $\check{E}_i$ be such that $\mu\left(\left[R_i\right]\right)\geq\gamma$
and $\min_{q\in R_i}E_{I_{n_i}}(q)\geq\check{E}_i$ for every $i$.
Then, for every $\beta>0$ we have:
\[
p_\mu(\beta) \leq \varliminf_{i\to\infty} 
\frac{\log_2\left(\abs{R_i}\right)-\beta\check{E}_i}{\abs{I_{n_i}}} \eqnsp .
\]

\begin{proof}
By the Shannon-McMillan-Breiman theorem (with convergence in probability),
for every ${\alpha_1>0}$, we have:
\[
\mu\left(\left\{x\in\Omega_\A,\abs[\bigg]{\frac{-\log_2\left( 
\mu\left(\left[x_{I_n}\right]\right)\right)}
{\abs{I_n}}-h(\mu)}>\alpha_1\right\}\right)\underset{n\to\infty}{\longrightarrow}0\eqnsp .
\]
Likewise, using the ergodic theorem (with convergence in probability),
for every $\alpha_2>0$, we have:
\[
\mu\left(\left\{x\in \Omega_\A,\abs[\bigg]{\frac{E_{I_n}(x)}{\abs{I_n}}-\mu\left(\phi\right)}
>\alpha_2\right\}\right)\underset{n\to\infty}{\longrightarrow} 0 \eqnsp .
\]
Combining the two, for every $\alpha>0$, we obtain:
\[
\mu\left(\left\{x\in \Omega_\A,\abs[\bigg]{\frac{
-\log_2\left(\mu\left(\left[x_{I_n}\right]\right)\right)
-\beta E_{I_n}(x)}{\abs{I_n}}-p_\mu(\beta)}
>\alpha\right\}\right)\underset{n\to\infty}{\longrightarrow} 0 \eqnsp .
\]

Now, let $0<\epsilon<\gamma$.
For $n$ large enough, the latter probability is at most $\epsilon$, thus,
with probability at least $1-\epsilon$, we have $\mu\left(\left[x_{I_n}\right]\right)
= 2^{\left(-p_\mu(\beta)\pm\alpha\right)\abs{I_n}-\beta E_{I_n}(x)}$.
Let $R_i'\subseteq R_i$ denote the subset of patterns
for which the latter estimate holds.
Then, for $i$ big enough, we have:
\[
0<\gamma-\epsilon\leq \mu\left(\left[R_i'\right]\right)
=\sum_{q\in R_i'} \mu([q])\leq \abs{R_i}\times
2^{-\left(p_\mu(\beta)-\alpha\right)\abs{I_{n_i}}-\beta \check{E}_i} \eqnsp .
\]
This can be rewritten as:
\[
p_\mu(\beta)\leq\frac{\log_2\left(\abs{R_i}\right)-\beta\check{E}_i}{\abs{I_{n_i}}}
-\frac{\log_2(\gamma-\epsilon)}{\abs{I_{n_i}}}+\alpha \eqnsp .
\]
We conclude by taking the $\varliminf$ as $i\to\infty$ and noticing that $\alpha>0$ is arbitrary.
\end{proof}
\end{lemma}

\subsection{Counting Templates} \label{sec:uniform-markers:templates}

We shall refer to the sets $R_i$ used in Lemma~\ref{lem:pressure-upper-bound} as \emph{templates}.
In order to apply the lemma, we will need appropriate choices for templates.
Furthermore, we need a lower bound $\check{E}_i$ for the energy contents of the elements of $R_i$,
and an upper bound for the entropy of $R_i$.
In this section, we introduce an appropriate family of templates in terms of markers,
and give a sharp upper bound for their entropies.
As it turns out, the lower bound $\check{E}_i=0$ for the energy contents will be sufficient for us.

\begin{definition}[Marker Templates] \label{def:templates}
Let $Q$ be a marker set on $I_\ell$, $\mathcal{W}$ a set of probability measures on $Q$,
$\epsilon>0$ and $n\geq\ell$.
Let us define $R_n\left[Q,\mathcal{W},\epsilon\right]$ as the set of patterns $p\in\A^{I_n}$ such that:
\begin{itemize}
    \item $p$ is $\epsilon$-covered by markers,
    that is, $\abs[\big]{\left\{k\in I_n,\exists i \text{ s.t.\ }
    k\in I_\ell+i\subseteq I_n,p_{I_\ell+i}\in Q\right\}}\geq (1-\epsilon)\abs{I_n}$,
    \item the empirical distribution of markers from $Q$
    covering $p$ belongs to $\mathcal{W}$.
\end{itemize}
We call $R_n\left[Q,\mathcal{W},\epsilon\right]$
the \emph{$(Q,\mathcal{W},\epsilon)$-template} on $I_n$.
\end{definition}

\begin{lemma}[Entropy of Marker Templates] \label{lem:template-counting}
Let $Q$ a marker set on $I_\ell$, and $\mathcal{W}$ a closed set of probability measures on $Q$.
Then, for every $0<\epsilon<\frac{1}{2}$ and $n$ sufficiently large:
\[
\frac{1}{\abs{I_n}}\log_2\left(\abs{R_n[Q,\mathcal{W},\epsilon]}\right)\leq
\epsilon\left(1+\log_2(\abs{\A})\right)+H(\epsilon)+\frac{H^*}{\abs{I_\ell}} \eqnsp ,
\]
where $H(\epsilon):=-\epsilon\log_2(\epsilon)-(1-\epsilon)\log_2(1-\epsilon)$ is the binary entropy function,
and $H^*=\max_{w\in\mathcal{W}}H(w)$ stands for the maximal entropy among the elements of~$\mathcal{W}$.

\begin{proof}
Note that the positions of each of the markers
can be directly deduced from the set
of coordinates not covered by markers.
An element of $R_n[Q,\mathcal{W},\epsilon]$ can thus
be described as a set of $r$ non-covered sites
(with $r\leq\epsilon\abs{I_n}$),
a symbol from $\A$ for each such site,
and a family of $M_r:=\frac{\abs{I_n}-r}{\abs{I_\ell}}$
markers from $Q$ whose empirical distribution belongs to~$\mathcal{W}$.
Hence:
\[
\abs{R_n[Q,\mathcal{W},\epsilon]} \leq
\sum_{r\leq\epsilon\abs{I_n}} \binom{\abs{I_n}}{r} \abs{\A}^r
\sum_{\rho\in\mathcal{W}}\abs{T_{M_r}(\rho)} \eqnsp ,
\]
with $T_m(\rho)$ the set of families of elements in $Q$ of length $m$
with empirical distribution $\rho$.
In order for $T_m(\rho)$ to be non-empty,
$\rho$ must be a rational measure  with common denominator $m$,
of which at most $(m+1)^{\abs{Q}}$ exist.
For such probability measures, a known upper bound~\cite[Theorem 11.1.3]{CoTho06}
gives $T_m(\rho)\leq 2^{m H(\rho)}$.
It follows that:
\[
\begin{split}
\abs{R_n[Q,\mathcal{W},\epsilon]} &\leq
\sum\limits_{r\leq\epsilon\abs{I_n}} \binom{\abs{I_n}}{r} \abs{\A}^r
\left(M_r+1\right)^{\abs{Q}}2^{M_r H^*} \\
&\leq \sum\limits_{r\leq\epsilon\abs{I_n}}\binom{\abs{I_n}}{r} 
\abs{\A}^{\epsilon\abs{I_n}} \left(M_0+1\right)^{\abs{Q}}2^{M_0 H^*} \\
&\leq 2^{H(\epsilon) \abs{I_n}}
\abs{\A}^{\epsilon\abs{I_n}} \left(M_0+1\right)^{\abs{Q}}2^{M_0 H^*} \eqnsp ,
\end{split}
\]
where, in the last step, we have used the general inequality
$\sum_{r\leq\epsilon m}\binom{m}{r}\leq 2^{H(\epsilon)m}$,
which holds whenever $\epsilon<\frac{1}{2}$ (see~\cite[Theorem 3.1]{Gal14}).
By taking the logarithm, we obtain:
\[
\frac{1}{\abs{I_n}}\log_2\left(\abs{R_n[Q,\mathcal{W},\epsilon]}\right)\leq
H(\epsilon)+\epsilon\log_2(\abs{\A})+\abs{Q}\frac{\log_2\left(\abs{I_n}+\abs{I_\ell}\right)
-\log_2\left(\abs{I_\ell}\right)}{\abs{I_n}}+\frac{H^*}{\abs{I_\ell}} \eqnsp .
\]
In particular, as $n\to \infty$, we have
$\abs{Q}\frac{\log_2\left(\abs{I_n}+\abs{I_\ell}\right)
-\log_2\left(\abs{I_\ell}\right)}{\abs{I_n}}\to 0$,
hence the $\epsilon\times 1$ term in the announced bound for $n$ large enough.
\end{proof}
\end{lemma}

\subsection{Equidistribution for Relative Frequencies of Markers}

Exploiting the results of Section~\ref{sec:uniform-markers:ubiquity}
and the bounds in Section~\ref{sec:uniform-markers:pressure-bounds},
we now want to show that, at moderately low temperatures,
markers cover almost all of the lattice, and with an almost uniform empirical distribution among themselves.

More specifically, we already know (from Lemmas~\ref{lem:covering} and~\ref{lem:relative-frequency})
that below a certain temperature, the markers cover most of the lattice and their empirical distribution
is close to the conditional measure $\mu_Q$ (recall that $\mu_Q:=\mu([q]|[Q])$).
We now identify a lower temperature above which the entropy of $\mu_Q$ is close to $\log_2(\abs{Q})$,
the entropy of the uniform distribution $\lambda_Q$.
Hence, within the designated temperature interval,
markers appear in a ``disordered'' fashion (\ie with more or less uniform frequencies).

Let us emphasize that at much lower temperatures,
markers group themselves into ``ordered'' blocks, hence their frequencies will be far from uniform.

\begin{theorem}[Equidistribution] \label{thm:equidistribution}
Let $\Phi$ be a non-negative, finite-range interaction (of range $r$)
with a non-empty set $Z_\Phi$ of null-energy configurations.
Suppose that $\Phi$ is not identically null, and let $\alpha=\min\left\{\Phi_I(x),\Phi_I(x)>0\right\}$.

Let $Q$ be a marker set on $I:=I_\ell$ with margin factor $\tau$ (see Definition~\ref{def:marker}),
and let $J:=I_m$ with $m=(2+\tau)\ell-1$.
Let $\kappa,\epsilon>0$, $n\geq 2r$, and
$\epsilon':=1-\frac{1-\epsilon}{(1+\tau)^d}$,
and assume $\kappa,\epsilon'<\frac{1}{2}$.
Suppose that the following two criteria are satisfied:
\begin{itemize}
    \item Entropy: $\frac{\log_2\left(\abs{G_{I_n}}\right)}
    {\abs{I_n}}\geq (1-\kappa)\frac{\log_2(\abs{Q})}{\abs{I}}$,
    \item Temperature:
    $\frac{2\log_2(\abs{\A})}{\alpha\epsilon}\abs{J}\leq\beta\leq
    \frac{\epsilon}{\norm{\phi}}\frac{
    \abs{I_n}}{\abs{I_n}-\abs{I_{n-2r}}}$.
\end{itemize}
Then, for every $\mu\in\G(\beta)$,
we have the following properties:
\begin{enumerate}
\item Covering: $\mu(\{x\in \Omega_\A,\exists i\in\Z^d,0\in I+i,x_{I+i}\in Q\})\geq 1-\epsilon'$,
\item Uniformity:
$H\left(\mu_Q\right)\geq (1-2\kappa)\log_2(\abs{Q})-
\left[\epsilon+\epsilon'\left(1+\log_2(\abs{\A})\right)
+H\left(\epsilon'\right)\right]\times\abs{I}-H(\kappa)$.
\end{enumerate}

\begin{proof}
The covering claim follows directly
from Lemma~\ref{lem:covering} for ergodic measures,
as long as the lower bound of the temperature criterion is satisfied.
The extension to all of $\G(\beta)$ then follows by ergodic decomposition,
once we recall that $\G(\beta)$ is a Choquet simplex
with the ergodic Gibbs measures as its extremal points~\cite[Corollary~3.14]{Ruelle04}.

To prove the second claim,
consider $\mathcal{W}:=\left\{w,
\norm{ w-\mu_Q}_{\mathrm{TV}}\leq \kappa\right\}$,
the $\kappa$-neighbourhood of $\mu_Q$
with respect to the total variation distance
$\norm{ w-w'}_{\mathrm{TV}}:=\frac{1}{2}\sum_{q\in Q}\abs{w'(q)-w(q)}$.
Observe that, if $\abs{w(q)-\mu_Q(q)}\leq \frac{2\kappa}{\abs{Q}}$
for every $q\in Q$, then $w\in \mathcal{W}$.

Let $0<\delta<1$.
Using the temperature criterion again,
we can apply Lemma~\ref{lem:covering}~and~\ref{lem:relative-frequency}
(for $\frac{2\kappa}{\abs{Q}}$ instead of $\kappa$).
Reformulating the result in the vocabulary of Section~\ref{sec:uniform-markers:templates},
for each $n$, we have $\mu(R_n)\geq 1-\delta$
where $R_n:=R_n[Q,\mathcal{W},\epsilon']$ is the $(Q,\mathcal{W},\epsilon')$-template.
Using these templates in Lemma~\ref{lem:pressure-upper-bound}, with $\gamma=1-\delta>0$ and $\check{E}_i=0$,
we have $p(\beta)=p_\mu(\beta)\leq \varliminf\frac{ \log_2\left(\abs{R_i}\right)}{\abs{I_i}}$.
Now, applying Lemma~\ref{lem:template-counting},
we obtain the bound
\[
p(\beta)\leq\frac{H^*}{\abs{I}}+ \epsilon'\left(1+\log_2(\abs{\A})\right)
+H\left(\epsilon'\right) \eqnsp , 
\]
where $H^*=\max_{w\in\mathcal{W}}H(w)$.

On the other hand, using Lemma~\ref{lem:pressure-lower-bound},
we have:
\begin{alignat*}{3}
& p(\beta)&&\geq \frac{\log_2\left(\abs{G_{I_n}}\right)}{\abs{I_n}}&&
-\beta\frac{\abs{I_n}-\abs{I_{n-2r}}}{\abs{I_n}}\norm{\phi}\\
& &&\geq(1-\kappa)\frac{\log_2(\abs{Q})}{\abs{I}}&&-\epsilon \eqnsp ,
\end{alignat*}
with the bound on the left term coming from the entropy criterion,
and the bound on the right term following from temperature criterion.

Comparing the two bounds on $p(\beta)$ we find that:
\[
H^*\geq (1-\kappa)\log_2(\abs{Q})-
\left[\epsilon+\epsilon'\left(1+\log_2(\abs{\A})\right)
+H\left(\epsilon'\right)\right]\times\abs{I} \eqnsp .
\]
The uniformity claim now follows from the general bound
$\abs{H(w)-H(w')}\leq H(\kappa)+\kappa\log_2(\abs{Q})$ which holds whenever
$w$ and $w'$ are two measures on $Q$ with $\norm{w-w'}_{\mathrm{TV}}\leq\kappa\leq\frac{1}{2}$~\cite{Zha07}.
\end{proof}
\end{theorem}

\begin{figure}
	\centering
	\begin{tikzpicture}
		\draw [very thick, ->] (-1,0) -- (9,0) ;
		\draw (9,0) node [right] {$\beta$} ;
		
		\draw (1,0) node [above] {$\beta\in T_1$} ;
		\draw (1,.3) node [above] {$\overbrace{\hspace{80pt}}$} ;
		\draw (1,.6) node [above]
		{$\G(\beta)\subseteq B\left(\lambda_1,\theta_1\right)$} ;
		
		\draw (3,0) node [below] {$\beta\in T_2$} ;
		\draw (3,-.3) node [below] {$\underbrace{\hspace{80pt}}$} ;
		\draw (3,-.6) node [below]
		{$\G(\beta)\subseteq B\left(\lambda_2,\theta_2\right)$} ;
		
		\draw (5,0) node [above] {$\beta\in T_3$} ;
		\draw (5,.3) node [above] {$\overbrace{\hspace{80pt}}$} ;
		\draw (5,.6) node [above]
		{$\G(\beta)\subseteq B\left(\lambda_3,\theta_3\right)$} ;
		
		\draw (7,0) node [below] {$\beta\in T_4$} ;
		\draw (7,-.3) node [below] {$\underbrace{\hspace{80pt}}$} ;
		\draw (7,-.6) node [below]
		{$\G(\beta)\subseteq B\left(\lambda_4,\theta_4\right)$} ;
	\end{tikzpicture}
	\caption{Overlapping temperature intervals for Theorem~\ref{thm:equidistribution}.}
	\label{fig:overlapping-intervals}
\end{figure}
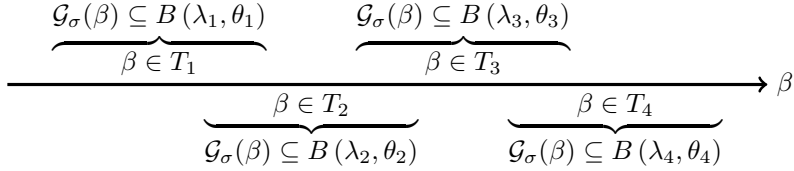

Our objective is to construct a model for which the above theorem can be applied
to a sequence of marker sets $Q_k$ at larger and larger scales, and with $\kappa_k,\epsilon_k\to 0$.
The model will be tuned in such a way that the temperature intervals
corresponding to consecutive scales overlap, as illustrated in Figure~\ref{fig:overlapping-intervals}.
This will provide us with control over the Gibbs measures at \emph{all} temperatures,
and in turn, allow us to engineer the set of ground states.

In the next section, we construct a class of interactions that achieve the above objective.

\section{Constructing Well-Behaved Simulating Tilesets} \label{sec:robinson}

In this section, the goal is to build step by step (and layer by layer)
a structure satisfying the requirements of Theorem~\ref{thm:equidistribution},
and in which a class of \emph{well-behaved} Turing machines can be encoded.
In the following sections, we will use these machines to force specific distributions on words.

A general sketch of the construction will be given
in Section~\ref{sec:robinson:sketch},
after which each following subsection
will focus on one of the layers of the structure.

\subsection{General Ideas of the Construction} \label{sec:robinson:sketch}

The final tiles have $6$ different interdependent layers.
At each step of the construction we obtain a non-empty
SFT $X_{\F_j}$ on an alphabet $\A_j$,
and each layer will be responsible for one aspect of the construction.

\begin{itemize}
\item First, we start with a $X_{\F_1}$ a variant of the Robinson tiling on the alphabet $\A_1$,
enhanced for local alignment of the macro-tiles, with an alternating bicoloured structure,
as was the case in a previous article~\cite{GaySa22}.
This tileset provides us with self-avoiding Red squares in a hierarchical arrangement,
which will allow us to encode increasingly large space-time diagrams of Turing machines,
in the sparse area of any Red square that avoids smaller squares.
Some particular scales of macro-tiles will later be chosen for the marker sets $Q_k$.

\item We add a three-phased layer
$\A_{\mathrm{phase}}:=\{\symb{F},\symb{B},\symb{H}\}$
(\ie Frozen, Blocking and Hot)
and we define the tileset
$\A_2\subseteq \A_1\times\A_{\mathrm{phase}}$ as follows.
Tiles with a Red line on the layer $\A_1$ can be anything,
other tiles must be either $\symb{F}$ or $\symb{H}$.
For the corresponding local rules $\F_2$, neighbouring tiles linked by a Red line on their interface,
or without any Red line on them, must have the same phase value.
For tiles with a Red straight line (setting aside Red corners),
as we have a clearly identifiable ``inside'' and ``outside'' faces,
from outside to inside we only allow the following phase transitions:
$\symb{F}\symb{F}\symb{F}$, $\symb{H}\symb{H}\symb{H}$ and $\symb{H}\symb{B}\symb{F}$.
This last transition represents cases where, in a globally admissible tiling,
the Red square is Blocking, so that all of its inside is Frozen while the outside is Hot.
In practice, in a typical (locally admissible) Robinson macro-tile,
we will have a connected area made of Hot tiles that blur communication,
with several Blocking squares,
each encoding some signal synchronised among all of its Frozen interior.
This behaviour contrasts with generic globally admissible tilings,
which will be entirely Frozen.

\item The entropy criterion of Theorem~\ref{thm:equidistribution}
is easier to satisfy (at increasingly large scales of markers $Q_k$
while maintaining $\kappa_k\to0$) for positive-entropy tilings
(or null-entropy tilings for which $\frac{1}{\abs{I_n}}H\left(G_{I_n}\right)\to h(X)=0$
as slowly as possible).
However, the most naive embeddings of Turing machines into the Robinson tiling
give a really tight structure with a fast-decreasing entropy.
Thus, to preserve the desired $\kappa_k\to0$ behaviour,
we need to make room for more entropy to spawn.
As stated right before,
the goal of Frozen squares in our construction is to encode a signal within,
and to synchronise it with neighbouring Frozen squares.
Because such tiles will only encode very little information
(about $n$ bits of internal information
and $2^n$ bits of outside communication with neighbouring squares for $16^n$ tiles in total),
they will be the biggest source of entropy decrease here.
To counterbalance this phenomenon, we add a layer $\A_{\mathrm{scale}}$ to obtain $\A_3$
with forbidden patterns $\F_3$ that will limit which scales
of Robinson macro-tiles are allowed be Blocking.
More precisely, a square \emph{is allowed} to be Blocking (it is \emph{Blockable}) if and only if
the scale at which it occurs ($N$ for a $(2N+1)$-macro-tile)
is a power of $3$ ($N=3^k$).
This will give us the $k$-th scale of markers:
\[
Q_k:=\left\{\omega\in G_{I_{l_{n_k}}},\pi_{\A_{\mathrm{Robinson}}}(\omega)
\text{ is a Robinson $\left(2\times 3^k+1\right)$-macro-tile}\right\} \eqnsp .
\]

\item Then, a new problem arises in that we eventually want to control
the uniform distribution $\lambda_{Q_k}$ on $k$-markers.
In particular, for the temperature intervals of Theorem~\ref{thm:equidistribution}
to be able to overlap from one scale to the next,
we need to keep the distributions $\lambda_{Q_k}$ and $\lambda_{Q_{k+1}}$
as close as reasonably possible.
To do so, in the rules $\F_4$ for the alphabet $\A_4$ including a new layer $\A_{\mathrm{odometer}}$,
we encode a 2D odometer between Red squares, so that only
a low density $\frac{1}{t_k}$ of $k$-markers within a bigger Hot $(k+1)$-marker
are indeed Blocking (and thus perform computations), while the rest stay Hot.
We want $t_k$ to be big-enough (to have $\lambda_{Q_{k+1}}$ as close to $\lambda_{Q_k}$),
but at the same time not too large
(so that configurations in $X_{\F_4}$ have a density $1$ of Frozen tiles).
Choosing $t_k\approx \log_2(k)$ will allow us to satisfy both wanted properties.

\item With this structure in mind, we add
the last general-purpose layer $\A_{\mathrm{signal}}=\left\{\mo,\symb0,\po\right\}$ to obtain $\A_5$,
which encodes a symbol in each Red line, synchronised in all directions,
so that two neighbouring Frozen tiles encode the same bit $\pmo$ in their central cross,
but Blocking and Hot lines transmit a $\symb0$
(and a Blocking square insulates the $\pmo$ signals on its inside
from synchronising with the $\symb0$ on its outside).
As a globally admissible tiling is generically Frozen everywhere, without an infinite cut,
it encodes a single infinite word in $\{\pmo\}^\N$.

Locally, this synchronising property of $\F_5$ will make it so that
the central Blocking Red square of a $k$-marker will encode
a well-defined binary word $w$ of length $3^k-1$,
which can be seen as a (read-only) Toeplitz encoding
(meaning that $123\dots$ reads as $1213121\dots$, each following element with a period twice as long)
for Turing machines operating inside the sparse space-time diagram.

\item At last, given a well-behaved Turing machine $M$ (see Definition~\ref{def:well-behaved}),
we explain how the tileset $\A_{6(M)}$ is obtained by adding
a layer $\A_{\mathrm{simul}(M)}$ to $\A_5$,
so that inside each Blocking Red square,
computations occur according to $M$ (and the rules $\F_{6(M)}$),
in order to force a given value $M(s)$ on the first bits of $w$
conditionally to each possible value
of the non-deterministic ``random'' seed $s$ given to the machine $M$ as an input.
\end{itemize}

The sum of the results presented in the rest of the section may be summarised as follows:
\begin{maintheorem}
Let $M$ be a \emph{well-behaved} Turing machine with two synchronous tapes.
We assume that, given a binary seed $s$ on the read-only input tape,
$M$ computes a binary word $M(s)$ (on the alphabet $\{\pmo\}$)
of length $\left\lfloor \log_2(\abs{s})\right\rfloor$
with a time complexity $o\left(2^{3^{\abs{s}}}\right)$.

Then there exists a simulating tileset $X_{\F_{6(M)}}$ on the alphabet $\A_{6(M)}$,
such that after some scale $k_0(M)$, a Blocking $k$-marker checks
that the prefix of the word it encodes is $M(s)$,
where $s$ is the free seed of the Blocking square.

For this tileset (and the potential associated to it),
Theorem~\ref{thm:equidistribution} applies to the marker sets $\left(Q_k\right)$,
with overlapping intervals for the temperature criterion,
and such that $\epsilon_k,\kappa_k\to 0$.

Furthermore, there is a common alphabet $\A_0\subseteq\A_{6(M)}$ independent of $M$
(corresponding to Frozen tiles in which no Turing computations occurs),
such that the restriction $\F_0$ of $\F_{6(M)}$ to $\A_0$ does not depend on $M$ either,
and that $\M_\sigma\left(X_{\F_{6(M)}}\right)=\M_\sigma\left(X_{\F_0}\right)$.
This common subshift is ``universal'', in the sense that we have
an affine bijection $\gamma:\M\left(\{\pmo\}^\N\right)\to\M_\sigma\left(X_{\F_0}\right)$.
\end{maintheorem}

\subsection{Layer 1: Robinson Tiles}

The Robinson tileset~\cite{Rob71} has been introduced to prove
the undecidability of the tileability of the plane
through a much smaller and simpler tileset than Berger's first example~\cite{Ber66}.
The Robinson tileset enforces an aperiodic hierarchical structure in the tilings,
in which one can embed Turing machine space-time diagrams
(see the lecture notes by Jeandel and Vanier~\cite{JeanVa20} for more details on the construction).

\begin{figure}
    \centering
    \includegraphics[width=.7\textwidth]{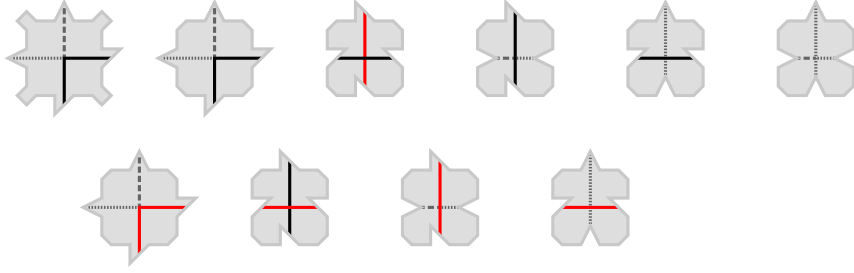}
    \caption{Variant of the Robinson tileset,\\
    using two colours for the full lines to obtain non-intersecting squares.\\
    The dotted and dashed lines help to strengthen the structure.}
    \label{fig:RobEnhanced}
\end{figure}

\begin{figure}
    \centering
    \includegraphics[width=.7\textwidth]{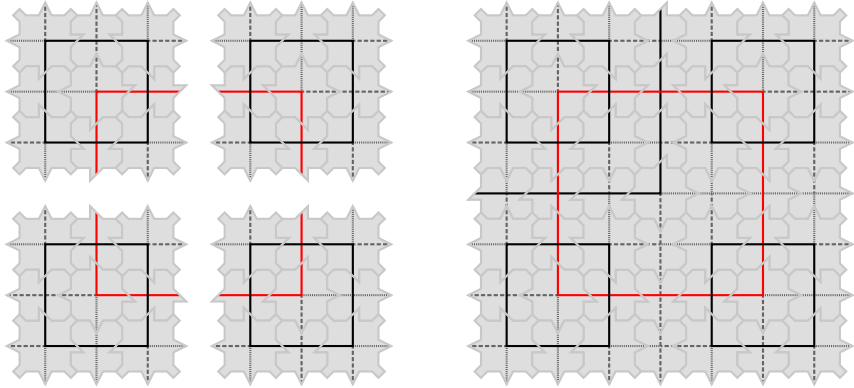}
    \caption{On the left, four $2$-macro-tiles form a square with their Red corners.\\
    On the right, the cross is filled to form a $3$-macro-tile,\\
    with a Red square and a Black corner.}
    \label{fig:macrorule}
\end{figure}

If we try to use the canonical Robinson tileset to define marker sets,
we will fail to achieve the covering property.
Instead, we will use here the variant tiles of Figure~\ref{fig:RobEnhanced} as well as their rotations
(and symmetries and exchanges of dashed and dotted lines,
except for the three leftmost ones), giving us a tileset $\A_1=\A_{\mathrm{Robinson}}$.
The forbidden patterns $\F_1$ are chosen so that
two neighbouring tiles must have matching shapes
and the same type and colour of line along their common edge,
and a square of four tiles must have exactly one tile with ``bumpy-corners''
(the rotations of the top-left tile in Figure~\ref{fig:RobEnhanced})
filling the small diamond-shaped hole in the middle.
Figure~\ref{fig:macrorule} illustrates a typical admissible pattern using these rules.

\begin{figure}
    \centering
    \begin{minipage}{.49\textwidth}
        \centering
        \includegraphics[width=.85\textwidth]{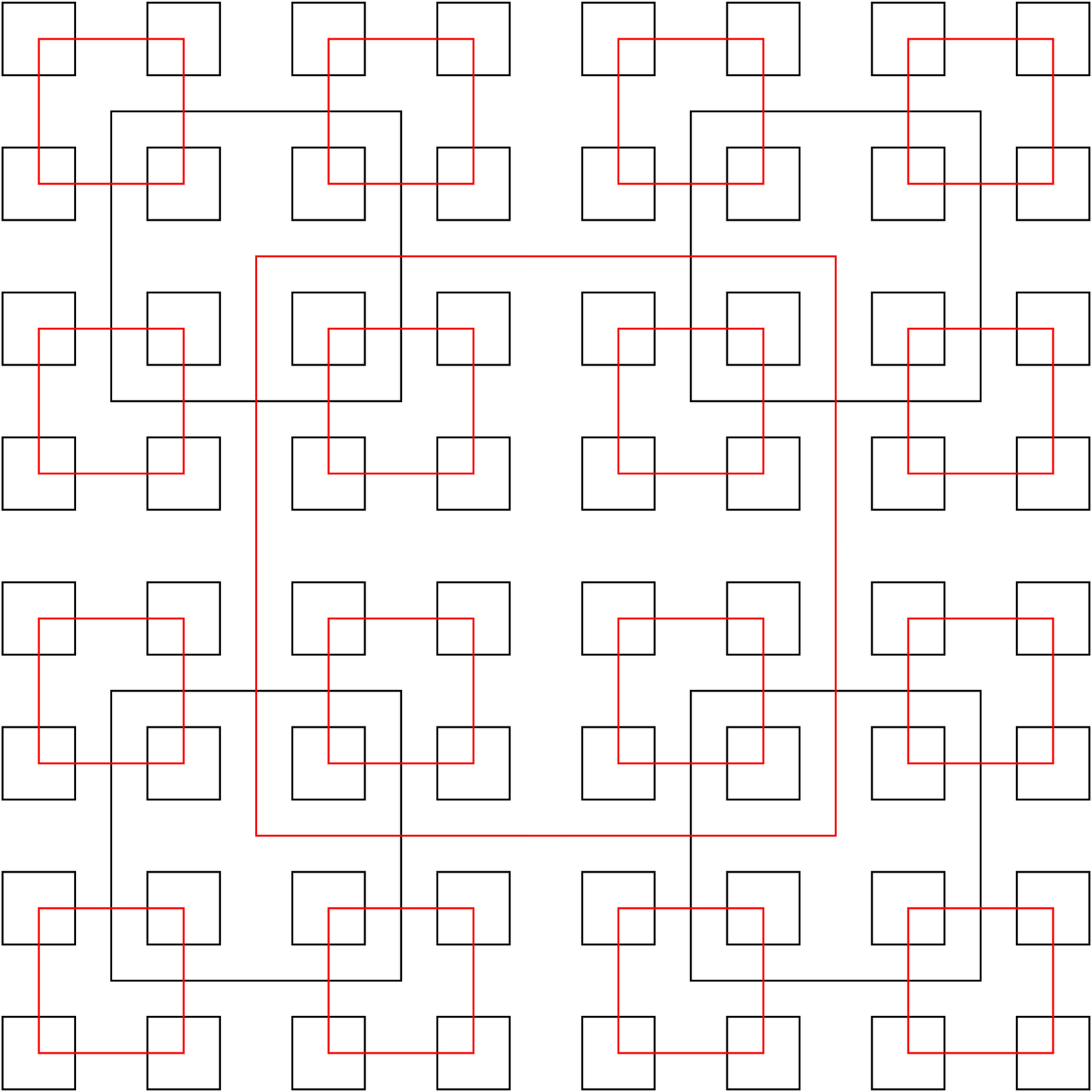}
        \caption{Structure of the Red\\and Black lines in a 5 macro-tile.}
        \label{fig:macrostruct}
    \end{minipage}\hfill
    \begin{minipage}{.49\textwidth}
        \centering
        \includegraphics[width=.85\textwidth]{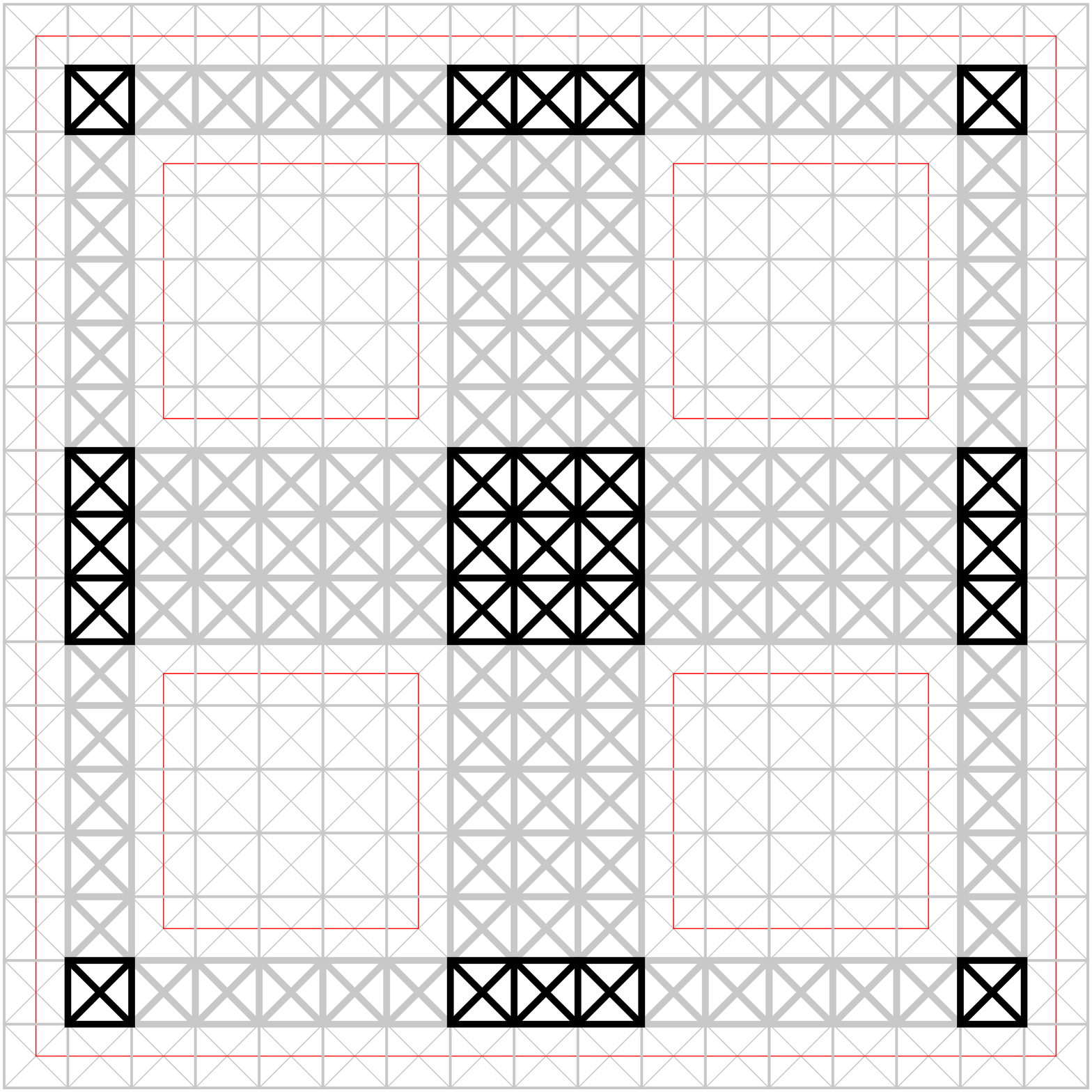}
        \caption{Sparse computation area\\inside the Red square of a 5 macro-tile.}
        \label{fig:sparse}
    \end{minipage}
\end{figure}

\begin{remark}
We call the bumpy-corner tiles $1$-macro-tiles.
In their middle, they have a cross with a Black corner, that can take four different orientations.
Their Black corners can be arranged in a square shape, around a bigger cross with a Red corner,
which we will call a $2$-macro-tile. Now, this Red corner can have four possible orientations,
and we can make a square with four of them around a bigger Black corner,
which gives us a $3$-macro-tile as in Figure~\ref{fig:macrorule}.
Likewise, by induction,
we can make a square of $n$-macro-tiles around a cross to form a $(n+1)$-macro-tile.
In particular, as is the case for the Robinson tiling itself,
$n$-macro-tiles are non-overlapping (as is required for marker sets in Definition~\ref{def:marker}).

A $n$-macro-tile is a square $I_{l_n}$ of tiles, with $l_n:=2^n-1$
(hence a $2\times2^n$-periodic behaviour for $n$-macro-tiles in a generic globally admissible tiling).
Its central square naturally has half the size, $2^{n-1}\pm1$
(depending on whether we include the border tiles in the square or not).

Each $n$-macro-tile has a big square in the middle of one colour,
and a central cross with a corner of the other colour.
The colours alternate with the parity of $n$.
In particular, whenever $n=2N+1$ is odd, then the $n$-macro-tile has
a big Red square and a Black corner in the middle.

At such scales, the Red square contains a $\left(2^N+1\right)\times \left(2^N+1\right)$ sparse area,
made of rectangle patches avoiding smaller Red squares,
connected to each other by communication channels,
as illustrated in Figure~\ref{fig:sparse}.
This sparse area will allow us to encode arbitrarily big spacetime diagrams,
which will be of use in one way or another in most of the next layers.
\end{remark}

\begin{lemma} \label{lem:ergodicity-robinson}
The SFT $X_{\F_1}$ is uniquely ergodic.

\begin{proof}
The proof uses the very same argument as for the usual Robinson tiling,
and we will just give a rough sketch here.
Inductively, a given macro-tile must be inside arbitrarily large(r) macro-tiles.
At the limit, we conclude that any tiling $\omega\in X_{\F_1}$
is made of one, two or four ``$\infty$-macro-tiles'' occupying the whole plane,
two half-planes or four quarter-planes, separated by a ``cut''.
In each such $\infty$-macro-tile,
small $n$-macro-tiles of any given scale $n\geq 1$ behave periodically,
hence we have well-defined frequencies for any finite pattern.
These frequencies are the same regardless of which kind of $\infty$-macro-tile we are considering,
and the ``cuts'' have a null density in $\Z^2$,
hence we always obtain the same ergodic measure $\mu_\omega\in\M_\sigma\left(\Omega_\A\right)$
regardless of the base configuration $\omega\in X_{\F_1}$.
\end{proof}
\end{lemma}

In a previous article~\cite[Proposition 7.7]{GaySa23},
we proved a reconstruction property on a similar tileset
(the proof of which applies verbatim to the current tilset),
which states that if we look at a locally admissible tiling $\omega$ of a square window,
then by forgetting up to $l_n$ tiles on each border we end up with a family of well-aligned
and well-oriented $n$-macro-tiles.
Notably, this property does not hold for the canonical Robinson tileset.

In particular, to obtain a marker set of macro-tiles,
we so far have a covering property that guarantees that
a $n$-macro-tile (defined on a ${l_n}$-square) can be encountered in a $3l_n$-square.
The previous reconstruction bound is optimal if we want a bound that works for \emph{any} square,
but by adapting the proof of the reconstruction lemma to the specific case
of a $\left(2l_n+5\right)$-square, we obtain the following result:

\begin{lemma}[Covering Lemma] \label{lem:coveringRobinson}
Let $n\geq 2$. For any locally admissible tiling $\omega\in\A_{\mathrm{Robinson}}^{I_{2l_n+5}}$
there is a translation $J\subseteq I_{2l_n+5}$ of $I_{l_n}$ such that $\omega_J$ is a $n$-macro-tile.

The result translates more broadly for admissible tilings on the alphabets $\A_j$
(with $1\leq j\leq 6$ and forbidden patterns $\F_j$)
if we project $\omega\in\A_j^{I_{2l_n+5}}$ onto its first coordinate $\A_{\mathrm{Robinson}}$.

\begin{proof}
Let us remind that, once we have made a square with four $n$-macro-tiles,
the only locally admissible way to fill the central cross is to produce a $(n+1)$-macro-tile.
Hence, if we have a $3\times3$ grid of (well-aligned and well-oriented) $n$-macro-tiles,
we are certain to find among them a square pattern that will form a $(n+1)$-macro-tile.
More generally, if we start with a grid of $(2u+1)\times(2u+1)$ $n$-macro-tiles,
then we can group them into squares so that we extract a $u\times u$ grid of $(n+1)$-macro-tiles.
By a direct induction, starting from a $\left(2^{n-1}-1\right)\times\left(2^{n-1}-1\right)$ grid
of (well-aligned and well-oriented) $2$-macro-tiles,
we can find in it a $n$-macro-tile (in the form of a $1\times 1$ grid).

Now, according to the Reconstruction Lemma~\cite[Proposition 7.7]{GaySa23},
we can always observe such a grid of $2$-macro-tiles in a locally admissible tiling of
$I_m$ with
\[
m=4\times \left(2^{n-1}-1\right)+7=2^{n+1}+3=2l_n+5 \eqnsp ,
\]
which concludes the proof.
\end{proof}
\end{lemma}

It follows that, if we consider $Q$ the set of $n$-macro-tiles,
then $Q$ is a marker set with margin factor $\tau=\frac{6}{l_n}$.
In particular, for reasons that will be explained later on, we define the following marker sets:
\[
Q_k:=\left\{\omega\in G_{I_{l_{n_k}}},\pi_{\A_{\mathrm{Robinson}}}(\omega)
\text{ is a Robinson $n_k$-macro-tile}\right\} \eqnsp ,
\]
with $N_k:=3^k$ and $n_k:=2N_k+1$, and with margin factor $\tau_k:=\frac{6}{l_{n_k}}$.
In other words, $Q_k$ is the set of locally admissible tilings of a square $I_{l_{n_k}}$,
on the alphabet $\A_j$ (with $j\in\llbracket 1,6\rrbracket$ depending on the context),
for which the layer $\A_{\mathrm{Robinson}}$ contains a well-formed $n_k$-macro-tile.
Hence, the $k$-th scale of markers correspond the the $3^k$-th scale of macro-tiles with a Red square.

\subsection{Layer 2: Frozen, Blocking, or Hot}

As said earlier, for the second layer,
we consider $\A_{\mathrm{phase}}=\{\symb{F},\symb{B},\symb{H}\}$ a three-phased alphabet,
or rather two phases, the Frozen tiles $\symb{F}$ and Hot tiles $\symb{H}$,
with Blocking tiles $\symb{B}$ playing the role of a thin interface between the other two.

\begin{figure}
    \centering
    \includegraphics[width=.4\textwidth]{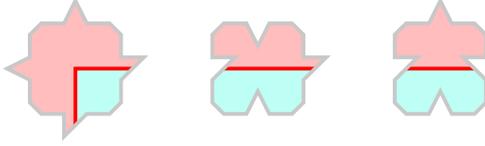}
    \caption{Schematic representation of Blocking tiles with a Red line,\\
    with a Frozen area on the inside and a Hot area on the outside.}
    \label{fig:Blocking}
\end{figure}

We define the tileset
$\A_2\subseteq \A_1\times\A_{\mathrm{phase}}$ as follows.
Tiles with a Red line on the layer $\A_{\mathrm{Robinson}}$ can be anything,
but other tiles must be either $\symb{F}$ or $\symb{H}$ and cannot be of type $\symb{B}$.
The $\symb{B}$ tiles are schematically represented in Figure~\ref{fig:Blocking},
with a red $\symb{H}$ area on their ``outer'' side and a blue $\symb{F}$ area on their ``inner'' side.

For the local rules, we want neighbouring tiles to have ``matching'' phases.
Hence, neighbouring tiles linked by a Red line on their interface,
or without any Red line on them, must have the same phase value.
For tiles with a Red straight line (setting aside Red corners),
as we have a clearly identifiable ``inside'' and an ``outside'' faces,
from outside to inside we only allow the following phase transitions:
$\symb{F}\symb{F}\symb{F}$, $\symb{H}\symb{H}\symb{H}$ and $\symb{H}\symb{B}\symb{F}$.
This last $\symb{H}\symb{B}\symb{F}$ case corresponds to the background colours
in Figure~\ref{fig:Blocking}.

\begin{figure}
    \centering
    \includegraphics[width=.4\textwidth]{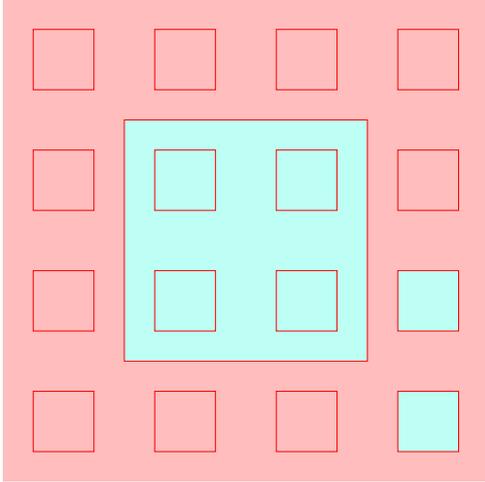}
    \caption{Example of admissible arrangement of Blocking squares in a macro-tile.}
    \label{fig:Puddle}
\end{figure}

Consequently, in a macro-tile (or a globally admissible tiling),
we will typically have several $\symb{B}$ Red squares, with all their inside being $\symb{F}$,
floating around in a connected $\symb{H}$ puddle, as seen in Figure~\ref{fig:Puddle}.
Alternatively, we may have an entirely $\symb{F}$ tiling.
If we halt the construction at this layer,
then just by assuming all the scales are Hot except the very first one,
we obtain a periodic grid of of squares in $3$-macro-tiles,
that can each be independently $\symb{B}$ of $\symb{H}$, hence a positive topological entropy.
The goal of the next two layers is to control which scales and which tiles can be $\symb{B}$,
resulting in a return to the zero-entropy case
(which makes the entropy criterion in Theorem~\ref{thm:equidistribution} harder to satisfy
while still having $\kappa_k\to 0$)
but gives us a much more rigid structure as a result
(which will allow us to quantify the distribution of the yet-to-be-encoded words).

\subsection{Layer 3: Forcing the Blockable Scales}

As stated in the introductory section,
in order for the entropy to decrease to $0$ slowly enough,
we want to control what scales of macro-tiles are \emph{allowed} to have a $\symb{B}$ square,
which we will call \emph{Blockable} scales.
This process is done in two steps,
which we will ultimately identify as the new layer $\A_{\mathrm{scale}}$.
First, we will use a \emph{sequential} cellular automaton~\cite{Wolfram02}
to ``obtain'' $N$ the step of computation (of a $(2N+1)$-macro-tile).
Second, we will use $N$ as an input for a Turing machine to check whether $N=3^k$ for some $k\in\N$.

\subsubsection{Computing the Scale}

\begin{figure}
    \centering
    \includegraphics[scale=3]{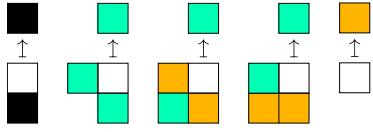}
    \caption{Rules of the sequential cellular automaton,\\
    applied from left to right, in a \emph{if/then/else} fashion.}
    \label{fig:cell_rules}
\end{figure}

For this subsection, forget about the previous tiling, and consider the sequential cellular automaton
with the transition rules from Figure~\ref{fig:cell_rules}.

A one-dimensional cellular automaton $C$ can be simulated in a two-dimensional SFT on $\A$,
with local rules that determine how to compute $x_{t+1}=C\left(x_t\right)$ with $x_t\in\A^\Z$.
A space-time diagram of the automaton thus corresponds to a tiling of the subshift.
Unlike a parallel cellular automaton where each cell's value
at time $t+1$ depends on its neighbourhood at time $t$,
for sequential automata we also need to take into account the left-neighbourhood at time $t+1$
to update sequentially all the cells, by doing a kind of left-to-right sweep.

\begin{figure}
    \centering
    \includegraphics[scale=3]{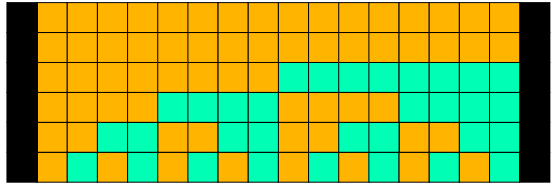}
    \caption{Example of spatially bounded space-time diagram, with $f_0=15$.}
    \label{fig:cell_diagram}
\end{figure}

In our specific case, illustrated in Figure~\ref{fig:cell_rules},
the value $x_{t+1}(k)$ of the position $k\in\Z$ at time $t+1$
depends on $x_t(k-1)$ and $x_t(k)$ at the previous time,
as well as its updated leftmost neighbour $x_{t+1}(k-1)$.
Informally, the automaton can be explained as follows.
With the leftmost rule, black cells represent static walls,
and will stay unchanged from time $t$ to $t+1$ regardless of the left-neighbourhood.
The rest of the rules translate the fact that, \emph{if} at time $t$,
the colour of two neighbouring cells goes from green to yellow,
\emph{then} at time $t+1$ we force a color switch at the corresponding position
(either green-to-yellow or yellow-to-green), \emph{else} the colour stays unchanged.
Note that, for a transition $x_{t+1}=C\left(x_t\right)$ to be well-defined,
one must force an ``initialisation'' on the left of each cell,
which is done through black walls which force their right neighbour
to be initialised as yellow at the next step.
Hence, if we have a finite window bounded by black cells on both ends,
then $f_t$ the number of colour-flips at time $t$ satisfies the recurrence relation 
$f_{t+1}=\left\lfloor\frac{f_t}{2}\right\rfloor$, as can be seen in Figure~\ref{fig:cell_diagram}.
The following proposition directly follows.

\begin{proposition}
Consider a bounded initial configuration with a yellow initial block,
such that $f_0=2^N-1$, with $N\geq 1$.
Then $N$ is written in unary in the rightmost column (which is green up to and including step $N-1$,
after which the whole configuration is a stable yellow block).
\end{proposition}

In particular, if we rotate the whole construction clockwise
(with time transitions from left to right),
adapt the transition rules to work within the sparse computation area 
highlighted in Figure~\ref{fig:sparse},
initialise the first cell as black and then alternate from yellow to green and forth,
we can obtain a unary coding of $N$ on the bottom row of the sparse area
of the $N$-th scale of Red squares, ready to be used by a Turing machine.

\subsubsection{Checking the Scale}

We have now a $\left(2^N+1\right)\times\left(2^N+1\right)$ sparse computing area
(in the $N$-th Red square), with a unary input $N$.
We want to check whether $N$ is a power of three, in $2^N+1$ steps at most.
According to Appendix~\ref{appendix:unary}, we can do so with time complexity $O(N\ln(N))$.
In particular, $N\ln(N)=o\left(2^N+1\right)$,
so we guarantee that for big enough values of $N$ the Turing machine will always terminate
(and either accept or reject $N$) fast-enough.
Once the machine terminates, we make it go idle so that the decision can then reach
the upper border of the bounded spacetime diagram.

Thus, we set-up the tileset so that a Red square is Blockable if and only if
the Turing machine \emph{accepts} the input
(and forbid it to be Blockable if the machine rejets
or simply has not enough time to terminate its computation).

As stated earlier, the set of $k$-markers $Q_k$ corresponds to the $N_k$-th scale of Red squares,
with a $n_k$-macro-tile on the Robinson layer and any locally admissible behaviour on the rest.
We will say that $Q_k$ corresponds to the $k$-th scale of Blockable Red squares.

This new layer has two effects on the \emph{number} of $k$-markers.
In one hand, this number is decreased by the mere fact that most Red square are \emph{not} blockable,
hence we cut out a source of combinatorial explosion.
On the other hand, while this layer acts in a deterministic way
inside a (locally admissible) Red square,
a $k$-marker still has a lot of area outside of Red squares,
that may be used by bigger markers for their own sparse computation areas,
and this adds a new source of entropy to take into account.
These effects will be taken into account later on when we estimate the size of $Q_k$.

\subsubsection{Partitioning the Marker Set}

From now on, we will in particular distinguish
$Q_k=Q_k^\symb{H}\sqcup Q_k^\symb{F} \sqcup Q_k^\symb{B}$,
depending on whether the Red square of a $k$-marker is $\symb{H}$ in the inside
(thus on the sparse area outside),
$\symb{F}$ (everywhere), or a $\symb{B}$ interface (Hot outside, Frozen inside).
We can already prove an upper bound on the cardinality of $Q_k^\symb{F}$:

\begin{lemma} \label{lem:frozen-counting}
Let $Q_k^\symb{F}$ be the set of Frozen $k$-markers, on the alphabet $\A_j$ (with $j\geq 2$ layers).
Then $\abs{ Q_k^\symb{F} } \leq \abs{\A_j}^{\perim_k}$
with $\perim_k:=4 \times \left(l_{n_k}-1\right)=8\left(4^{3^k}-1\right)$ the perimeter of a $k$-marker.

\begin{proof}
The hypothesis $j\geq 2$ means that $\A_j$ contains in particular the second layer
$\A_{\mathrm{phase}}$, so that the subset $Q_k^\symb{F}\subseteq Q_k$ is well-defined.
By definition of $Q_k^\symb{F}$,
the layer $\A_{\mathrm{phase}}$ is here fixed, entirely equal to $\symb{F}$.

For the first layer $\A_{\mathrm{Robinson}}$,
the only degree of freedom is the orientation of its central cross,
which is fixed by outline of the pattern.

In the case $j\geq 3$, on the layer $\A_{\mathrm{scale}}$,
we distinguish two areas.
If we are looking within a given complete Red square,
in which case the odometer performs a complete computation in a deterministic fashion
regardless of the marker $q\in Q_k^{\symb{F}}$ we consider.
If we are looking at the sparse area outside of the Red squares,
then it may be used to perform odometer computations for Red squares at higher scales $k'\geq k$,
but then the computations are deterministic conditionally to a labelling of the outline of the marker.

This argument of \emph{fixed conditionally to the outline}
(either for the sparse computation area, or for some information encoded in Red squares
in a deterministic way or synchronised with neighbouring Red squares at the same scale)
will likewise hold for the layers $\A_{\mathrm{odometer}}$,
$\A_{\mathrm{signal}}$ and $\A_{\mathrm{simul}(M)}$ introduced in the following subsections,
hence the upper bound $\abs{A_j}^{\perim_k}$
with $\perim_k$ the number of tiles in the outline of a $k$-marker.
\end{proof}
\end{lemma}

We now want to obtain similar upper bounds for $Q^\symb{B}_k$ and $Q^\symb{H}_k$, thus on $Q_k$.
Likewise, we want a \emph{lower} bound for $Q^\symb{H}_k$,
which will allow us to conclude that
$\frac{\abs{Q^\symb{H}_k}}{\abs{Q_k}}$ goes to $1$ fast-enough,
that $\lambda_{Q_k}\approx\lambda_{Q_k^\symb{H}}$.
Computing these will require a much more intricate analysis (and an accordingly intricate structure),
which is why we need to wait until all the layers are properly defined (\ie the case $j=6$)
until we can make sense of it.

\subsection{Layer 4: Controlling the Density of Blocking Squares}

\begin{figure}
    \centering
    \includegraphics[width=.85\textwidth]{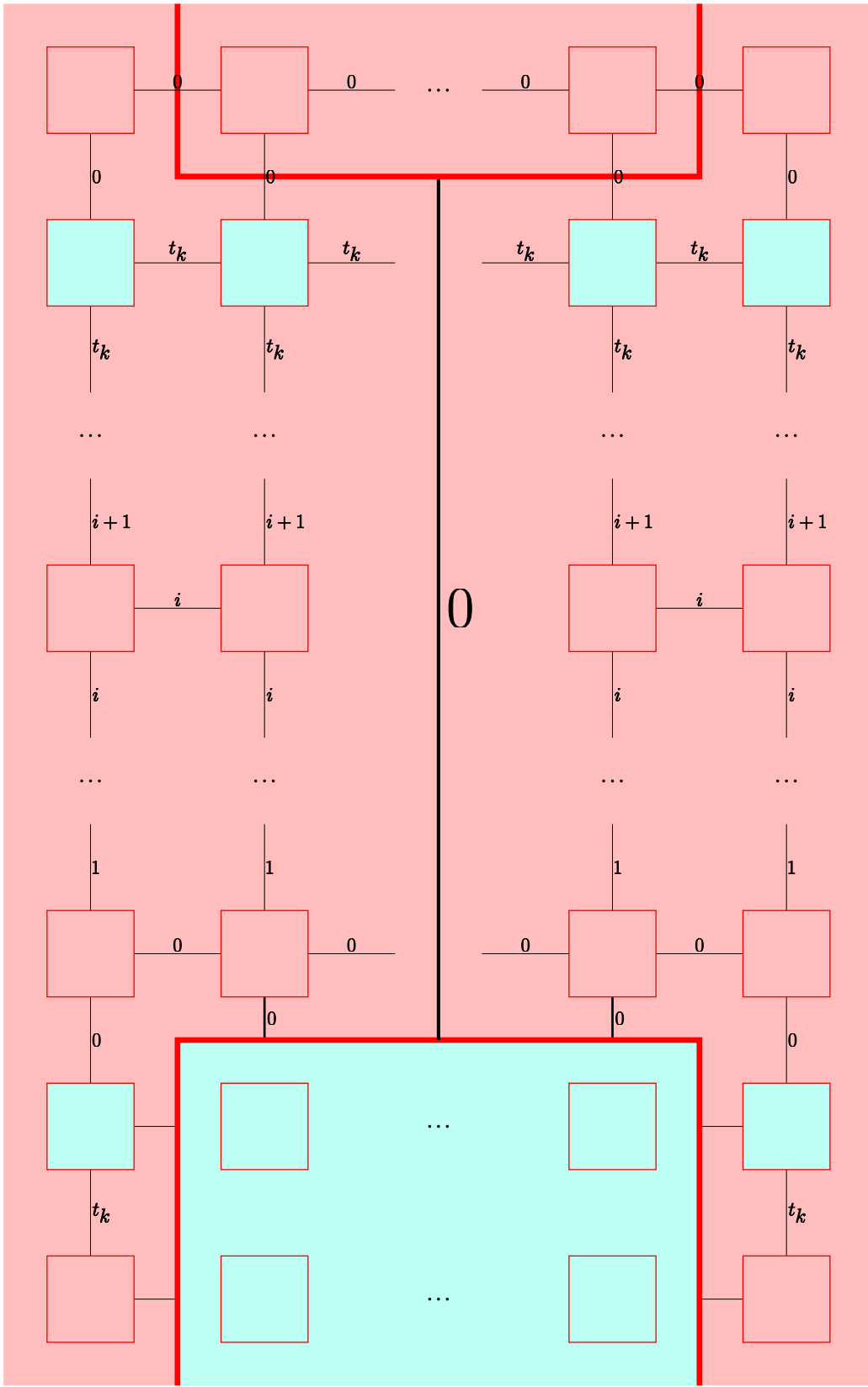}
    \caption{Structure of the odometer at the scale of $k$-markers.}
    \label{fig:Odometer}
\end{figure}

On this new layer,
we want to set-up a communication at the scale of $k$-markers between \emph{neighbouring} Red squares,
in order to control precisely which Blockable tiles are in $\symb{H}$ or $\symb{B}$,
in particular within a bigger marker.
To do so, we will implement an odometer with
period $t_k:=2^{\left\lfloor\log_2(\left\lfloor\log_2(k)\right\rfloor)\right\rfloor}-1$
at the scale of $k$-markers,
using the sparse communication channels \emph{in-between} neighbouring squares of a given scale,
in all four directions, in such a way that:
\begin{itemize}
\item A Red Blockable non-$\symb{F}$ square receives on the bottom a unary coding
of an integer $i\in\N$ (within the sparse area not already used by smaller markers).
\item This square computes $t_k$ using the unary input $N_k$ from the previous layer (also sparse),
and then compares it to the input $i$ so that:
\begin{itemize}
    \item if $i<t_k$, then the square is forced to be $\symb{H}$, and the square outputs $i+1$ above,
    \item if $i=t_k$, then the square is forced to be $\symb{B}$, and outputs $0$ above,
    \item if $i>t_k$, then the square \emph{rejects} the input, which results in an invalid tiling.
\end{itemize}
\item This unary integer $i$ is ``mirrorred'' on the left and right sides of the square,
and must match the input of the left and right neighbouring non-Frozen tiles.
\item The Blockable square of a non-$\symb{F}$ $k$-marker forcefully initialises at $0$
the input of any smaller $l$-marker (for scales $l<k$) directly next to its top border.
\item Nothing can be transmitted \emph{through} $\symb{F}$ markers,
except in their sparse area that may be used by higher-scale $\symb{B}$ markers.
\end{itemize}
All these properties are illustrated in Figure~\ref{fig:Odometer} where we see,
schematically, a $\symb{B}$ $(k+1)$-marker on the bottom,
which both transmits the integer $0$ to the next $(k+1)$-marker above, forcing it to be $\symb{H}$,
as well as initialising the $k$-markers next to its top side to $0$ too.
This signal then cycles periodically modulo $\left(t_k+1\right)$ within each column of $k$-markers.
The signal also synchronises laterally, so that the columns on the sides behave correctly
even though they do not have an initialisation to $0$ within them.

In order for this process to work correctly, we thus need to be able to subdivide a $(k+1)$-marker
into a grid of $\left(t_k+1\right)\times\left(t_k+1\right)$ blocks of $k$-markers,
that align well with the Red square.
As the Red square itself is a grid of $k$-markers of length
$\frac{2^{2\times 3^{k+1}}}{2^{3^k+1}}=2^{5\times 3^k-1}$,
it suffices to have a smaller power of $2$ for $t_k$ in order to obtain the wanted periodic behaviour.
Hence, the choice
$t_k=2^{\left\lfloor\log_2(\left\lfloor\log_2(k)\right\rfloor)\right\rfloor}-1\approx \log_2(k)$
is compatible with these constraints.
Assume for now that the tiles can compute $t_k$ and compare it to $i$ fast-enough.
The following lemma directly follows.

\begin{lemma}
Consider $\omega\in Q_{k+1}^\symb{H}$ a $\symb{H}$ $(k+1)$-marker on $\A_j$ ($j\geq 4$).
It can be seen as a grid of $k$-markers,
with a proportion exactly $\frac{1}{t_k+1}$ of $\symb{B}$ $k$-markers
and all the rest being $\symb{H}$ ones.
\end{lemma}

\begin{proposition} \label{prop:frozen-density}
Denote $\freq{\symb{F},l}{\symb{B},k}$ (resp. $\freq{\symb{F},l}{\symb{H},k}$)
the proportion of $l$-markers in a $\symb{B}$ (resp. $\symb{H}$) $k$-marker that are $\symb{F}$.
We naturally have the initialisation $\freq{\symb{F},k}{\symb{H},k+1}=0$,
as a $\symb{H}$ $(k+1)$-marker is a grid of $\symb{H}$ and $\symb{B}$ $k$-markers.
Then, for $k\geq l$:
\[
\left\{
\begin{aligned}
&\freq{\symb{F},l}{\symb{H},k+1}=
\frac{1}{t_k+1}\freq{\symb{F},l}{\symb{B},k}+\frac{t_k}{t_k+1}\freq{\symb{F},l}{\symb{H},k} \eqnsp , \\
&\freq{\symb{F},l}{\symb{B},k}=\frac{1}{4}+\frac{3}{4}\freq{\symb{F},l}{\symb{H},k} \eqnsp .
\end{aligned}
\right.
\]
It follows that $1-\freq{\symb{F},l}{\symb{H},k+1} 
=\left(1-\frac{1}{4\left(t_k+1\right)}\right)\left(1-\freq{\symb{F},l}{\symb{H},k}\right)$.
Thus, by induction,
\[
\freq{\symb{F},l}{\symb{H},l+j+1}
=1-\prod_{i=1}^j \left(1-\frac{1}{4\left(t_{l+i}+1\right)}\right) \eqnsp .
\]
Furthermore:
\[
\begin{split}
\prod_{i=j}^{k-1} \left(1-\frac{1}{4\left(t_i+1\right)}\right) &\leq
\exp\left(-\frac{1}{4}\sum_{i=j}^{k-1}\frac{1}{t_i+1}\right)\leq
\exp\left(-\frac{1}{4}\int_j^k \frac{1}{\log_2(x)}\mathrm{d}x\right)\\
&\leq\exp\left(-\frac{1}{4}\left(\frac{k}{\log_2(k)}-j\right)\right) \to 0 \eqnsp .
\end{split}
\]
Hence $\freq{\symb{F},l}{\symb{H},k}=1-o(1)$ when $k\to\infty$ for a fixed $l$
(and even when $l\to\infty$ as long as we stay
in the asymptotic regime $l=o\left(\frac{k}{\log_2(k)}\right)$).
\end{proposition}

\begin{proposition} \label{prop:genericFrozen}
Generically, a (globally admissible) tiling of $X_{\F_j}$ (with $j\geq 4$) is entirely Frozen.

\begin{proof}
Consider $\omega\in X_{\F_j}$ and fix a scale of small $l$-markers.
Structurally, for any choice of $k$, we can cut $\omega$ into an infinite grid of $k$-markers.
It follows from the previous proposition that the proportion of $\symb{F}$ $l$-markers
in their own infinite grid is $\lim_{k\to\infty}\freq{\symb{F},l}{*,k}=1$.
Then, notice that the density of this grid of $l$-markers in $\Z^d$ goes to $1$ as $l\to\infty$,
so that overall,
$\omega$ has a density $1$ of tiles equal to $\symb{F}$ on the layer $\A_{\mathrm{phase}}$.
\end{proof}
\end{proposition}

\begin{corollary}
The tileset $X_{\F_4}$ is uniquely ergodic.

\begin{proof}
The previous proposition tells us that the layer $\A_{\mathrm{phase}}$ is
generically constant equal to $\symb{F}$.
Likewise, Lemma~\ref{lem:ergodicity-robinson} tells us that
the layer $\A_{\mathrm{Robinson}}$ is uniquely ergodic
(in a way that is compatible with the constant behaviour of $\A_{\mathrm{phase}}$).
Because any part of the plane is generically inside the Red square
of a $\symb{F}$ marker for big-enough scales,
we conclude that likewise nothing happens on the layer $\A_{\mathrm{odometer}}$ generically.
The argument for the remaining (third) layer $\A_{\mathrm{scale}}$ is a bit more intricate,
but the general idea is that computations within each scale of Red squares are identical
regardless of the square we consider, hence we end up with the same well-defined frequencies
regardless of the tiling $\omega\in X_{\F_4}$ we consider, which concludes the argument.
\end{proof}
\end{corollary}

\begin{remark} \label{rmk:common-subshift}
The notable consequence of this proof is that,
generically, we only observe a strict subset $\A'\varsubsetneq \A_4$ in configurations,
made of $\symb{F}$ tiles without any odometer structure.
Hence, if we restrict $\F_4$ to $\F'$ which uses only forbidden patterns with tiles in $\A'$,
then we can conclude that $\M_\sigma\left(X_{\F'}\right)=\M_\sigma\left(X_{\F_4}\right)$
is the same singleton.
This remark will later be improved to hold for a whole family of non-uniquely-ergodic tilesets.
\end{remark}

This generic behaviour is in stark contrast to what we expect from a typical $k$-marker,
to the highly-intricate arrangement of $\symb{B}$ squares of many scales
guided by all the odometers initialised by the big Red square in the center of the marker.
This mismatch between the ``typical'' (uniform) behaviour for $k$-markers
and their behaviour within globally admissible tilings is precisely what will allow us
to force highly complex signals in the presence of noise,
through computations that will disappear once the amount of noise goes to $0$,
leaving only the signals at the limit.

We now need to actually justify that non-$\symb{F}$ $k$-markers have enough computation time
to compute $t_k$ from $N_k=3^k$ and then compare it to the odometer input $i$.
According to the time complexities in Appendix~\ref{appendix:unary},
we can step by step compute $N_k\mapsto k=\left\lfloor \log_3\left(N_k\right)\right\rfloor$,
then $k\mapsto k':=\left\lfloor \log_2(k)\right\rfloor$,
and finally $k'\mapsto t_k = 2^{\left\lfloor \log_2(p)\right\rfloor}-1$.
If $k\leq 1$, we cannot actually have a well defined $t_k\geq 0$ in this case,
so we just manually deal with it by setting $t_0=t_1=t_2=0$.
Each step of this process adds a new layer of logarithm on the complexity,
so that $N=3^k\mapsto t_k$ still has an overall complexity of order $N\log_3(N)$.
As long as $4N\leq 2^N+1$ (\ie $N\geq 3$, thus $k\geq 1$),
we can guarantee that the Turing machine has enough time not only to compute $t_k$,
but also compare it with $i$ (in at most $N$ steps)
and then update the value of $i$ to either $i+1$ or $0$ (in at most $N$ steps).
We can simply manually deal with the case $k=1$ by making this scale non-Blockable instead.
Another notable detail is that we will back-propagate the computed value of $k$ within the sparse area,
so that the last layer $\A_{\mathrm{simul}(M)}$ can in turn use it as an input
for another Turing machine.

With all this being said and done, we are back to a uniquely ergodic zero-entropy situation,
but now with some structure allowing computation
at carefully chosen places (the $\symb{B}$ squares) in the typical ($\symb{H}$) $k$-markers.
We now need to encode words in the tiles, so that there is \emph{something} to be computed.
This will also serve as a source of ``local'' entropy
(still resulting in zero-entropy tilings at the limit but with a slower decrease rate,
with some amount of combinatorial explosion).

\subsection{Layer 5: Encoding Words}

On this new layer, we want a $\symb{F}$ marker to encode a binary word,
and to synchronise it with neighbouring $\symb{F}$ markers.
This way, we will have many ``independent'' short words in $\symb{B}$ squares
in a big $\symb{H}$ marker, but only \emph{one} infinite word in generic $\symb{F}$ configurations.

To do so, we add a signal $\pmo$ or $\symb0$ onto each Red line of the Robinson,
with the following rules.
Red lines on $\symb{F}$ tiles encode a bit $\pmo$,
while Red lines on non-$\symb{F}$ tiles encode a $\symb0$.
The signals are synchronised in all four directions
(so that two neighbouring Red $\symb{F}$ squares encode the same bit),
but when a Red line crosses a $\symb{B}$ interface,
the $\symb{H}$ side will naturally be equal to the neutral symbol $\symb0$ while
the $\symb{F}$ side will take a binary $\pmo$ value.

Consequently, a $\symb{B}$ $k$-marker in $Q_k^\symb{B}$ encodes a well-defined binary word
(on the alphabet $\{\pmo\}$) of length $\btot(k)=3^k-1$ inside its central Red square.
This word can even be accessed by a Turing machine within the sparse area
as a Toeplitz encoding~\cite[Lemma 5.8]{GaySa22}
(\ie the word $abcde\dots$ reads as $abacabadabac\dots$,
each letter with a doubled period of occurrence).

Of course, this encoding spans the whole length of the finite input tape of the Turing machine here,
so we cannot even reasonably decode all of it (and write the result at the beginning of the input tape)
without running out of time.
Hence, in the next section, for the last layer, we will consider Turing machines that 
only read the first $\bread(k)$ bits of the signal in $k$-markers.
For now, the only thing we require is $\bread(k)\to\infty$
(so that in the thermodynamical framework, asymptotically when $\beta\to\infty$,
as the typical scale of $k$-markers goes to $\infty$,
we control more and more bits of an infinite binary string).
However, we want $\bread(k)$ to stay small-enough,
so that computing $\bread(k)$ and then decoding these first bits of the Toeplitz input
takes only a negligible proportion of the $2^{3^k}+1$ space-time horizon
of the $\symb{B}$ square of a $k$-marker).

The construction up to the previous layer gave us a uniquely ergodic structure.
Now, with this layer, the encoded words add a degree of freedom that breaks unique ergodicity.
However, we still have a well-understood structure.
This time, we do not have a uniquely ergodic system,
but we may identify ergodic measures with infinite binary strings.
This can be formalised as follows:

\begin{proposition} \label{prop:affine-bijection}
Let $\M\left(\{\pmo\}^\N\right)$ be the set of probability measures on infinite binary words.
We have an affine injective map
$\gamma:\M\left(\{\pmo\}^\N\right)\to\M_\sigma\left(\Omega_{\A_5}\right)$
such that $\mathrm{Im}(\gamma)=\M_\sigma\left(X_{\F_5}\right)$.

In particular, we have a formal correspondence between (ergodic) extremal measures 
in $\M_\sigma\left(X_{\F_5}\right)$ and extremal measures $\M\left(\{\pmo\}^\N\right)$
(\ie Dirac measures $\delta_w$ with $w\in\{\pmo\}^\N$).

\begin{proof}
Let $\omega\in X_{\F_5}$.
Following the tracks of Remark~\ref{rmk:common-subshift}, generically,
$\omega$ is all $\symb{F}$, and induces an ergodic measure when we project it to $\A_4$.

Now, as neighbouring $\symb{F}$ $k$-markers synchronise their $\btot(k)$ bits $\pmo$,
it follows that the all-$\symb{F}$ generic tiling $\omega$
encodes a well-defined infinite word $w\in\{\pmo\}^\N$.
Thus, $\omega$ has well-defined frequencies
and induces a unique shift-invariant measure $\mu_\omega$.
Furthermore, $\mu_\omega$ actually only depends on the \emph{word} $w$ encoded in $\omega$,
so we can denote it $\mu_w$.

This gives us a bijective correspondence $\gamma:\delta_w\mapsto\mu_w$
between extremal points of $\M\left(\{\pmo\}^\N\right)$ and $\M_\sigma\left(X_{\F_5}\right)$.
Furthermore, $\gamma$ is continuous, using the weak-* topology on both ends.
As $\M\left(\{\pmo\}^\N\right)$ is a Bauer simplex,
it follows~\cite[Theorem II.4.1]{Alf71} that we can extend $\gamma$ as the announced convex map
$\gamma:\M\left(\{\pmo\}^\N\right)\to\M_\sigma\left(\Omega_{\A_5}\right)$.
\end{proof}
\end{proposition}

At this point, we are done introducing the structural general-purpose layers,
shared by all the class of simulating tilesets we are about to define.
At last, we need to properly define the simulating tileset corresponding to a Turing machine $M$,
and quantify its effect on the entropy.

\subsection{Layer 6: Controlling Words Through Entropy}

For a given Turing machine $M$,
we want to add a new layer to $\A_5$, to obtain our complete alphabet $\A_{6(M)}$.
We will define the behaviour on the new layer $\A_{\mathrm{simul}(M)}$ as follows.
Suppose that $M$, being given a binary string $s$ as an input,
returns a binary string $M(s)$ of length $\bread(\abs{s})$.

Then, using the sparse area in a $\symb{B}$ Red square, we perform the following computations,
using four synchronised tapes, three of them being read-only.
On the first tape, we can access a unary integer $k$
(given by the layer $\A_{\mathrm{odometer}}$ in the tiling).
On the second one, we have access to a binary ``random'' seed $s$
(random in the sense that it is not fixed by the local rules of the simulating tileset,
and be uniformly distributed for the measure $\lambda_{Q_k}$).
On the third one, $M$ can access the Toeplitz encoding of some binary string
(given by the layer $\A_{\mathrm{signal}}$ in the tiling).
We then use the fourth tape to perform the following tasks:
\begin{itemize}
    \item Check that the seed $s$ is of length $k$ (break local rules if not),
    \item Decode the Toeplitz input into a prefix $u$ of length $\bread(k)$,
    \item Compute a word $M(s)$ (of length $\bread(k)$),
    \item Check if $M(s)$ is equal to $u$ (break local rules if not).
\end{itemize}
Within $\symb{F}$ squares, we setup an ``empty'' computation,
and likewise for $\symb{H}$ tiles,
so that we see this blank symbol on most of the layer $\A_{\mathrm{simul}(M)}$
\emph{except} the sparse computation area of $\symb{B}$ squares.

Remember that at the $k$-th scale of computations,
we have a space-time horizon of size $\left(2^{3^k}+1\right)$.
Following the ideas of Appendix~\ref{appendix:unary},
the first task is just a back-and-forth, and takes more or less $2k=o\left(2^{3^k}\right)$ steps.
Likewise, by a similar back-and-forth,
the last task takes about $\bread(k)=o\left(2^{3^k}\right)$ steps.
By using a copy of the unary string $k$,
we may decode the first $\bread(k)=\left\lfloor \log_2(k)\right\rfloor$ bits
of the Toeplitz input in time $2k\log_2(k)=o\left(2^{3^k}\right)$,
by eliminating one digit of $k$ out of two at each step.
Hence, as long as $M(s)$ can be computed in $o\left(2^{3^{\abs{s}}}\right)$ steps,
then up to hardcoding the behaviour of the computations
in the simulating layer $\A_{\mathrm{simul}(M)}$ up to a finite scale $k_0$,
we can guarantee that any $\symb{B}$ $k$-marker can perform all the previous tasks
before running out of time.
We will explain why this time complexity constraint is not an actual obstruction
in Section~\ref{sec:measures}.

\begin{definition}[Well-Behaved Machine] \label{def:well-behaved}
In the context of this paper, we will say that a Turing machine $M$ is \emph{well-behaved}
if it satisfies the previous assumptions,
\ie it computes a binary signal $M(s)$ of length $\bread(\abs{s})$
on a read-only input binary seed $s$ (written on a synchronous tape),
with time complexity $o\left(2^{3^{\abs{s}}}\right)$.
\end{definition}

To summarise, this new layer serves to make sure that,
in a marker $q\in Q_k^\symb{B}$ on the alphabet $\A_{6(M)}$,
the central Red square encodes a (non-fixed) binary seed $s$ of length $k$
and forces $M(s)$ as the prefix (of length $\bread(k)$)
of the word encoded on the layer $\A_{\mathrm{signal}}$.
Assuming $s$ follows the uniform distribution $U_k$ on $\{\pmo\}^k$,
we obtain a prefix $M(s)$ distributed according to $M^*\left(U_k\right)$.
In particular, setting aside the matter of computation time for now,
we can realise dyadic measures on $\{\pmo\}^{\bread(k)}$
with precision $\frac{1}{2^k}$ with this process.
Again, the study of which measures can actually be obtained will be done in Section~\ref{sec:measures}.

\begin{theorem} \label{thm:affine-bijection}
Regardless of the machine $M$ we consider,
there is a common alphabet $\A_0\subseteq \A_{6(M)}$,
such that $\F_{6(M)}$ always restricts to
the same set of forbidden patterns $\F_0$ on the alphabet $\A_0$,
and $\M_\sigma\left(X_{\F_{6(M)}}\right)=\M_\sigma\left(X_{\F_0}\right)$.

Then, we have an affine injection
$\gamma:\M\left(\{\pmo\}^\N\right)\to\M_\sigma\left(\Omega_{\A_0}\right)$
with $\mathrm{Im}(\gamma)=\M_\sigma\left(X_{\F_0}\right)$.

\begin{proof}
Let $\A_0$ be the common subset of $\symb{F}$ tiles that contain an empty symbol
on the layers $\A_{\mathrm{odometer}}$ and $\A_{\mathrm{simul}(M)}$,
and $\F_0$ the restriction of $\F_{6(M)}$.
Using once again the ergodicity argument from Remark~\ref{rmk:common-subshift},
generically, a tiling $\omega\in X_{\F_{6(M)}}$ is entirely $\symb{F}$,
so the new layer $\A_{\mathrm{simul}(M)}$ only performs empty computations.
In other words, generically, $\omega\in X_{\F_0}$.
The rest of the proof follows \emph{exactly} that of Proposition~\ref{prop:affine-bijection}.
\end{proof}
\end{theorem}

Notably, as was the case for the odometer structure,
the new Turing machine computations can only occur \emph{locally},
in finite markers, but vanish as the scale goes to infinity
until we obtain globally admissible tilings without any computation.

At long last, let us bring up again the question of the cardinality
of $Q_k^\symb{H}$ and $Q_k^\symb{B}$, thus of the set $Q_k$.

\begin{definition}
Consider a well-behaved machine $M$ and the corresponding tileset $\A_{6(M)}$.
Let $\widetilde{Q_k^\symb{H}}$ (resp. $\widetilde{Q_k^\symb{B}}$) be
the set of tuples of seeds $s$
(of length $l$ in a $\symb{B}$ square of scale $l$) and compatible words $u$
(of length $\btot(l)$ in the same square, such that $M(s)$ is a prefix of $u$ of length $\bread(l)$),
for \emph{all} the $\symb{B}$ squares in a $\symb{H}$ (resp. $\symb{B}$) $k$-marker.
\end{definition}

\begin{lemma} \label{lem:factor-bound}
Let $*\in\{\symb{H},\symb{B}\}$. If $t_k>0$, we have the following bounds:
\[
\abs{\widetilde{Q_k^*}}\leq\abs{Q_k^*}\leq
\abs{\widetilde{Q_k^*}}\times\abs{\A_{6(M)}}^{\perim_k}\eqnsp .
\]

\begin{proof}
Structurally, the behaviour within each $\symb{B}$ square of a given marker is independent,
\ie we can freely change the value of the compatible seed-signal $(s,u)$ pair it encodes
in a locally admissible way without affecting the rest of the marker.
Complementarily, we can change the information flowing \emph{outside} of $\symb{B}$ squares
(\eg the value of the odometer for the central square of a $\symb{H}$ marker),
regardless of the seed-signal pairs encoded in each $\symb{B}$ square.
If we denote $Q_k^{*,\text{sparse}}$ this complementary information,
then we have the product decomposition $Q_k^* =\widetilde{Q_k^*}\times Q_k^{*,\text{sparse}}$.

The assumption $t_k>0$ guarantees that both $\symb{H}$ and $\symb{B}$ $k$-markers exist,
hence $1\leq \abs{ Q_k^{*,\text{sparse}}}$.
The upper bound $\abs{Q_k^{*,\text{sparse}}}\leq \abs{\A}^{\perim_k}$
comes from the same argument as in Lemma~\ref{lem:frozen-counting},
from the fact that any non-fixed information of the $k$-marker (except seed-signal pairs)
represents a computation spanning the whole tile,
fixed by the boundary condition on the perimeter of the $k$-marker.
\end{proof}
\end{lemma}

Henceforth, we can shift the focus from $Q_k^*$ to $\widetilde{Q_k^*}$,
which has the added interest of encoding \emph{only} highly-structured hierarchical behaviour,
enforced by the odometers, without any dependence on the alphabet $\A_{6(M)}$.
In cases where $t_k=0$, by construction, there cannot be $\symb{H}$ $k$-markers,
so the previous lower bound on $Q_k^\symb{H}$ cannot hold,
as we want the hierarchical structure of $\widetilde{Q_k^\symb{H}}$ to be defined nonetheless
in the following upper bound (\ie have $\widetilde{Q_k^\symb{H}}\neq \emptyset$ in any case,
by encoding the tuples $\symb{H}$ markers \emph{would} encode were they to actually exist).
Contrary to what was previously said, for the sake of having simpler computations,
we will assume that there is a one bit seed at the scale of $0$-markers (instead of none),
so that $\abs{\widetilde{Q_0^\symb{B}}}=2$.
However, the overall order of the following bound still holds
with the initial definition for $\A_{6(M)}$
(and more broadly as long $t_k\approx \log_2(k)$ asymptotically, regardless of what happens
on the first scales).

\begin{proposition} \label{prop:marker-cardinal-induction}
We have the following inductive behaviour:
\begin{itemize}
    \item $\abs{ \widetilde{Q_{k+1}^\symb{H}} } =
    \abs{ \widetilde{Q_k^\symb{H}} }^{256^{3^k}
    \frac{t_k}{t_k+1}} \times 
    \abs{ \widetilde{Q_k^\symb{B}} }^{256^{3^k}
    \times\frac{1}{t_k+1}}$,
    \item $\abs{ \widetilde{Q_k^\symb{B}} } = 
    \abs{\widetilde{Q_k^\symb{H}}}^{\frac{3}{4}}\times
    2^{\rho(k)}$, with $\rho(k)=k+\btot(k)-\bread(k)$
    the total number of bits used by the seed-signal pair
    in the $\symb{B}$ square of a $k$-marker.
\end{itemize}

\begin{proof}
For the first point,
notice that a given $\symb{H}$ $(k+1)$-marker is just
a $16^{3^k}\times 16^{3^k}$ grid of smaller $k$-markers, either $\symb{H}$ or $\symb{B}$,
each one with its own family of $\symb{B}$ squares encoding independent seed-signal pairs.
Hence, we have a total of $256^{3^k}$ independent $k$-markers.
By construction, the odometer forces a proportion $\frac{1}{t_k+1}$ of them to be $\symb{B}$,
and the rest $\symb{H}$, from which the formula follows.

Likewise, for the second equality,
notice that the area outside of the central $\symb{B}$ square
can be subdivided in well-behaved grids of smaller $\symb{B}$ and $\symb{H}$ markers,
in the same proportions as in a $\symb{H}$ marker.
This outside area occupies $\frac{3}{4}$ of the marker,
hence the $\abs{\widetilde{Q_k^\symb{H}}}^{\frac{3}{4}}$ factor.
The second factor follows directly from the definition of $\rho(k)$,
of the fact that an input seed $s$ in the $\symb{B}$ square (of length $k$)
forces the value of the prefix $M(s)$ of length $\bread(k)$ out of the $\btot(k)$ bits of the signal,
hence $2^{\rho(k)}$ seed-signal pairs in total.
\end{proof}
\end{proposition}

Asymptotically, most of the entropy comes from the unchecked bits of the signal,
so that $\rho(k)\approx 3^k$.
Later on, we will use the brutal bound $\rho(k)\leq l_{n_k}$,
which will still prove good-enough for our purposes,
and also holds regardless of the length of the seed
(both the seed and the signal must fit on the border of the $\symb{B}$ square).

\begin{proposition} \label{prop:marker-cardinal-bound}
We have $\abs{\widetilde{Q_k^\symb{H}}}= C_k^{16^{3^k}}$,
with $2^{4^{-k}}\leq C_k\leq 2$.

\begin{proof}
Denote
$h_k=\log_2\left(\abs{\widetilde{Q_k^\symb{H}} }\right)$ 
and 
$b_k=\log_2\left(\abs{\widetilde{Q_k^\symb{B}}}\right)$.
The sets are non-empty by definition so that the logarithms are always well-defined, non-negative.
We have the system:
\[
\left\{
\begin{aligned}
&h_{k+1} = 256^{3^k}\left( \frac{t_k}{t_k+1}h_k+\frac{1}{t_k+1}b_k\right) \eqnsp ,\\
&b_k=\frac{3}{4}h_k+\rho(k) \eqnsp .
\end{aligned}
\right.
\]
By replacing $b_k$ in the first line, we obtain a first-order recursion scheme for $h_k$.
In particular, if we normalise $u_k:=\frac{h_k}{16^{3^k}}$, then we obtain the induction:
\[
u_{k+1}=u_k \left(1-\frac{1}{4 \left(t_k+1\right)}\right) +
\frac{\rho(k)}{\left(t_k+1\right) 16^{3^k}} \eqnsp .
\]
In particular, using $C_k:=2^{u_k}$,
we will obtain the desired expression for $\abs{\widetilde{Q_k^\symb{H}}}$.

In order to get the upper bound on $C_k$,
we just notice that
$\frac{4t_k+3}{4\left(t_k+1\right)}\leq 1$,
and $u_0=0$,
so $u_k$ can be bounded by
$u=\sum_{n=1}^\infty \frac{\rho(n)}{\left(t_n+1\right)
16^{3^n}} \leq \sum_{n=1}^\infty
\frac{4^{3^n}}{1\times 16^{3^n}}\leq 1<\infty$.

For the lower bound,
we use the fact that
$u_{k+1}\geq \left(1-\frac{1}{4\left(t_k+1\right)}\right) u_k$,
so by a direct induction
$u_k\geq u_1\times \prod_{n=1}^{k-1}
\left(1-\frac{1}{4\left(t_n+1\right)}\right)$.
Notice that $b_0=\rho(0)= 1$ and $t_0=0$ so that $u_1=16b_0\geq 1$.
Now, using the lower bound
$\ln(1-\epsilon)\geq -4\ln(4) \epsilon$
for $\epsilon\in \left[0,\frac{1}{4}\right]$,
we can obtain:
\[
\begin{split}
\prod\limits_{n=1}^{k-1}
\left(1-\frac{1}{4\left(t_n+1\right)}\right)
&=\exp\left(\sum\limits_{n=1}^{k-1}
\ln\left(1-\frac{1}{4\left(t_n+1\right)}\right)\right)\\
&\geq\exp\left(\left(-4\ln(4)
\sum\limits_{n=1}^{k-1} \frac{1}{4\left(t_n+1\right)}\right)\right)\\
&= 4^{-\sum\limits_{n=1}^{k-1} \frac{1}{t_n+1}}\\
&\geq 4^{-k} \eqnsp ,
\end{split}
\]
hence $2^{4^{-k}}$ works for the lower bound.
\end{proof}
\end{proposition}

Note that for an optimal lower bound, we would have to use
$\sum_{n=2}^{k-1} \frac{1}{\ln(n)} \sim \mathrm{li}(k)\sim \frac{k}{\ln(k)}$,
which would complicate both the proof and the expression of the bound,
without improving further results.

This is where we close this section.
We will continue studying the size of $Q_k$ in the next section,
but this time by relating it explicitly to the hypotheses of Theorem~\ref{thm:equidistribution}
from the previous section.

\section{Uniform Marker Distribution for Simulating Tilesets} \label{sec:chaos}

The goal of this section is to intertwine together
the results on the tileset $\A_{6(M)}$, from Section~\ref{sec:robinson},
with the results on Gibbs measures from Section~\ref{sec:uniform-markers},
using the potential $\phi_M$ associated to $\F_{6(M)}$.
As in the previous section, we suppose here that the machine $M$ is well-behaved
(see Definition~\ref{def:well-behaved}).

\subsection{More Marker Bounds for the Entropy Criterion}

In this subsection, we state all the counting arguments we will need
so that Theorem~\ref{thm:equidistribution} from Section~\ref{sec:uniform-markers}
applies to the $k$-markers from Section~\ref{sec:robinson} in a useful way.

In the previous section,
we computed estimations for $\abs{Q_k^\symb{H}}$,
$\abs{Q_k^\symb{B}}$ and $\abs{Q_k^\symb{F}}$.
These estimates will allow us to derive some important bounds relating to markers.

First, let us prove that, for the uniform distribution $\lambda_{Q_k}$,
the typical $k$-marker is $\symb{H}$.
This will then allow us to rewrite the entropy criterion
of Theorem~\ref{thm:equidistribution} in a more usable way.

\begin{lemma}
Denote $p_k^\symb{H}:=\lambda_{Q_k}\left(Q_k^\symb{H}\right)$.
Assume $t_k>0$, so that Lemma~\ref{lem:factor-bound} applies. Then:
\[
p_k^\symb{H} \geq 1-2^{-\frac{16^{3^k}}{4^{k+1}}+O\left(4^{3^k}\right)} \eqnsp .
\]

\begin{proof}
We have:
\[
p_k^\symb{H}=\frac{\abs{Q_k^\symb{H}}}{\abs{Q_k^\symb{H}}+\abs{Q_k^\symb{B}}+\abs{Q_k^\symb{F}}}
\geq 1-\frac{\abs{Q_k^\symb{B}}+\abs{Q_k^\symb{F}}}{\abs{Q_k^\symb{H}}}
\geq 1-\frac{\abs{\A_{6(M)}}^{l_{n_k}}\abs{\widetilde{Q_k^\symb{B}}}
+\abs{\A_{6(M)}}^{4l_{n_k}}}{\abs{\widetilde{Q_k^\symb{H}}}} \eqnsp .
\]
The first inequality comes from $\abs{Q_k}=\abs{Q_k^\symb{H}}
+\abs{Q_k^\symb{B}}+\abs{Q_k^\symb{F}}\geq \abs{Q_k^\symb{H}}$,
and for the second one we use Lemma~\ref{lem:factor-bound} to bound $\abs{Q_k^\symb{B}}$
and Lemma~\ref{lem:frozen-counting} to bound $\abs{Q_k^\symb{F}}$.
Then, we use Propositions~\ref{prop:marker-cardinal-induction}~and~\ref{prop:marker-cardinal-bound}:
\[
p_k^\symb{H}\geq 1-\abs{\A_{6(M)}}^{l_{n_k}}\frac{\abs{\widetilde{Q_k^\symb{H}}}
^{\frac{3}{4}}\times2^{\rho(k)}}{\abs{\widetilde{Q_k^\symb{H}}}}
-\frac{\abs{\A_{6(M)}}^{4l_{n_k}}}{\abs{\widetilde{Q_k^\symb{H}}}}
\geq 1-\frac{2\abs{\A_{6(M)}}^{4 l_{n_k}}2^{l_{n_k}}}{2^{4^{-(k+1)}\times 16^{3^k}}} \eqnsp ,
\]
hence the announced bound.
\end{proof}
\end{lemma}

Let us now look at the entropy criterion of Theorem~\ref{thm:equidistribution}.

\begin{lemma} \label{lem:kappa-value}
Assume that $t_k,t_{k+1},t_{k+2}>0$.
Let $B_k:=I_{l_{n_k}}$ be the window where $k$-markers are defined,
$J_k:=I_{l_{n_{k+2}}}$ the window for $(k+2)$-markers,
and $G_k$ the set of locally admissible tilings of $J_k$.
We have:
\[
\frac{\log_2\left(\abs{G_k}\right)}{\abs{J_k}}\geq \left(1-\kappa_k\right) 
\frac{\log_2\left(\abs{Q_k}\right)}{\abs{B_k}} \eqnsp ,
\]
with $\kappa_k=\frac{2+o(1)}{\log_2(k)}$.

\begin{proof}
Let us first obtain an upper bound for the term on the right,
using in particular Lemma~\ref{lem:factor-bound}
and Proposition~\ref{prop:marker-cardinal-bound}:
\begin{alignat*}{3}
&\log_2\left(\abs{Q_k}\right) &&=
\log_2\left(\abs{Q_k^\symb{H}}\right) &&+\log_2\left(1+
\frac{\abs{Q_k^\symb{B}}+\abs{Q_k^\symb{F}}}{\abs{Q_k^\symb{H}}}\right)\\
& &&=\log_2\left(\abs{Q_k^\symb{H}}\right)&&+\log_2\left(1+
\frac{1-p_k^\symb{H}}{p_k^\symb{H}}\right)\\
& &&\leq h_k+\perim_k\log_2\left(\abs{\A_{6(M)}}\right)
&&+\frac{1-p_k^\symb{H}}{\ln(2) p_k^\symb{H}} \eqnsp .
\end{alignat*}

This window $J_k$ can notably contain an admissible $\symb{H}$ $(k+2)$-marker
(as $t_{k+2}>0$).
This marker is a $16^{3^{k+1}}\times 16^{3^{k+1}}$ grid of $(k+1)$-markers,
a proportion $\frac{t_{k+1}}{t_{k+1}+1}$ of which are $\symb{H}$.
Likewise, each such $\symb{H}$ $(k+1)$-marker contains
$256^{3^k}\times \frac{t_k}{t_k+1}$ $\symb{H}$ $k$-markers,
each of them encoding independent seed-signal pairs on their coordinate $\widetilde{Q_k^\symb{H}}$.
It follows that:
\[
\begin{split}
\log_2\left(\abs{G_k}\right) &\geq 256^{3^{k+1}}\frac{t_{k+1}}{t_{k+1}+1}\times
256^{3^k}\frac{t_k}{t_k+1}\times h_k \\
&= 256^{4\times 3^k}\times
\left(1-\frac{1}{t_{k+1}+1}\right)\left(1-\frac{1}{t_k+1}\right)h_k\\
&\geq 256^{4\times 3^k}\times \left(1-\frac{1}{t_k+1}\right)^2h_k \eqnsp .
\end{split}
\]
Now, regarding the ratio of the window sizes:
\[
\begin{split}
\frac{\abs{B_k}}{\abs{J_k}}
&=\left(\frac{2^{2\times 3^k+1}-1}{2^{2\times 3^{k+1}+1}-1}\right)^2\\
&\geq \left(\frac{2\times 4^{3^k}}{2\times 4^{3^{k+2}}}\right)^2\times
\left(1-\frac{1}{2^{2\times 3^k+1}}\right)^2\\
&=\frac{1}{256^{4\times 3^k}}\times\left(1-\frac{1}{2\times 4^{3^k}}\right)^2 \eqnsp .
\end{split}
\]

By putting together all of these bounds, we obtain:
\[
\frac{\frac{\log_2\left(\abs{G_k}\right)}{\abs{J_k}}}
{\frac{\log_2\left(\abs{Q_k}\right)}{\abs{B_k}}}\geq
\left(1-\frac{1}{t_k+1}\right)^2
\left(1-\frac{\perim_k\log_2\left(\abs{\A_{6(M)}}\right)+o(1)}{h_k}\right)
\left(1-\frac{1}{2\times 4^{3^k}}\right)^2 \eqnsp .
\]
One can check that the slowest factor here is by orders of magnitude the leftmost one,
hence we conclude the announced $\kappa_k=\frac{2+o(1)}{\log_2(k)}$.
\end{proof}
\end{lemma}

In other words, using the family of marker sets $\left(Q_k\right)$
will allow us to satisfy the entropy criterion of Theorem~\ref{thm:equidistribution}
with $\kappa_k\to 0$.

\subsection{Injecting the Marker Bounds in the Equidistribution Result}

Here, we provide a restating of Theorem~\ref{thm:equidistribution}
into Theorem~\ref{thm:equidistribution-robinson}, for our specific class of tilesets.
First we give a conditional uniformity result which will serve us later on.

\begin{proposition} \label{prop:UniformWords}
Let $\mu\in\mathcal{G}(\beta)$ be a shift-invariant Gibbs measure,
and $\mu_k$ the induced conditional measure on $Q_k^\symb{H}$.
Denote $\widetilde{\mu_k}$ be the push-forward measure on $\widetilde{Q_k^\symb{H}}$.
Then $\widetilde{\mu_k}$ is the uniform distribution.

\begin{proof}
We use here the factorisation
$Q_k^\symb{H}=\widetilde{Q_k^\symb{H}}\times Q_k^{\symb{H},\text{sparse}}$
as in Lemma~\ref{lem:factor-bound}.
Once the coordinate $Q_k^{\symb{H},\text{sparse}}$ is fixed,
all possible choices of a marker in $Q_k^\symb{H}$
(\ie of seed-signal pairs in $\widetilde{Q_k^\symb{H}}$) have the same energy content,
so $\widetilde{\mu_k}$ is uniform conditionally to this choice.
This is true regardless of the choice, hence $\widetilde{\mu_k}$ is uniform.
\end{proof}
\end{proposition}

Quite notably, this structural result follows from the way we defined the tileset,
and holds at \emph{any} temperature, regardless of the measure $\mu$ we consider.
The role of Theorem~\ref{thm:equidistribution} is then to provide us
with a temperature range where $\mu_k$ represents some typical behaviour.

With the associated potential $\phi_M$ we use here,
we can somewhat simplify some of the notations used in Theorem~\ref{thm:equidistribution},
for the condition on $\beta$ in particular.
As our interactions only take the values $0$ and $1$, with range $r=1$,
we have $\alpha=1$ and $\norm{\phi_M}=2$.
Assuming we look at the marker set $Q_k$,
then what was called $A$ is here the window $B_k$.
What was called $I_n$ is here the window $J_k$.
As $r=1$, $\abs{I_n}-\abs{I_{n-2r}}$ is here
the perimeter of $J_k=B_{k+2}$, which we denoted $\perim_{k+2}\leq 4 l_{n_{k+2}}$.
It follows that we can replace $\frac{\abs{I_n}}{\abs{I_n}-\abs{I_{n-2r}}}$
by the slightly tighter upper bound $\frac{l_{n_{k+2}}}{4}$.
At last, if we denote $C:=2\log_2\left(\abs{\A_{6(M)}}\right)$,
we obtain the following statement:

\begin{theorem} \label{thm:equidistribution-robinson}
Let $M$ be a well-behaved Turing machine, and $\phi_M$ the associated potential on $\Omega_{\A_{6(M)}}$.
Assume that $\epsilon_k$ and $\beta$ are such that $\beta \in T_k:=
\left[\frac{C\times l_{n_k}^2}{\epsilon_k},\frac{\epsilon_k\times l_{n_{k+2}}}{8} \right]$.
Denote $0\triangleleft Q_k^\symb{H}$ the event where the origin is covered by a $\symb{H}$ $k$-marker.
Then, for any $\mu\in\G(\beta)$:
\[
\mu\left(\left\{0\triangleleft Q_k^\symb{H}\right\}\right)
\geq 1-O\left(4^k \epsilon_k'\left(1-\log_2\left(\epsilon_k'\right)\right)\right)
-\underset{k\to\infty}{o}(1) \eqnsp .
\]

\begin{proof}
As proven in the previous section, $Q_k$ satisfies Definition~\ref{def:marker},
it is a marker set with margin factor $\tau_k=\frac{6}{l_{n_k}}$.
Hence, it satisfies the entropy criterion with the $\kappa_k$ computed in Lemma~\ref{lem:kappa-value},
and the temperature criterion rewritten as $\beta\in T_k$,
so that Theorem~\ref{thm:equidistribution} applies here.
Using the first part of the result of the theorem, we have:
\[
\begin{split}
\mu\left(\left\{0\triangleleft Q_k^\symb{H}\right\}\right)
&=\mu\left(\left\{0\triangleleft Q_k\right\}\right)\times\mu_{Q_k}\left(Q_k^\symb{H}\right)\\
&\geq \left(1-\epsilon_k'\right)\times\mu_{Q_k}\left(Q_k^\symb{H}\right)\\
&\geq 1-\epsilon_k'-\left(1-\mu_{Q_k}\left(Q_k^\symb{H}\right)\right) \eqnsp .
\end{split}
\]

To conclude, we now need a bound on this last term.
We will use the fact that the convergence $\lambda_{Q_k}\left(Q_k^\symb{H}\right)=p_k^\symb{H}\to 1$ 
happens \emph{really} fast.
We want to estimate $q_k^\symb{H}:=\mu_{Q_k}\left(Q_k^\symb{H}\right)$.
We can rewrite:
\begin{alignat*}{4}
&H\left(\mu_{Q_k}\right)
&&= H\left(\mathcal{B}\left(q_k^\symb{H}\right)\right) &&+
q_k^\symb{H} H\left(\mu_{Q_k^\symb{H}}\right)
&&+ \left(1-q_k^\symb{H}\right) H\left(\mu_{Q_k^\symb{B}\sqcup Q_k^\symb{F}}\right) \\
& &&\leq O(1) &&+ q_k^\symb{H} \log_2\left(\abs{Q_k^\symb{H}}\right)
&&+ \left(1-q_k^\symb{H}\right) \left(\log_2\left(\abs{Q_k^\symb{B}}\right)+o(1)\right)\\
& &&\leq O(1) &&+ q_k^\symb{H} \log_2\left(\abs{Q_k^\symb{H}}\right)
&&+\left(1-q_k^\symb{H}\right)\log_2\left(\abs{Q_k^\symb{H}}\right)\left(\frac{3}{4}+o(1)\right)\eqnsp .
\end{alignat*}

As $H\left(\lambda_{Q_k}\right)=\log_2\left(\abs{Q_k}\right)
\geq \log_2\left(\abs{Q_k^\symb{H}}\right)$,
we obtain $H\left(\mu_{Q_k}\right)\leq 
\frac{3+q_k^\symb{H}}{4}H\left(\lambda_{Q_k}\right)(1+o(1))$.
On the other hand, using the second part
of the result of Theorem~\ref{thm:equidistribution}:
\[
\begin{split}
\frac{H\left(\mu_{Q_k}\right)}{H\left(\lambda_{Q_k}\right)}
&\geq 1-2\kappa_k-\frac{H\left(\kappa_k\right)}{H\left(\lambda_{Q_k}\right)}
-\left(\epsilon_k+O\left(\epsilon_k'\right)+H\left(\epsilon_k'\right)\right) 
\frac{l_{n_k}^2}{H\left(\lambda_{Q_k}\right)}\\
&\geq 1-o(1)-O\left(\epsilon_k'\times\left(1-\log_2\left(\epsilon_k'\right)\right)\right)
\times O\left(4^k\right) \eqnsp ,
\end{split}
\]
hence the announced bound.
\end{proof}
\end{theorem}

In particular, to obtain a full control on the low-temperature behaviour of the system,
we need the intervals $T_k$ from the temperature criterion for $k$-markers
to overlap from one scale to the next (for this, the bigger $\epsilon_k$ is, the better)
while still having $\epsilon_k\to 0$ fast-enough to guarantee the bound goes to $0$ too.
A careful study allows us to propose the following tuning of $\epsilon$:

\begin{corollary}
Let $\epsilon_k:=\sqrt{\frac{1}{l_{n_k}}}$.
Then $4^k \epsilon_k'\left(1-\log_2\left(\epsilon_k'\right)\right)\to 0$.
Moreover, the intervals $T_k$ overlap in the sense that $\min T_{k+1} \leq \max T_k$ after a rank,
so that $\bigcup T_k$ contains an asymptotic interval $\left[\beta_0,\infty\right[$,
as illustrated in Figure~\ref{fig:overlapping-intervals}.

\begin{proof}
Let us remind that
$\epsilon_k'=1-\frac{1-\epsilon_k}{\left(1+\tau_k\right)^2}\leq \epsilon_k+2\tau_k$.
In particular, $\tau_k = \frac{6}{l_{n_k}} =o\left(\epsilon_k\right)$ so $\epsilon_k'\sim\epsilon_k$,
thus we can instead study the asymptotic behaviour of $-4^k\epsilon_k\log_2\left(\epsilon_k\right)$.
We have:
\[
-\log_2\left(\epsilon_k\right)\sim
\frac{1}{2}\log\left(l_{n_k}\right)\sim \frac{n_k}{2}\sim 3^k\eqnsp .
\]
On the other hand, $\epsilon_k=\frac{1}{\sqrt{2\times4^{3^k}+1}}\sim\frac{1}{\sqrt{2}\times2^{3^k}}$.
Thus, we conclude that whole term is equivalent to $\frac{12^k}{\sqrt{2}\times 2^{3^k}}=o(1)$.
For the other part of the result, we have:
\[
\frac{\max T_k}{\min T_{k+1}}=\frac{\epsilon_k \times l_{n_{k+2}}}{8}\times
\frac{\epsilon_{k+1}}{C\times l_{n_k}^2}=\frac{1}{8C}\times\epsilon_k\epsilon_{k+1} 
\frac{l_{n_{k+2}}}{l_{n_{k+1}}^2}\sim \frac{1}{8C} 2^{n_{k+2}-\frac{n_k+5n_{k+1}}{2}}
\sim\frac{4^{3^k}}{32C} \eqnsp .
\]
This equivalent goes to $\infty$, so in particular $\max T_k\geq\min T_{k+1}$ after a rank.
\end{proof}
\end{corollary}

\begin{remark}
The asymptotic behaviour $\frac{\max T_k}{\min T_{k+1}}\to\infty$
(and thus the overlapping property) still holds
if we change the common multiplicative constants used to define the intervals
(\ie $\alpha$ and $\norm{\phi}$).
It follows that we can actually generalise the result to a larger class of potentials associated
to the forbidden patterns $\F_{6(M)}$,
as long as we still give a positive penalty to each forbidden pattern in the local interactions,
but not necessarily zero-one valued.
\end{remark}

We now want to use this theorem to prove that,
in each of the overlapping temperature intervals $T_k$,
we have an increasingly tight bound on the distance of Gibbs measures
to some well-behaved measure $\lambda_k\in\M_\sigma\left(X_{\F_{6(M)}}\right)$,
induced by the computations of the Turing machine $M$.
This will give us a contracting ``tube'' around Gibbs measures as $\beta\to\infty$,
so that $\G(\infty):=\acc[\beta\to\infty]{\G(\beta)}=\acc{\lambda_k}$.

Before doing so, let us make a brief detour through the question of the induced measures on words,
which will hopefully give along the way a mental picture of the multiple behaviours that co-exist
at multiple scales in a tiling.

\subsection{Frequencies of Signals in Markers in Markers in Markers}

Denote $m_k=M^*\left(U_k\right)$ the image measure on $\{\pmo\}^{\bread(k)}$
induced by the uniform distribution $U_k$ on $\{\pmo\}^k$,
which is notably a dyadic measure with precision $\frac{1}{2^k}$.

Theorem~\ref{thm:affine-bijection} tells us that, at temperature zero,
ground states can be entirely described by a corresponding measure
on infinite binary words $\{\pmo\}^\N$.
What we need now is a more general way to relate measures
in $\M_\sigma\left(\Omega_{\A_{6(M)}}\right)$,
and Gibbs states in particular, to measures in $\M_\sigma\left(X_{\F_0}\right)$
(\ie measures on words).

In practice, we will end up with measures on \emph{finite} binary words of various lengths.
In order to compare such measures, we may either shorten the length of the encoded words
(through projection) or lengthen it (by adding a deterministic string of $\po$ symbols at the end).

More precisely, to follow a physical analogy,
we are interested at what happens at the \emph{microscopic} scale of the system.
At this scale, what we typically observe is a highly ordered grid of small $\symb{F}$ markers,
encoding the same signal, of length $\bread(l)$, the distribution of which we want to quantify.
At the \emph{macroscopic} end of the spectrum, what we have is a kind of loosely structured
(but still dense) arrangement of $\symb{H}$ $k$-markers.
While such markers do not align or synchronise with their neighbours,
they have a highly intricate structure,
with a complex arrangement of $\symb{B}$ markers of various sizes
at an intermediate \emph{mesoscopic} scale.
Within each such $\symb{B}$ area, the Turing machine forces the distribution $m_k$ on the signal.
There is a non-zero amount of non-$\symb{F}$ $l$-markers,
for which the encoded signal is not well-defined, but this will not be an issue asymptotically,
as the scale $k$ goes to $\infty$.
This idea formalises as follows:

\begin{proposition} \label{prop:measure-weights}
Let $w_{l,k}$ be the empirical signal distribution (on binary words of length $\bread(l)$)
for $l$-markers in a $\symb{H}$ $k$-marker with uniformly distributed seeds.
Note that $w_{l,k}$ is not a \emph{probability} distribution,
as there is a positive frequency of $H$ $l$-markers that do not encode a well-defined signal.
Then we initialise $w_{l,l}=0$ (the \emph{null} measure) and:
\[
w_{l,k+1} = \frac{1}{4\left(t_k+1\right)} m_k + \frac{4t_k+3}{4\left(t_k+1\right)}w_{l,k} \eqnsp .
\]

\begin{proof}
The idea of the proof is the same as in Proposition~\ref{prop:frozen-density}:
we see the $\symb{H}$ $(k+1)$-marker as a grid of $k$-markers,
with a frequency $\frac{1}{t_k+1}$ of $\symb{B}$ ones,
and only a quarter of their area actually being inside the $\symb{B}$ square
where the computations force the distribution $m_k$ induced by its uniform seed.
\end{proof}
\end{proposition}

We may translate $w_{l,k}$ into a well-defined probability distribution,
of total mass $1$, by adding a ``trash'' state of weight 
$1-w_{l,k}\left(\{\pmo\}^{\bread(l)}\right)$.
However, as $m_k$ is a probability distribution with total mass $1$,
we have $1-w_{l,k+1}\left(\{\pmo\}^{\bread(l)}\right)=\prod_{i=l}^k 
\frac{4t_i+3}{4\left(t_i+1\right)}\underset{k\to\infty}{\longrightarrow} 0$,
\ie the trash state plays no role asymptotically in any case.

\begin{remark}[Diagonal Convergence]
Assume there is a diagonal extraction $\theta$ such that,
for any $l\in\N$, $\left(w_{l,\theta(k)}\right)_k$ converges to a limit $w_l$.
Then the family $\left(w_l\right)$ is compatible with projections,
and can uniquely be extended as a measure $w_\infty\in\M\left(\{\pmo\}^\N\right)$.
If the measures $m_k$ are constant after some rank
(\ie are all projections of some measure on infinite words $m_\infty$),
then naturally $w_\infty=m_\infty$.
\end{remark}

With this multi-scale picture in mind,
we now want to actually relate the Gibbs measures to similar measures on the admissible tilings.

\subsection{Distance Bounds for Gibbs Measures and Random Tilings} \label{sec:chaos:distance}

To prove the desired bounds, we use here the weak-$*$ distance
$d^*=\sum_{n=1}^\infty \frac{d_n}{2^{n+1}}$ with:
\[
d_n\left(\mu,\mu'\right)=\sum_{\omega\in\A^{\llbracket 0,n-1\rrbracket^d}}
\abs{\mu([\omega])-\mu'([\omega])} \eqnsp .
\]
In particular, as $d_n\leq d_{n+1}\leq 2$, we obtain $d\leq d_n+\frac{1}{2^n}$.

To obtain the desired tube of measures $\lambda_k$,
we will step by step jump from Gibbs measures to measures
on a grid of $\symb{H}$ macroscopic $k$-markers
(the scale $k(\beta)$ of which will be given by Theorem~\ref{thm:equidistribution-robinson},
so that $\beta\in T_k$ for the Gibbs measure),
then on a grid of $\symb{F}$ microscopic $l$-markers
(at a much lower scale $l(k)$, such that $l\to\infty$),
and finally on admissible tilings.
In doing so, we will need several times the following lemma:

\begin{lemma} \label{lem:HighCondProbLowDist}
Let $X$ be a family of disjoint events,
and $d_X\left(\mu,\mu'\right):=\sum_{A\in X}\abs{\mu(A)-\mu'(A)}$.
Consider two probability measures $\mu$ and $\mu'$, with events $U$ and $V$ such that,
for any $A\in X$, $\mu(A\abs{U)=\mu'(A}V)$.
Then $d_X\left(\mu,\mu'\right)\leq 2\left( \mu\left(U^c\right)
+\mu'\left(V^c\right)\right)$.

\begin{proof}
For any $A\in X$, we have:
\[
\begin{split}
\abs{\mu(A)-\mu'(A)} &= \abs{\mu(U)\mu(A|U)-\mu'(V)\mu'(A|V)
+\mu\left(U^c\right)\mu\left(A\middle|U^c\right)
-\mu'\left(V^c\right)\mu\left(A\middle|V^c\right) }\\
&\leq \abs{\mu(U)-\mu'(V)}\mu(A|U)
+\mu\left(U^c\right)\mu\left(A\middle|U^c\right)
+\mu\left(U^c\right)\mu\left(A\middle|V^c\right) \eqnsp .
\end{split}
\]
As the events $A\in X$ are disjoint, by summing over $X$
for each of the three conditional distributions $\mu(\cdot\abs{U)=\mu'(\cdot}V)$,
$\mu\left(\cdot\middle|U^c\right)$ and $\mu'\left(\cdot\middle|V^c\right)$,
we conclude that:
\[
d\left(\mu,\mu'\right)\leq \abs{\mu(U)-\mu'(V)}+\mu\left(U^c\right)+\mu\left(U^c\right)=
\abs{\mu\left(U^c\right)-\mu'\left(V^c\right)}+\mu\left(U^c\right)+\mu'\left(V^c\right) \eqnsp ,
\]
thus the announced result.
\end{proof}
\end{lemma}

The informal interpretation of this result is that,
if two measures are identical conditionally to high-probability events,
then they are weak-$*$ close.

Consider a Gibbs measure $\mu\in\G(\beta)$.
Define then $\overline{\mu_k}$ the probability distribution where
we have a grid of well-aligned iid $k$-markers of law $\mu_k=\mu_{Q_k^\symb{H}}$,
with tiles chosen at random in the one-tile thick grid around the markers,
all of this being averaged over the periodic translational orbit.

\begin{proposition}
Let $\mu\in\G(\beta)$, and $\overline{\mu_k}$ the induced distribution
on a grid of $\symb{H}$ $k$-markers. Then for any $i\leq l_{n_k}$ we have:
\[
d^*\left(\mu,\overline{\mu_k}\right) \leq 
\frac{1}{2^i} + 4\left(1-\left(\frac{l_{n_k}-i+1}{l_{n_k}}\right)^2
\mu\left(0\triangleleft Q_k^{\symb{H}}\right)\right) \eqnsp .
\]

\begin{proof}
Let us denote $U_i^k$ the event where the whole window $I_i=\llbracket 0,i-1\rrbracket^2$
is covered by a single marker from $Q_k^\symb{H}$.
By shift-invariance of $\mu$, conditionally to $U_i^k$,
the content of the window $I_i$ is a uniform compatible translation inside a $k$-marker
(among the $\left(\frac{l_{n_k}-i+1}{l_{n_k}}\right)^2$ such that the square fits in the marker),
which itself is chosen with law $\mu_k$ independently.

By shift-invariance, for any $0\leq x,y<l_{n_k}$ (with $l_{n_k}$ the length of a $k$-marker),
we have the same probability of having the upper-right corner of the marker at position $(x,y)$
conditionally to $0\triangleleft Q_k^\symb{H}$, so by counting the number of positions such that
$\llbracket 0,i-1\rrbracket^2\triangleleft Q_k^\symb{H}$ we conclude that,
as long as $i\leq l_{n_k}$,
we have $\mu\left(U_i^k\right)=\left(\frac{l_{n_k}-i+1}{l_{n_k}}\right)^2 \mu\left(U_1^k\right)$.

Likewise, for $\overline{\mu_k}$,
the event $U_i^k$ is realised whenever the window $I_i$ does not overlap with the period grid,
with probability $\left(\frac{l_{n_k}-i+1}{l_{n_k}}\right)^2$,
and induces the same conditional distribution.

By using $X=\left\{ [w],w\in \A_{6(M)}^{I_i}\right\}$ in Lemma~\ref{lem:HighCondProbLowDist},
so that $d_X=d_i$, it follows that:
\[
\begin{split}
d^*\left(\mu,\overline{\mu_k}\right)&\leq \frac{1}{2^i}+d_i\left(\mu,\overline{\mu_k}\right)\\
&\leq\frac{1}{2^i}+2\left(2-\left(\left(\frac{l_{n_k}-i+1}{l_{n_k}}\right)^2
\mu\left(U_1^k\right)-\left(\frac{l_{n_k}-i+1}{l_{n_k}}\right)^2\right)\right)\\
&\leq \frac{1}{2^i}+
4\left(1-\left(\frac{l_{n_k}-i+1}{l_{n_k}}\right)^2\mu\left(U_1^k\right)\right)\eqnsp ,
\end{split}
\]
thus announced bound.
\end{proof}
\end{proposition}

Informally, if $\beta\in T_k$, then Theorem~\ref{thm:equidistribution-robinson} applies,
so that $k$-markers represent the appropriate macroscopic scale for $\mu\in\G(\beta)$.
Likewise, we then need a similar argument to compare $\mu$ to its microscopic scale, and forth.

\begin{theorem} \label{thm:Gibbs-accumulation}
Let $k(\beta)$ be a non-decreasing scale parameter such that $\beta\in T_{k(\beta)}$
in Theorem~\ref{thm:equidistribution-robinson}.
Then we have a sequence of distributions $\lambda_k\in\M_\sigma\left(X_{\F_0}\right)$
such that $d^*\left(\G(\beta),\lambda_{k(\beta)}\right)\underset{\beta\to\infty}{\longrightarrow} 0$.
It follows that $\G(\infty)=\acc{\lambda_k}=\acc[\beta\to\infty]{\mu_\beta}$
for any trajectory $\left(\mu_\beta\right)_{\beta>0}$.

\begin{proof}
First, consider $\beta\in T_k$, so that Theorem~\ref{thm:equidistribution-robinson} applies
to $\mu\in\G(\beta)$ at the scale of $k$-markers.
Following the notations of the previous proposition,
the event $U_1^k$ (\ie $0\triangleleft Q_k^\symb{H}$) has a high probability.
In particular, if $i=\underset{k\to\infty}{o}\left(l_{n_k}\right)$ but $i\to\infty$ nevertheless,
then the upper bound on $d\left(\mu,\overline{\mu_k}\right)$ goes to $0$ as $k\to\infty$.

Likewise, denote $\mu_k^m$ the empirical distribution of $\symb{F}$ $m$-markers
inside a $\symb{H}$ $k$-marker itself chosen at random with law $\mu_k$.
This time, we define the shift-invariant measure $\overline{\mu_k^m}$,
with independent $m$-markers aligned on a smaller grid.
Then, we consider the event $V_i^m$ such that $I_i\triangleleft Q_m^\symb{F}$.
According to Proposition~\ref{prop:frozen-density},
which computes the proportion of $\symb{F}$ $m$-markers in a $\symb{H}$ $k$-marker,
we conclude that $V_i^m$ has high probability for both $\overline{\mu_k}$ and $\overline{\mu_k^m}$
in the asymptotic regime $m=o\left(\frac{k}{\log_2(k)}\right)$
(we will use $m\approx\log_2\left(\log_2(k)\right)$,
for reasons made clear in Proposition~\ref{prop:repetition-slowdown} in the last section).
Then, with $i=m=o\left(l_{n_m}\right)$,
we have an asymptotic regime such that $d^*\left(\overline{\mu_k},\overline{\mu_k^m}\right)\to 0$.

Now, the important thing to notice is that, following Proposition~\ref{prop:UniformWords},
the law $\widetilde{\mu_k}$ induced by $\mu_k$ on $\widetilde{Q_k^\symb{H}}$ stays the same
regardless of the initial measure $\mu$.
In turn, this implies that the measure on $\{\pmo\}^{\bread(m)}$ induced by $\mu_k^m$
is always the same (renormalised) measure $w_{m,k}$ from Proposition~\ref{prop:measure-weights}.
After extending $w_{m,k}$ as a probability measure on $\{\pmo\}^\N$
(by adding infinitely many $\po$ bits), using Proposition~\ref{prop:affine-bijection},
we can define the corresponding random tiling
$\lambda_k^m=\gamma\left(w_{m,k}\right) \in \M_\sigma\left(X_{\F_0}\right)$,
uniquely defined as a function of $(m,k)$ regardless of the initial choice of $\mu$.

Notice how the globally admissible tiling used for $\lambda_k^m$ may
affect the content of the sparse communication area of a $m$-marker
in a different way than it does for $\overline{\mu_k^m}$:
assuming both $m$-markers encode the same \emph{word},
the only part where we can guarantee they are identical is inside their Red squares.
Thankfully, we know~\cite[Lemma 5.4]{GaySa22}
that such Red squares form a Sierpiński-carpet-like shape,
with a proportion of tiles outside of the Red squares
of order $O\left(\left(\frac{\sqrt{3}}{2}\right)^n\right)$ in a Robinson $n$-macro-tile.
It follows that, in the right asymptotic regime,
a small square window $I_i$ is inside a Red square of a $m$-marker with high probability
for both $\overline{\mu_k^m}$ and $\lambda_k^m$,
and conditionally to this event we observe the very same empirical behaviour once again.
Thus, in the right asymptotic regime we have $d^*\left(\overline{\mu_k^m},\lambda_k^m\right)\to 0$.

To conclude, in the asymptotic regime,
with $\lambda_k:=\lambda_k^m$ that does not depend on the initial choice of the Gibbs measure $\mu$,
we have $d^*\left(\G(\beta),\lambda_{k(\beta)}\right)\to 0$.
\end{proof}
\end{theorem}

In this theorem, we used the equivalence between measures on words $w_{m,k}$
(which we will simply denote $w_k$ from now on)
and invariant measures on tilings $\lambda_k=\gamma\left(w_k\right)$.
In particular, because $\gamma$ is a continuous map,
we can now focus on the accumulation set of random words
$\acc{w_k}\subseteq\M\left(\{\pmo\}^\N\right)$
instead of $\acc{\lambda_k}$.

\section{Forcing \texorpdfstring{$\Pi_2$}{Π₂} Sets of Ground States to Appear} \label{sec:measures}

In Section~\ref{sec:constraints},
we established that for a computable potential inducing a uniform model,
the ground states have a set complexity $\Pi_2$.
Such sets can in particular be described as accumulation sets of computable sequences.
In Section~\ref{sec:chaos},
we established that the class of simulating tilesets from Section~\ref{sec:robinson}
always induces a uniform model
(stable if $\G(\infty)$ is a singleton, uniformly chaotic otherwise).

In this final section, our goal is to use this class of simulating tilesets
to realise any connected $\Pi_2$ set, through a corresponding computable sequence,
as a set of ground states.
Such matters are usually studied in the context of computability, not complexity,
the question revolving around whether it \emph{can} be done,
not if it can be done \emph{rapidly}.
Hence, we will use several slowdown arguments,
to guarantee that we can not only get increasingly close to the target measures,
but also that they can be computed fast-enough, by a well-behaved Turing machine.
After this, we will at last conclude with the main result of the article.

\subsection{Slowdown Through Repetition}

Remark that we define the weak-$*$ distance $d^*$ on $\M\left(\{\pmo\}^\N\right)$
using the same formula as in Section~\ref{sec:chaos:distance} for $\M_\sigma\left(\Omega_\A\right)$,
so that the idea of Lemma~\ref{lem:HighCondProbLowDist} still applies here.

In a direct sense, the measures on words \emph{simulated} by the Turing machine $M$
are the measures $m_k$ forced within $\symb{B}$ squares.
However, as proven in Theorem~\ref{thm:Gibbs-accumulation},
the measure \emph{induced} by $M$,
the behaviour we typically observe in Gibbs measures at temperatures in the range $T_k$,
is $w_k$ (which is a convex combination of the computed measures)
and we have no guarantee that $d^*\left(m_k,w_k\right)\to 0$.
However, by adding repetitions of each instance of $m_k$, we obtain the following proposition.

\begin{lemma} \label{lem:connected-accumulation}
Assume that $d^*\left(m_k,m_{k+1}\right)\to 0$.
Then $\acc{\left[m_k,m_{k+1}\right]}=\acc{m_k}$.
\end{lemma}

\begin{proposition} \label{prop:repetition-slowdown}
Let $M$ be a well-behaved Turing machine, simulating a sequence $\left(m_k\right)$,
and such that $d\left(m_k,m_{k+1}\right)\to 0$.
Then there exists a well-behaved machine $M'$,
simulating $\left(m_k'\right)$ and inducing $\left(w_k'\right)$,
such that $\acc{m_k}=\acc{w_k'}$.

\begin{proof}
To do so, consider the sequence of scales $s_j=2^{2^j}$,
and define $j(k):=\left\lfloor\log_2\left(\left\lfloor\log_2(k)\right\rfloor\right)\right\rfloor$,
such that $s_j\leq k < s_{j+1}$.
We want a machine $M'$ for which the induced measures $\left(m_k'\right)$
satisfy $m_k'=m_{j(k)}$ (on words of length $\bread(k)$,
by adding $\bread(k)-\bread(j)$ bits $\po$ at the end).
If we have an input seed $s$, then $\abs{s}=k$ so $s$ can act as a read-only unary input to compute $j$,
which can be done in time $o\left(2^{3^k}\right)$ according to Appendix~\ref{appendix:unary}.
After this, we use $s_{\leq j}$ (the first $j$ bits of the seed $s$) to compute $M(s_{\leq j})$,
and finally return $M'(s):=M \left(s_{\leq j}\right)\cdot\left(\po\right)^{\bread(k)-\bread(j)}$.
The machine $M'$ thus defined is indeed well-behaved and adds repetitions
to the sequence $\left(m_k\right)$.

With these repetitions, we now have $d^*\left(m_k,w_{s_{k+1}}'\right)\to 0$.
Indeed, using the same computations as in Proposition~\ref{prop:measure-weights},
we conclude that on the first $\bread(j)$ bits we have the following measure inequality:
\[
w_{j,s_{j+1}}'\geq \left(1-
\exp\left(-\frac{1}{4}\left(\frac{s_{j+1}}{2^{j+1}}-s_j\right)\right)\right)m_j \eqnsp .
\]
As $\frac{s_{j+1}}{2^{j+1}}-s_j=s_j\left(2^{2^j-j}-1\right)\to\infty$,
this gives us a high-probability event where $w_{j,s_{j+1}}'$ behaves as $m_j$,
thus using Lemma~\ref{lem:HighCondProbLowDist} we conclude that their distance goes to $0$.
Thus, in the asymptotic regime $m=j(k)$ in Theorem~\ref{thm:Gibbs-accumulation}
(and $w_k':=w_{j(k),k}'$),
$\acc[j\to\infty]{m_j}=\acc[j\to\infty]{w_{s_j}'}\subseteq\acc[k\to\infty]{w_k}$.

Furthermore, the measures $\left(w_k'\right)_{s_j\leq k\leq s_{j+1}}$ are all in the interval
$\left[w_{s_j}',m_j\right]$ by induction,
and $\left[w_{s_j}',m_j\right]\approx\left[m_{j-1},m_j\right]$,
so $\acc[k\to\infty]{w_k'}\subseteq \acc[j\to\infty]{\left[m_j,m_{j+1}\right]}$.

By hypothesis, $d\left(m_k,m_{k+1}\right)\to 0$ so the previous lemma applies,
the chain of inclusions collapses into equalities,
thus $\acc[k\to\infty]{w_k'}=\acc[j\to\infty]{m_j}$.
\end{proof}
\end{proposition}

\begin{remark}
If we remove the $d\left(m_k,m_{k+1}\right)\to 0$ hypothesis from the previous proposition,
then we conclude instead that $\acc{w_k'}=\acc{\left[m_k,m_{k+1}\right]}$,
because as $j\to\infty$,
the block $\left(w_k'\right)_{s_j\leq k\leq s_{j+1}}$ goes through
increasingly more tiny steps across $\left[w_{s_j}',m_j\right]\approx\left[m_{j-1},m_j\right]$.
\end{remark}

In particular, as long as $\left(m_j\right)$ does not converge,
$M'$ induces a uniformly chaotic model.

\begin{theorem} \label{thm:existence-uniform-chaos}
There exists a (computable, associated to a well-behaved Turing machine) potential
inducing a uniformly chaotic model.

\begin{proof}
Consider the sequence of words $u_{2n}=\left(\po\right)^{\bread(2n)}$
and $u_{2n+1}=\left(\mo\right)^{\bread(2n+1)}$.
By simply counting the parity of $\abs{s}$ in linear time and then simply
writing this parity bit $\bread(\abs{s})$ times,
we obtain a well-behaved machine $M$ that simulates $m_k=\delta_{u_k}$.
Notably $m_{2k}\to\delta_+:=\delta_{(\po)^\N}$ and $m_{2k+1}\to \delta_-:=\delta_{(\mo)^\N}$,
and $d^*\left(\delta_-,\delta_+\right)>1$,
so we actually fall out of the scope of the previous proposition.
However, by following the argument of the remark,
we conclude that for the system induced by the machine $M'$,
we have a uniform accumulation (according to Theorem~\ref{thm:Gibbs-accumulation})
to the non-singleton set $\G(\infty)\cong \left[\delta_-,\delta_+\right]$.
\end{proof}
\end{theorem}

In some sense, this proof transposes the initial idea of Chazottes and Hochman~\cite{ChaHo10},
\ie forcing the system to oscillate between two distant incompatible states,
in our own framework.

\subsection{Complements on Computable Analysis for 
\texorpdfstring{$\M\left(\{\pmo\}^\N\right)$}{M(\{±1\}\^{}ℕ)}}

In Section~\ref{sec:constraints},
we established that any computable potential that induces a uniform model
has a $\Pi_2$-computable compact and connected set of ground states $\G(\infty)$.
Then, we established the existence of a class of computable potentials
$\left(\phi_M\right)_{M\text{ well-behaved}}$ that induce uniform models,
with their ground states as a connected subset
of $\M_\sigma\left(X_{\F_0}\right)\cong \M\left(\{\pmo\}^\N\right)$.

To obtain a kind of completeness result,
we would like to justify that \emph{any} connected $\Pi_2$ subset of $\M_\sigma\left(X_{\F_0}\right)$
can be realised as a set of ground states.
However, by using well-behaved Turing machines,
what we may be able to control is the limit set $\acc{m_j}\subseteq\M\left(\{\pmo\}^\N\right)$.

In Corollary~\ref{cor:injection-pi2-equivalence}
it is established that, whenever two computable spaces are in a bicomputable bijection,
they have the same classes of (connected) $\Pi_2$-computable compact subsets.
Hence, if the map $\gamma:\M\left(\{\pmo\}^\N\right)\to \M_\sigma\left(X_{\F_0}\right)$
from Theorem~\ref{thm:affine-bijection} is bicomputable,
then we will indeed be able to focus on connected $\Pi_2$ sets of measures on binary strings.

First, to avoid ambiguity,
let us make our notion of computation on $\M\left(\{\pmo\}^\N\right)$ explicit.
We use the dyadic base $\dense=\bigcup\dense_k$,
with $\dense_k$ the combinations of Dirac measures 
$\left(\delta_{w\cdot(\po)^\infty}\right)_{w\in\{\pmo\}^{\bread(k)}}$,
with weights $\frac{i}{2^k}$.
We have a monotonous inclusion $\dense_k\subseteq \dense_{k+1}$ and,
as both the length of the words $\bread(k)$ and the power $k$ go to infinity,
any dyadic measure is indeed in the sets $\dense_k$ after a rank.
Thus defined, $\left(\M\left(\{\pmo\}^\N\right),d^*,\dense\right)$ is a computable space,
and is (locally) recursively compact.

\begin{proposition} \label{prop:bicomputable-map}
Let $M$ be a well-behaved machine.
The map $\gamma:\M\left(\{\pmo\}^\N\right)\to \M_\sigma\left(\Omega_{\A_{6(M)}}\right)$
from Theorem~\ref{thm:affine-bijection} is computable.
Furthermore, we have $\gamma':\M_\sigma\left(\Omega_{\A_{6(M)}}\right)\to\M\left(\{\pmo\}^\N\right)$
a computable pseudo-inverse, such that $\gamma'\circ\gamma$ is the identity map.

\begin{proof}
As is often the case for computability, doing an exhaustive formal proof
would be quite long without giving much insight, so we will focus on the key ideas here.

Notice that the distances $d^*$ are vector norm, and $\gamma$ is convex,
so we just need to computably map Dirac measures
$\delta_{w\cdot(\po)^\infty}\in\dense_k$ (with $k=\abs{w}$)
to periodic points $f(w,r)\in\per$
(such that $d^*\left(\gamma\left(\delta_{w\cdot(\po)^\infty}\right),f(w,r)\right)=o(r)$
uniformly in $w$).
Then, we extend $f(\cdot,r)$ by convexity on each set $\dense_k$.
This will give us dyadic combinations of periodic points instead of elements of $\per$ directly,
but we can then computably send those onto elements of $\per$ at distance $r$.
Now, to compute $f(w,r)$, we just need to return a periodic grid of $\symb{F}$ $(2\abs{w}+1)$-macro-tiles
encoding the signal $w$.

For the other direction, we want $\gamma'$ to be a computable ``projection''
onto $\M_\sigma\left(X_{\F_0}\right)$.
To do so, starting from a periodic point $p\in\per$, with a target precision $\epsilon$ in mind,
we first need to find a value of $k$ high-enough such that the $\epsilon$-neighbourhood of $\dense_k$
covers $\M\left(\{\pmo\}^\N\right)$.
Then, for each element $x\in\dense_k$,
we can (uniformly) compute $d^*(\gamma(x),p)$,
so we just need to find an approximate minimiser of this distance in $\dense_k$
and return it as an approximation of $\gamma'(p)$.
\end{proof}
\end{proposition}

The proposition also applies if we consider the common alphabet $\A_0$ and subshift $X_{\F_0}$
instead of those specific to a Turing machine $M$.

Now, using Corollary~\ref{cor:injection-pi2-equivalence},
we can indeed equivalently consider connected $\Pi_2$-computable
subsets of $\M\left(\{\pmo\}^\N\right)$,
instead of connected $\Pi_2$-computable subsets of $\M_\sigma\left(X_{\F_{6(M)}}\right)$
(computable as subsets of $\M_\sigma\left(\Omega_{\A_{6(M)}}\right)$).
Using Proposition~\ref{prop:pi2-connected-accumulation}, in both cases,
we know that these connected $\Pi_2$-computable compact sets $K$ can be realised as
$K=\acc{x_n}$ for a computable $\left(x_n\right)$ such that $d^*\left(x_n,x_{n+1}\right)\to 0$.

In order to realise any $\Pi_2$-computable set,
the only gap left is one of complexity.
Indeed, we can realise a connected $\Pi_2$ subset $X\subseteq\M\left(\{\pmo\}^\N\right)$
as an accumulation set of a computable sequence $\left(m_k\right)\in\dense^\N$,
but we have not explained yet why this sequence may be simulated by a well-behaved machine $M$.
That will be the topic of the next subsection.

\subsection{Faster Computations Through Inductive Repetitions}

Another subtlety we glossed over is the difference between a measure $m$ being
simulated into $\symb{B}$ squares using a well-behaved machine $M$,
and $m$ being explicitly computed, encoded as an element of $\dense$.

To encode $m\in\dense_k$, \ie a dyadic measure on binary words of length $\bread(k)$
with precision $\frac{1}{2^k}$,
we can use the lexicographical order on
$\{\pmo\}^{\bread(k)}\equiv\left\llbracket 0,2^{\bread(k)}-1\right\rrbracket$
(with $2^{\bread(k)}\approx k$),
and write down the corresponding sequence of integer weights in $\left\llbracket0,2^k\right\rrbracket$
(using a binary string of length $k+1$ bits for each).
Hence, we can encode $m\in\dense_k$ using a string of length $\abs{m}\approx k^2$.

Then, being given this binary string $m\in\dense_k$ and a seed $s$ (with $\abs{s}=k$),
we can simply compute the partial sums of weights to compute a corresponding signal prefix $M(s)$,
in polynomial time (thus a $o\left(2^{3^k}\right)$),
thus simulating $m$ when $s$ is uniformly distributed in $\{\pmo\}^k$.

\begin{proposition} \label{prop:word-accumulation}
Let $X\subseteq \M\left(\{\pmo\}^\N\right)$ be a connected $\Pi_2$-computable compact set.
Then there is a well-behaved Turing machine $M$ that simulates measures $\left(m_k\right)$
(in $\symb{B}$ squares) such that $X=\acc{m_k}$ and $d^*\left(m_k,m_{k+1}\right)\to 0$.

\begin{proof}
As $X$ is a connected $\Pi_2$ set, Proposition~\ref{prop:pi2-connected-accumulation} tells us that
there is a Turing machine $T:\N\to\dense$ such that $X=\acc{T(k)}$
and $d^*\left(T(k),T(k+1)\right)\to 0$.

Without loss of generality, we may assume that $T(k)\in\dense_k$.
Indeed, if $T(k)\in\dense_j$ with $j< k$, as $\dense_j\subseteq \dense_k$,
we just have to translate its encoding.
If $j>k$ then we can computably ``project'' $T(k)\in\dense_j$ back into $\dense_k$,
by first summing the weights of words in $\{\pmo\}^{\bread(j)}$
with the same prefix in $\{\pmo\}^{\bread(k)}$ and then finding the nearest distribution
with precision $\frac{1}{2^k}$ instead of $\frac{1}{2^j}$).
This projection to a measure in $\dense_k$ does not affect the accumulation set $\acc{T(k)}$
by density of $\dense$.
Remark also that the both the re-encoding and the projection can be done in polynomial time.

Then, let $T'(k)$ be the machine defined inductively,
initialised at $T'(0)=\left(\delta_{(\po)^\infty},0\right)$
(that returns the only measure of $\dense_0$).
To compute $T'(k+1)$, we begin by computing $T'(k)=(m,j)$,
and then simulate $T(j+1)$ for $k$ steps.
If the new computation terminates, we return $(T(j+1),j+1)$ and else $T'(k)$.
This way, $T'$ induces the same accumulation set $T$ does (with distances going to $0$),
by adding repetitions to the sequence $(T(k))$ but still visiting at least one time each element,
and now with a time complexity $O\left(k^2\right)$.

Thence, we can define $M$ that, given a read-only binary seed $s$,
first computes $k=\abs{s}$, then $T'(k)\in\dense_j$ (with $j\leq k$), re-encodes it into $\dense_k$,
and finally computes the corresponding signal-prefix $M(s)$ of length $\bread(k)$.
It follows that $M$ has a time complexity $o\left(2^{3^{\abs{s}}}\right)$,
\ie it is a well-behaved machine that simulates the sequence $\left(m_k:=T'(k)\right)_{k\in\N}$
such that $\acc{m_k}=\acc{T(k)}=X$ and $d^*\left(m_k,m_{k+1}\right)\to 0$.
\end{proof}
\end{proposition}

\begin{theorem} \label{thm:pi2-lower-bound}
Let $K\subseteq\M_\sigma\left(X_{\F_0}\right)\cong\M\left(\{\pmo\}^\N\right)$
be a connected $\Pi_2$-computable compact subset of $\M_\sigma\left(\Omega_{\A_0}\right)$.
Then there exists a well-behaved Turing machine $M$ such that,
on the extended alphabet $\A_{6(M)}\supseteq\A_0$,
for the associated potential $\phi_M:\Omega_{\A_{6(M)}}\to \R$,
we have $K=\G(\infty)$.

\begin{proof}
First, by using the bicomputable map from Proposition~\ref{prop:bicomputable-map},
Corollary~\ref{cor:injection-pi2-equivalence}
tells us that the set $X:=\gamma'(K)\subseteq\M\left(\{\pmo\}^\N\right)$
is also a connected $\Pi_2$-computable compact.
Then, Proposition~\ref{prop:word-accumulation} says that
$X=\acc{m_j}$ is the accumulation of a sequence $\left(m_j\right)$ simulated by
a well-behaved machine $M$, such that $d^*\left(m_j,m_{j+1}\right)\to 0$.
Likewise, Proposition~\ref{prop:repetition-slowdown} allows us to obtain the same statement
but with the sequence of empirically induced measures $\left(w_k\right)$ instead.
Thus, with $\lambda_k=\gamma\left(w_k\right)$,
we have $K=\gamma(X)=\acc{\lambda_k}$.
Finally, Theorem~\ref{thm:Gibbs-accumulation} from the previous tells us that,
using the associated potential, we have $\G(\infty)=\acc{\lambda_k}$, which concludes the proof.
\end{proof}
\end{theorem}

\subsection{Undecidability of Chaoticity}

Let us conclude this section, and the whole article, by a small application of the previous framework,
that of \emph{undecidability} of the question of chaoticity.
By the question of chaoticity, we here mean the decision problem of whether
a given computable potential induces a chaotic model, as defined in Section~\ref{sec:framework:chaos}.
Among uniform models, such as our class of potentials induced by well-behaved machines,
deciding if the system is chaotic is equivalent to deciding if $\G(\infty)$ is not a singleton.

\begin{theorem}
\label{thm:chaos:complexity}
The question of chaoticity is $\Sigma_3$-complete. 
\begin{proof}
Let us begin with the upper bound.
Consider $\phi:\Omega_\A\to\R$ a computable potential.
A ground state $\mu\in\G(\infty)$ is stable if and only if we have $d^*\left(\mu,\G(\beta)\right)\to 0$.
Likewise, when $\mu\notin\G(\infty)$ we clearly have $d^*\left(\mu,\G(\beta)\right)\not\to 0$.
Thus, a model is \emph{chaotic} if and only if, for any measure $\mu$, we have
$D(\mu):=\varlimsup_{\beta\to\infty} d^*\left(\mu,\G(\beta)\right) > 0$.
The maps $\mu\mapsto d^*\left(\mu,\G(\beta)\right)$ are all $1$-Lipschitz, thus so is $D$.
By compactness of the space of measures, if follows that the model is chaotic
\emph{iff} $\inf_{\mu\in\M_\sigma\left(\Omega_\A\right)} D(\mu)>0$,
and we can then restrict this infimum to the dense family $\per$:
\[
\begin{array}{rl}
& \exists \epsilon>0,\forall x\in\per,\forall \beta_0>0,\exists \beta\geq \beta_0, 
\G(\beta)\cap\overline{B(x,\epsilon)}=\emptyset \\
\Leftrightarrow &\exists \epsilon\in\Q^{+*},\forall x\in\per,\forall \beta_0\in\N,
\exists \beta',\beta''\in\Q\cap[\beta,+\infty[, 
\G([\beta',\beta''])\cap\overline{B(x,\epsilon)}=\emptyset  \eqnsp . \\
\end{array}
\]
The equivalent formula comes from the fact that we can replace $\epsilon$
and $\beta_0$ by countable parameters by monotonicity,
and $\G(\beta)\cap\overline{B(x,\epsilon)}=\emptyset$ if and only if
it holds on a neighbourhood of $\beta$ by Lemma~\ref{lem:weak-continuity}.
Since by Proposition~\ref{prop:semi-decidable}
it is semi-decidable to know if $\G([\beta',\beta''])\cap\overline{B(x,\epsilon)}=\emptyset $,
we deduce that decide the chaoticity is $\Sigma_3$-computable.

Now, to prove that the problem is $\Sigma_3$-complete, 
we will make a computable reduction from the \emph{cofinality} problem 
(\ie whether $M$ does not halt on \emph{finitely} many inputs,
which is a known $\Sigma_3$-complete problem~\cite[Theorem 4.3.3]{Soa16}) to chaoticity.

Let $M$ a Turing machine (which takes encoded integers as inputs),
and denote $\tau(k)\in\overline{\N}$ the time it takes $M$ to compute $M(k)$.
First, fix a computable bijective map $f:\N\to\N^3$ such that each coordinate stays bounded by $i$.
Define the computable sequence of measure $(m_i)\in\dense^\N$ such that, with $f(i)=(k,\ell,t)$:
\[
m_i = \left\{
\begin{aligned}
&\frac{1}{2^k}\delta_{(\mo)^i(\po)^\infty}+\left(1-\frac{1}{2^k}\right)\delta_{(\po)^\infty}
&&\text{ if $\max_{i\in\llbracket k,k+\ell\rrbracket}\tau(i)=t$,}\\
&\delta_{(\po)^\infty}&&\text{ else.}
\end{aligned}
\right.
\]
Then, we can replace each measure $m_i$ by a path from $\delta_{(\po)^\infty}$ to $m_i$ and forth,
using incremental steps of weight $\frac{1}{2^i}$
(which divides $\frac{1}{2^k}$ as $k(i)\leq i$) for the case
$m_i=\frac{1}{2^k}\delta_{(\mo)^i(\po)^\infty}+\left(1-\frac{1}{2^k}\right)\delta_{(\po)^\infty}$.
This defines a computable sequence of measures $\left(m_j'\right)$
such that $d\left(m_j',m_{j+1}'\right)\to 0$,
and so that the accumulation set $K:=\acc{m_j'}=\acc{\left[\delta_{(\po)^\infty},m_i\right]}
\subseteq\M_\sigma\left(\{\pmo\}^\N\right)$ is a connected $\Pi_2$-computable compact.
Notably, we always have $\delta_{(\po)^\infty}\in K$.

If $M$ has a cofinite support, then after some rank $k$ it always halt.
Thus, for any interval length $\ell$,
with $i_\ell:=f^{-1}\left(k,\ell,\max_{i\in\llbracket k,k+\ell \rrbracket}\tau(i)\right)
\underset{\ell\to\infty}{\longrightarrow}\infty$, we have
$m_{i_\ell}\to \frac{1}{2^k}\delta_{(\mo)^\infty}+\left(1-\frac{1}{2^k}\right)\delta_{(\po)^\infty}$,
so $K$ is not a singleton.

Conversely, assume $M$ does not have a cofinite support.
Then, for any choice of $j$ we have only finitely many intervals $\llbracket k,k+l\rrbracket$
such that $k\leq j$ and $M$ halts on all the entries.
After some scale $i$, we thus have $m_i \geq \left(1-\frac{1}{2^j}\right)\delta_{(\po)^\infty}$.
It follows that $m_i\to\delta_{(\po)^\infty}$ and $K$ is a singleton.

The process that maps $M$ to the sequence $\left(m_j'\right)$,
and then to the corresponding simulating potential $\phi_M$ 
using Proposition~\ref{prop:repetition-slowdown},
is thus a computable reduction from the cofinality problem to the question of whether
$K$ is \emph{not} a singleton, \ie of whether the corresponding (uniform) model is chaotic.
\end{proof}
\end{theorem}

\appendix

\newpage

\section{Unary Computing} \label{appendix:unary}

In the construction of the tileset in Section~\ref{sec:robinson},
we used several Turing machines to do computations on integers
with a \emph{unary coding} with a good-enough time complexity.
Such machines are relatively easy to describe, through a kind of ``stick-based'' computing,
but precisely because of this are not well documented within the literature
except as exercises.

This appendix aims at giving a short yet clear introduction to how each such machine we used works,
and its time complexity.

\begin{proposition}[Integer Comparison]
Given two integers $i,j\in\N$, testing $i=j$ (resp. $i<j$) can be done in exactly $\min(i,j)+1$ steps.
\end{proposition}

\begin{proposition}[Power Checking]
Let $b\geq 2$ and $n$ be two integers.
We want here to decide whether $n=b^k$ for some power $k\in\N$.
If we assume $b$ is fixed and hardcoded within the machine,
there exists a Turing machine $M_b$ with an asymptotic time complexity of order $n\log_b(n)$.

\begin{proof}
To do so, we use a Turing machine that does \emph{back-and-forth} passes on the area of length $n$.
At each pass, the machine erases the first $b-1$ sticks out of $b$, cyclically.
At the end of a pass, if we have only seen one stick in total, we accept $n$.
Else, if we have not seen a multiple of $b$ sticks on the whole line
(\emph{i.e.} we are in the middle of a cycle of erasure), we reject it.
In the remaining case, we were able to divide by $b$ the number of sticks,
and go on for another pass in the other direction.

Because each pass (which from end to end takes about $n$ steps) divides the number of sticks by $b$,
we can do at most $\log_b(n)$ passes before we trigger a halting condition,
hence the announced complexity.
\end{proof}
\end{proposition}

Consider now a really similar matter,
but instead of studying a decision problem we want to compute a function.
In order to able to compose functions, we suppose here that $M$ terminates its computation
only once the reading head reaches the starting position.

\begin{proposition}[Rounded Logarithm]
Let $b\geq 2$ and $n$ be two integers.
We want here to compute $\left\lfloor \log_b(n)\right\rfloor$.
If we assume $b$ is fixed and hardcoded within the machine,
there exists a Turing machine $M_b$ that computes this function
with an asymptotic time complexity of order $n\log_b(n)$.

\begin{proof}
Just like for the Power Checking, we do several passes
of length (at most) $n$,
so that the number of sticks on the tape goes
from $n_t$ to $n_{t+1}=\left\lfloor \frac{n_t}{b}\right\rfloor$,
and for each pass such that $n_t>1$,
we add a stick to a second coordinate during the next pass.
We then simply conclude by erasing what remains on the first coordinate.
\end{proof}
\end{proposition}

\begin{proposition}[Nearest Power]
Let $b\geq 2$ and $n$ be two integers.
We want here to compute $b^{\left\lfloor \log_b(n)\right\rfloor}$,
the biggest power of $b$ that is lower or equal to $n$.
If we assume $b$ is fixed and hardcoded within the machine,
there exists a Turing machine $M_b$ that computes this function
with an asymptotic time complexity of order $2n\log_b(n)$.

\begin{proof}
We follow here the same general idea,
but we do instead a full back-and-forth for each big step.
At each such step, we begin by keeping the first stick
and rewriting the following $b-1$ we encounter using another symbol \texttt{\$}.
If at the end of the pass we found only one stick,
then while going back we revert to its original value every symbol \texttt{\$} we encounter.
Else, if we did not encounter a multiple of $b$ sticks during the pass,
then while going back, we erase entirely every symbol \texttt{\$} we encounter
until we reach a stick which we also erase, and then simply go for another back-and-forth.

To put it in another way, at the $t$-th step we check the $t$-th digit
expansion of $n$ in base $b$, and set it to $0$ if needed.
\end{proof}
\end{proposition}

\begin{remark}[Speeding-up the Machine for Small Inputs]
Assume that the unary decision machine $M$ has an complexity of $f(n)$ or less after a rank.
Then we have a machine $M'$ for the same problem that \emph{always} terminates in time $f(n)+2n$.
To do so, consider $X_A$ (resp. $X_R$) the finite set of inputs for which $M$ takes
more than $f(n)$ steps to accept (resp. reject).
We can hardcode $X_A \sqcup X_R$ into a deterministic automaton.
Then we can run the automaton first, which takes exactly $n$ steps,
and if $n\notin X_A \sqcup X_R$ then we simply go back at the start of the input in $n$ steps
and run $M$ on it, which takes $f(n)$ supplementary steps at most,
hence $2n+f(n)$ steps in total in this case.

Following the same track of thought, if $M$ takes at most $f(n)$ steps to decide on $n$,
and we have $g$ such $f(n)+2n\leq g(n)$ after a rank (and $g(n)\geq n$ always holds),
then by defining $X_A\sqcup X_R$ as the set of inputs for which $f(n)+2n>g(n)$,
we can obtain a machine $M'$ that always terminates in $g$ steps at most.

If the machine $M$ does not answer a \emph{decision problem} but \emph{computes} a function instead,
still in $f(n)$ steps at most, the general idea is the same
with the set $X$ such that $f(n)+2n>g(n)$ (there is no distinction between accepting and rejecting now)
but we must now assume that the inequality $g(n)\geq 2\max(n,M(n))$ always hold instead.
Indeed, if $n\notin X$, we take $2n+f(n)$ steps at most just like before,
but in the case $n\in X$,
it must then either add some sticks on the tape (if $M(n)>n$) and
then go back at the beginning in $M(n)$ steps,
hence $2M(n)$ steps in total,
or erase the last sticks and then go back (if $M(n)<n$) which takes $2n$ steps in total,
hence the condition $g\geq2\max(M,n)$.

In both situations, the lower bound on $g$ represents the minimal time
needed to be able to read the input,
and if needed to write the output.
\end{remark}

\newpage
\section{Notations} \label{appendix:notations}

\subsection*{Section~\ref{sec:framework}:}

\begin{tabular}{p{\indexalign}p{\pagealign}}
$I\Subset\Z^d$ & finite window \\
$I_n=\llbracket 0,n-1\rrbracket^d$ & $n$-square window \\
$\Omega_\A=\A^{\Z^d}$ & full-shift on a finite alphabet $\A$ \\
$[w]$ & cylinder set with a base $w\in\A^I$ on a window $I$ \\
$\M(X)$ & set of probability measures on a space $X$ \\
$X_\F$ & subshift of finite type induced by forbidden patterns $\F$ \\
$\M_\sigma(X)$ & shift-invariant probability measures \\
$\Phi$ & finite-range interaction with range $r$ \\
$E_I$ & energy content of $I$ induced by $\Phi$ \\
$\phi:\Omega_\A\to\R$ & continuous (or computable) potential \\
$h(\mu)$ & Kolmogorov-Sinai entropy \\
$p_\mu(\beta)=h(\mu)-\beta\mu(\phi)$ & pressure associated to $\phi$ \\
$p(\beta)=\max_\mu p_\mu(\beta)$ & topological pressure \\
$\G(\beta)$ & shift-invariant Gibbs measures at inverse temperature $\beta$ \\
$\acc{\cdots}$ & set of accumulation points associated to $\left(\cdots\right)$ \\
$\G(\infty)=\acc{\G(\beta)}$ & asymptotic ground states as $\beta\to\infty$ \\
\end{tabular}

\subsection*{Section~\ref{sec:constraints}:}

\begin{tabular}{p{\indexalign}p{\pagealign}}
$\dense$ & countable dense basis of a metric set $(X,d)$ \\
$K$ & compact subset of $X$ \\
$\mathcal{N}(K)$ & intersecting neighbourhood set of $K$ \\
$\Pi_k$ & complexity class in the arithmetical hierarchy \\
$\per$ & countable dense basis of periodic points for $\M_\sigma\left(\Omega_\A\right)$ \\
$\G\left(\left[\beta^-,\beta^+\right]\right)$ & union of Gibbs measures on a temperature interval \\
\end{tabular}

\subsection*{Section~\ref{sec:uniform-markers}:}

\begin{tabular}{p{\indexalign}p{\pagealign}}
$Z_\Phi$ & null-energy ground configurations \\
$G_I$ & ground patterns on $I$, such that $E_I(x)=0$ \\
$\alpha$ & minimal positive energy of forbidden patterns \\
$Q$ & marker set on a window $I$ with margin factor $\tau$ \\
$\mu_Q$ & conditional distribution induced by $\mu$ on $Q$ \\
$\lambda_Q$ & uniform distribution on $Q$ \\
\end{tabular}

\subsection*{Section~\ref{sec:robinson}:}

\begin{tabular}{p{\indexalign}p{\pagealign}}
$\A_1=\A_{\mathrm{Robinson}}$ &
Robinson structural layer, with forbidden patterns $\F_1$ \\
$\A_2\subseteq\A_1\times\A_{\mathrm{phase}}$ &
freezing structure, with $\A_{\mathrm{phase}}=\left\{\symb{F},\symb{B},\symb{H}\right\}$
and forbidden patterns $\F_2$ \\
$\A_3\subseteq\A_2\times\A_{\mathrm{scale}}$ & structure with Blockable scales,
with forbidden patterns $\F_3$ \\
$\A_4\subseteq\A_3\times\A_{\mathrm{odometer}}$ &
odometer structure at Blockable scales, with forbidden patterns $\F_4$ \\
$\A_5\subseteq\A_4\times\A_{\mathrm{signal}}$ &
structure encoding binary signals in $\symb{F}$ areas, with patterns $\F_5$ \\
$\A_{6(M)}\subseteq\A_5\times\A_{\mathrm{simul}(M)}$ &
structure simulating a machine $M$ in $\symb{B}$ squares,
with patterns $\F_{6(M)}$ \\
$\A_0\subset \A_{6(M)}$ & common alphabet for all machines $M$,
with $\F_0$ the projection of $\F_{6(M)}$\\
$Q_k$ & set of $k$-markers, $k$-th scale of simulation \\
$Q_k^* $ & set of Hot, Frozen or Blocking $k$-markers
(depending on $*\in\A_{\mathrm{phase}}$) \\
$\widetilde{Q_k^*}$ & set of tuples of words encoded in a $*$ $k$-marker
(with $*\in\left\{\symb{H},\symb{B}\right\}$) \\
$l_n=2^n-1$ & side-length of a Robinson $n$-macro-tile \\
$N_k=3^k$ & step of computation of a $k$-marker \\
$n_k=2N_k+1$ & scale of the Robinson macro-tile in a $k$-marker \\
$\perim_k=4\left(l_{n_k}-1\right)$ & perimeter of a $k$-marker \\
$t_k=2^{\left\lfloor \log_2\left(\left\lfloor \log_2(k)\right\rfloor\right)\right\rfloor}-1$ &
period of the odometer at the scale of $k$-markers \\
$\btot(k)=3^k-1$ & number of bits encoded in
the central square of a $\symb{B}$ $k$-marker \\
$\bread(k)=\left\lfloor\log_2(k)\right\rfloor$ &
number of controlled bits in a $\symb{B}$ square, length of the output of $M$ \\
\end{tabular}

\subsection*{Section~\ref{sec:chaos}:}

\begin{tabular}{p{\indexalign}p{\pagealign}}
$p_k^\symb{H}= \lambda_{Q_k}\left(Q_k^\symb{H}\right)$ &
proportion of $\symb{H}$ markers among $k$-markers \\
$m_l$ & distribution on strings of length $\bread(l)$ simulated by $M$ in a $\symb{B}$ $l$-marker \\
$w_{l,k}$ & empirical distribution of strings of length $\bread(l)$ in a $\symb{H}$ $k$-marker \\
\end{tabular}

\newpage
\phantomsection
\addcontentsline{toc}{section}{References}

\def\MR#1{\href{http://www.ams.org/mathscinet-getitem?mr=#1}{MR#1}}
\bibliographystyle{amsplain-nodash}
\bibliography{main}

\end{document}